\documentclass[rmp,aps,twocolumn,epsfig,eqsecnum,floatfix,amsmath,amssymb,superscriptaddress,bibnotes]{revtex4}

\usepackage{graphicx}
\usepackage{dcolumn}
\usepackage{bm}
\usepackage{epsfig}

\begin{document}
\ifx\href\undefined\else\hypersetup{linktocpage=true}\fi 
\bibliographystyle{apsrev}


\title{Angle-Resolved Photoemission Spectroscopy on Electronic Structure and
Electron-Phonon Coupling in Cuprate Superconductors
\\}

\author{X. J. Zhou}
\affiliation{Dept. of Physics, Applied Physics and Stanford
Synchrotron Radiation Laboratory, Stanford University,  Stanford,
CA 94305} \affiliation{Advanced Light Source, Lawrence Berkeley
National Lab, Berkeley, CA 94720} \affiliation{State Key
Laboratory for Superconductivity, Beijing National Laboratory for
Condensed Matter Physics and Institute of Physics, Chinese Academy
of Sciences, Beijing 100080, China}

\author{T. Cuk}
\affiliation{Dept. of Physics, Applied Physics and Stanford
Synchrotron Radiation Laboratory, Stanford University,  Stanford, CA
94305}

\author{T. Devereaux}
\affiliation{Department of Physics, University of Waterloo, Ontario,
Canada N2L 3GI}

\author{N. Nagaosa}
\affiliation{CREST, Department of Applied Physics, University of
Tokyo, Bunkyo-ku, Tokyo 113-8656, Japan}

\author{Z.-X. Shen}
\affiliation{Dept. of Physics, Applied Physics and Stanford
Synchrotron Radiation Laboratory, Stanford University,  Stanford, CA
94305}




\maketitle

\tableofcontents

\section{Introduction}

In addition to the record high superconducting transition
temperature (T$_c$), high temperature cuprate
superconductors\cite{BednorzMuller,MKWuYBCO} are characterized by
their unusual superconducting properties below T$_c$, and anomalous
normal state properties above T$_c$.  In the superconducting state,
although it has long been realized that superconductivity still
involves Cooper pairs\cite{Gough}, as in the traditional BCS
theory\cite{Schrieffer,McMillan,Marsiglio}, the experimentally
determined {\it d}$-$wave pairing\cite{Tsuei} is different from the
usual {\it s}$-$wave pairing found in conventional
superconductors\cite{Scalapino,DPines}. The identification of the
pairing mechanism in cuprate superconductors remains an outstanding
issue\cite{MagneticModePairing}. The normal state properties,
particularly in the underdoped region, have been found to be at odd
with conventional metals which is usually described by Fermi liquid
theory; instead, the normal state at optimal doping fits better with
the marginal Fermi liquid phenomenology\cite{MFLVarma}. Most notable
is the observation of the pseudogap state in the underdoped region
above T$_c$ \cite{Pseudogap}.  As in other strongly correlated
electrons systems, these unusual properties stem from the interplay
between electronic, magnetic, lattice and orbital degrees of
freedom. Understanding the microscopic process involved in these
materials and the interaction of electrons with other entities is
essential to understand the mechanism of high temperature
superconductivity.

Since the discovery of high-T$_c$ superconductivity in
cuprates\cite{BednorzMuller}, angle-resolved photoemission
spectroscopy (ARPES) has provided key experimental insights in
revealing the electronic structure of high temperature
superconductors\cite{ShenDessau,DamascelliReview,CampuzanoReview}.
These include, among others, the earliest identification of
dispersion and a large Fermi surface\cite{Olsen}, an anisotropic
superconducting gap suggestive of a {\it d}$-$wave order
parameter\cite{ShenSCGap}, and an observation of the pseudogap in
underdoped samples\cite{Marshall}. In the mean time, this technique
itself has experienced a dramatic improvement in its energy and
momentum resolutions, leading to a series of new discoveries not
thought possible only a decade ago. This revolution of the ARPES
technique and its scientific impact result from dramatic advances in
four essential components: instrumental resolution and efficiency,
sample manipulation, high quality samples and well-matched
scientific issues.

The purpose of this treatise is to go through the prominent results
obtained from ARPES on cuprate superconductors. Because there have
been a number of recent reviews on the electronic structures of
high-T$_c$
materials\cite{ShenDessau,DamascelliReview,CampuzanoReview}, we will
mainly present the latest results not covered previously,  with a
special attention given on the electron-phonon interaction in
cuprate superconductors. What has emerged is rich information about
the anomalous electron-phonon interaction well beyond the
traditional views of the subject.  It exhibits strong doping,
momentum and phonon symmetry dependence, and shows complex interplay
with the strong electron-electron interaction in these materials.

\section{Angle-Resolved Photoemission Spectroscopy}

\subsection{Principle}

\begin{figure}[tbp]
\begin{center}
\includegraphics[width=0.65\linewidth,angle=-90]{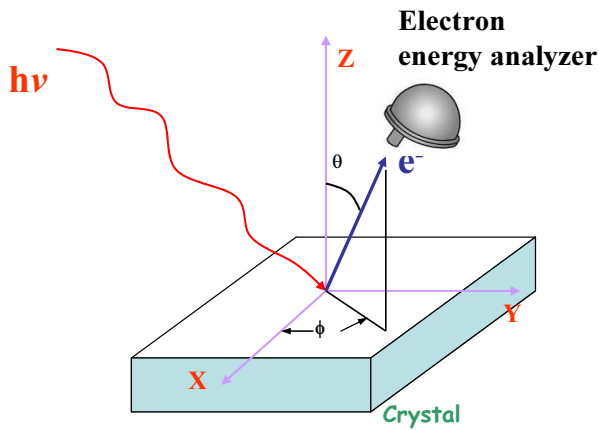}
\end{center}
\caption{Schematic of angle-resolved photoemission
spectroscopy.}\label{ARPESSchematic}
\end{figure}

    Angle-resolved photoemission spectroscopy is a powerful
technique for studying the electronic structure of materials(Fig.
\ref{ARPESSchematic})\cite{SHuefner}.  The information of interest,
i.e., the energy and momentum of electrons in the material, can be
inferred from that of the photoemitted electrons. This conversion is
made possible through two conservation laws involved in the
photoemission process:\\
(1). Energy conservation:  E$_B$=h$\nu$-E$_{kin}$-$\Phi$;\\
(2). Momentum conservation: K$_{||}$=k$_{||}$+{\bf G}.\\
where E$_B$ represents the binding energy of electrons in the
material; h$\nu$ the photon energy of incident light; E$_{kin}$ the
kinetic energy of photemitted electrons; $\Phi$ work function;
k$_{||}$ momentum of electrons in the material parallel to sample
surface; K$_{||}$ projected component of momentum of photoemitted
electrons on the sample surface which can be calculated from the
kinetic energy by $\hbar$K$_{||}$=$\sqrt{2mE_{kin}}$sin$\theta$ with
$\hbar$ being Planck constant;  {\bf G} reciprocal lattice vector.
Therefore, by measuring the intensity of the photoemitted electrons
as a function of the kinetic energy at different emission angles,
the electronic structure of the material under study, i.e., energy
and momentum of electrons,  can be probed directly\cite{SHuefner}.

For 3-dimensional materials, the electronic structure also relies on
k$_{\bot}$, the momentum perpendicular to the sample surface.
Because of the symmetry breaking near the sample surface, the
momentum perpendicular to the sample surface is not conserved. In
order to obtain k$_{\bot}$, one has to consider the inner potential
which can be obtained in various ways\cite{SHuefner}. For strictly
2-dimensional materials or quasi-2-dimensional materials such as the
cuprate superconductors discussed in this treatise, to the first
approximation, one may treat k$_{\bot}$ as a secondary effect.
However, one should always be wary about the residual
3-dimensionality in these materials and its effect on photoemission
data\cite{BansilKzEffect}.

The photoemission process can be understood intuitively in terms of
a ``three step model"\cite{Spicer}: (i) Excitation of the electrons
in the bulk by photons. (ii) Transport of the excited electrons to
the surface. (iii) Emission of the photoelectrons into vacuum. Under
the ``sudden approximation" (described below), photoemission
measures the single$-$particle spectral function A(k,$\omega$),
weighted by the matrix element M and Fermi function f($\omega$):
I$\sim$A(k,$\omega$)$\mid$M$\mid$$^2$f($\omega$)\cite{Hedin,Randeria}.
The matrix element $\mid$M$\mid$$^2$ term indicates that, besides
the energy and momentum of the initial state and the final state,
the measured photoemission intensity is closely related to some
experimental details, such as energy and polarization of incident
light, measurement geometry and instrumental resolution. The
inclusion of the Fermi function accounts for the fact that the
direct photoemission measures only the occupied electronic states.

The single-particle spectral function A(k,$\omega$) can be written
in the following way using the Nambu-Gorkov formalism:

\begin{equation}\label{Hamiltonian}
A(k,\omega)= -(1/\pi) Im G_{11}(k,\omega)
\end{equation}
\begin{equation}
\widehat{G}(k,\omega) = {\frac {Z(k,\omega)\omega\tau_{0} +
(\varepsilon(k) + \chi(k,\omega))\tau_{2} +
\phi(k,\omega)\tau_{1}}{{(Z(k,\omega)\omega)}^{2} -{(\varepsilon(k)
+ \chi(k,\omega))}^{2} - {\phi(k,\omega)}^{2}}} \label{eq:GreenG}
\end{equation}

\noindent where Z, $\chi$, and $\phi$ represent a renormalization
due to either electron-electron or electron-phonon interactions and
$\varepsilon(k)$ is the bare-band energy. $\tau_{0}, \tau_{1},
\tau_{2}$ are the matrices,  and G$_{11}$ represents the Pauli
electronic charge density channel measured in photoemission.  In the
weak coupling case, Z=1, $\chi=0$, and $\phi = \Delta$, the
superconducting gap. The same formalism can be extended to the
normal state by setting $\phi=0$. In the normal state, the spectral
function can be written in a more compact way\cite{Hedin,Randeria},
in terms of the real and imaginary parts of the electron self
energies Re$\Sigma$ and Im$\Sigma$:

\begin{equation}
A(k,\omega) = \frac{1}{\pi}\frac {|{\rm
Im}\Sigma(k,\omega)|}{(\omega - \varepsilon(k)-{\rm
Re}\Sigma(k,\omega))^{2} + ({\rm Im}\Sigma(k,\omega))^{2}}
\label{eq:SpecFunc}
\end{equation}

\noindent where {\rm Re}$\Sigma$ describes the renormalization of
the dispersion and {\rm Im}$\Sigma$ describes the lifetime.

In relating the photoemission process in terms of single particle
spectral function A(k,$\omega$), it is helpful to recognize some
prominent assumptions involved:\\

(1). The excited state of the sample (created by the ejection of the
photo-electron) does not relax in the time it takes for the
photo-electron to reach the detector. This so-called
``sudden-approximation" allows one to write the final state
wave-function in a separable form, $\Psi_{f}^{N} = \Phi_{f}^{k}
\Psi_{f}^{N-1}$, where $\Phi_{f}^{k}$ denotes the photoelectron and
$\Psi_{f}^{N-1}$ denotes the final state of the material with N-1
electrons. If the system is non-interacting, then the final state
overlaps with a single eigenstate of the Hamiltonian describing the
N-1 electrons, revealing the band structure of the single electron.
In the interacting case, the final state can overlap with all
possible
eigenstates of the N-1 system.\\
(2) In the interacting case, A($k,\omega$) describes a
``quasiparticle" picture in which the interactions of the electrons
with lattice motions as well as other electrons can be treated as a
perturbation to the bare band dispersion, $\varepsilon(k)$, in the
form of a self energy, $\Sigma(k,\omega)$. The validity of this
picture as well as (1) rests on whether or not the spectra can be
understood in terms of well-defined peaks representing poles in the
spectral function.\\
(3). The surface is treated no differently from the bulk in this
A($k,\omega$).  In reality surface states are expected and are
observed and can lead to confusion in the data
interpretation\cite{DamascelliReview}.  Surface termination also
affects photoemission process\cite{BansilMatrixBi2212}.

In addition to the matrix element M, there are other extrinsic
effects which contribute to measured photoemission spectrum, e.g.,
the contribution from inelastic electron scattering. On the way to
get out from inside the sample, the photoemitted electrons will
experience scattering from other electrons, giving rise to a
relatively smooth background in the photoemission spectrum.

\subsection{Technique}

\begin{figure}[tbp]
\begin{center}
\includegraphics[width=1.0\linewidth,angle=0]{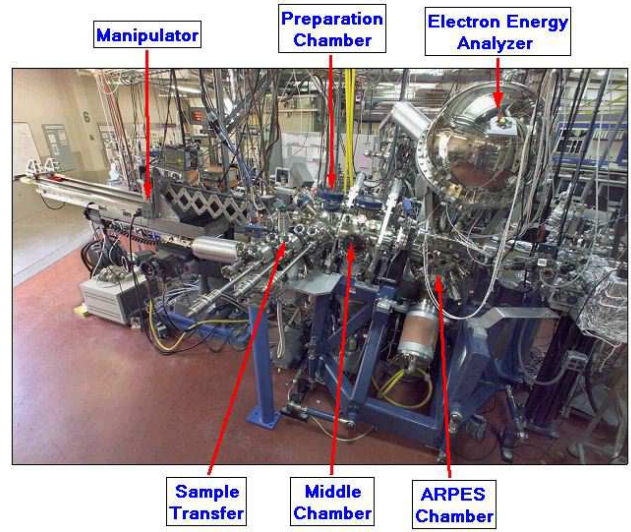}
\end{center}
\caption{A representative ARPES system on Beamline 10.0.1 at the
Advanced Light Source, Lawrence Berkeley National Lab.}
\label{ARPESSetup}
\end{figure}

As seen in Fig.\ref{ARPESSchematic}, an ARPES system consists of a
light source, chamber and sample manipulation and characterization
systems, and an electron energy analyzer. Fig.\ref{ARPESSetup} is an
example of a modern ARPES setup with the following primary components:\\

(1). Light source: Possible light sources for angle-resolved
photoemission are X-ray tubes, gas-discharge lamps, synchrotron
radiation source and VUV lasers. Among them, the synchrotron
radiation source is the most versatile in that it can provide
photons with continuously tunable energy, fixed or variable photon
polarization, high energy resolution and high photon flux. The
latest development of the VUV laser is significant as a result of
its super-high energy resolution and super-high photon flux. In
addition, the lower photon energy achievable by the VUV lasers makes
the measured electronic structure more bulk-sensitive in certain
materials\cite{TKissLaser}. However, the strong final state effect
may limit its application to certain
material systems. \\

\begin{figure}[b!]
\begin{center}
\includegraphics[width=1.0\linewidth,angle=0]{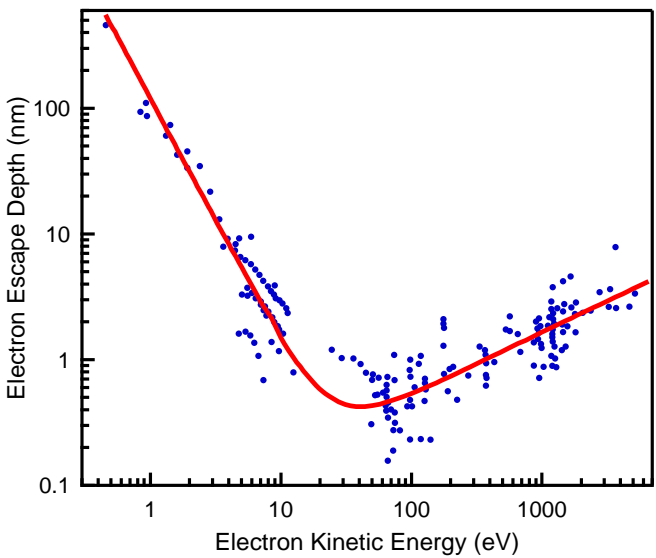}
\end{center}
\caption{Escape depth of photoemitted electron as a function of
kinetic energy\cite{SeahDench}. For elements and inorganic
compounds, the escape depth is found to follow the "universal curve"
(red solid line).}\label{EscapeDepth}
\end{figure}

(2). Chambers and sample manipulation and characterization systems:
In most of the photon energy range commonly used (20$\sim$100 eV),
the escape depth of photoemitted electrons is on the order of
5$\sim$20 $\AA$, as seen in Fig.\ref{EscapeDepth}\cite{SeahDench}.
This means that photoemission is a surface-sensitive technique.
Therefore, obtaining and retaining a clean surface during
measurement is essential to probe the intrinsic electronic
properties of the sample. To achieve this, the ARPES measurement
chamber has to be in ultra-high vacuum, typically better than
5$\times$10$^{-11}$ Torr. A clean surface is usually obtained either
by cleaving samples {\it in situ} in the chamber if the samples are
cleavable or by sputtering and annealing process if the sample is
hard to cleave. The quality of the surface can be characterized by
low energy electron diffraction (LEED) or other techniques such as
scanning tunneling microscopy (STM). The sample transfer system is
responsible for quickly transferring samples from air to UHV
chambers while not damaging the ultra-high vacuum. The manipulator
is responsible for controlling the sample position and orientation,
it also holds a cryostat that can change the sample temperature
during the measurement. An advanced low temperature cryostat which
can control the sample temperature precisely and has multiple
degrees of translation and rotation freedoms is critical to an ARPES
measuremnet. \\

\begin{figure}[tbp]
\begin{center}
\includegraphics[width=1.0\linewidth,angle=0]{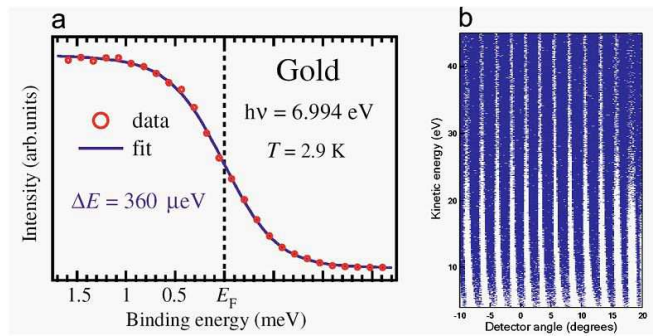}
\end{center}
\caption{(a). Ultrahigh-resolution photoemission spectrum of an
evaporated gold film measured using Scienta R4000 analyzer at a
temperature of 2.9 K (red circles), together with the Fermi-Dirac
function at 2.9 K convolved by a Gaussian with full width at half
maximum of 360 $\mu$eV (a blue line). Total energy resolution of 360
$\mu$eV was confirmed from the very good match between the
experimental and calculated spectra\cite{TKissLaser}. The energy
resolution from the VUV laser is estimated to be 260 $\mu$eV. (b).
Angle mode testing image of Scienta R4000 electron analyzer. the
test was performed using ``wire-and-slit" setup, with the angle
interval between adjacent slits being 1 degrees. In this particular
angular mode, the analyzer collects emission angle within 30 degrees
simultaneously.} \label{EKResolution}
\end{figure}

(3). Electron energy analyzer:  An analyzer measures the intensity
of photoemitted electrons as a function of their kinetic energy,
i.e., Energy Distribution Curve(EDC),  at a given angle relative to
the sample orientation. The dramatic improvement of the ARPES
technique in the last decade is in large part due to the advent of
modern electron energy analyzer, in particular, the Scienta series
hemisphere analyzers. The
enhancement of the performance lies in mainly three aspects:\\
(i). Energy resolution improvement. \\
The energy resolution of the electron energy analyzer improves
steadily over time. The upgrade of the one-dimensional multichannel
detection scheme of the VSW analyzer allows efficient measurement
with $\sim$20meV energy resolution. Among others, it enabled the
discovery of the {\it d}-wave superconducting gap
structure\cite{ShenSCGap}.  The first introduction of the Scienta
200 analyzer in the middle 1990's dramatically improved the energy
resolution to better than 5 meV. The latest Scienta R4000 analyzer
has improved the energy resolution further to better than 1 meV, as
seen in Fig.\ref{EKResolution}\cite{TKissLaser}.

\begin{figure}[tbp]
\begin{center}
\includegraphics[width=0.9\linewidth,angle=0]{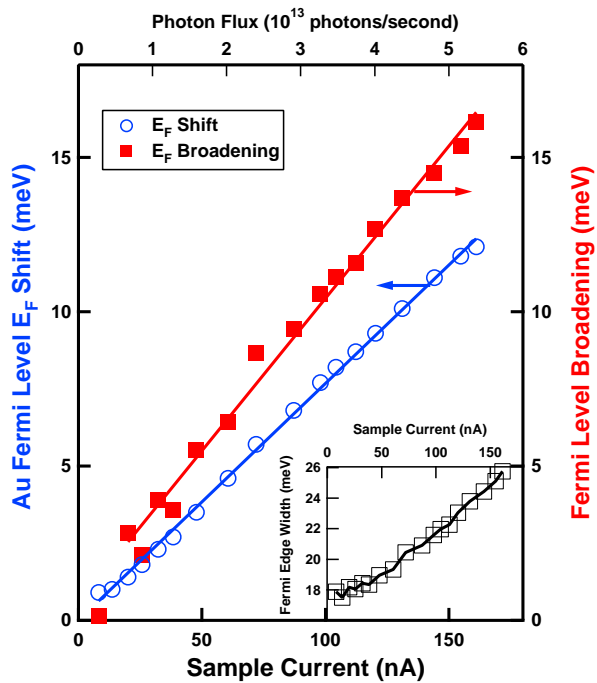}
\end{center}
\caption{Space charge and mirror charge effect in
photoemission\cite{ZhouSCE}. Fermi edge broadening (solid square)
and the Fermi edge shift (open circle) as a function of sample
current. The beam spot size is $\sim$0.43mm$\times$0.30 mm. The
inset shows the measured overall Fermi edge width as a function of
the sample current, which includes all contributions including the
beamline, the analyzer and the temperature broadening. The net
broadening resulting from pulsed photons is obtained by
deconvolution of the measured data, taking the width at low photon
flux as from all the other contributions. }\label{SpaceCE}
\end{figure}

We note that the total experimental energy resolution relies on both
the analyzer resolution and the light source resolution. Sample
temperature can also cause thermal broadening which is a limitation
in some cases. The necessity of multiple degrees of rotation
controls as well as the exposure of the surface during an ARPES
measurement often puts a lower limit on the sample temperature. In
addition, one should be aware of some intrinsic effects associated
with the photoemission process, i.e., space charge effect and mirror
charge effect\cite{ZhouSCE}. When pulsed light is incident on a
sample, the photoemitted electrons experience energy redistribution
after escaping from the surface because of the Coulomb interaction
between them (space charge effect) and between photo-emitted
electrons and the distribution of mirror charges in the sample
(mirror charge effect). These combined Coulomb interaction effects
give rise to an energy shift and a broadening whose magnitude
depends on the photon energy, photon flux, beam spot size, emission
angles and etc. For a typical third-generation synchrotron light
source, the energy shift and broadening  can be on the order of 10
meV (Fig.\ref{SpaceCE})\cite{ZhouSCE}. This value is comparable to
many fundamental physical parameters actively studied by
photoemission spectroscopy and should be taken seriously in
interpreting photoemission data and in designing next generation
experiments.

\begin{figure}[tbp]
\begin{center}
\includegraphics[width=1.0\linewidth,angle=0]{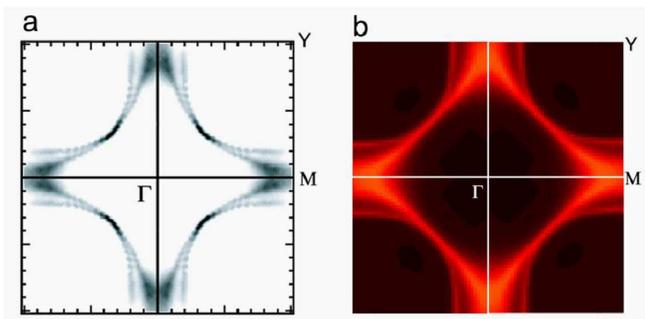}
\end{center}
\caption{(a). Experimentally measured Fermi surface in Pb-doped
Bi2212\cite{PashaBi2212}. (b). Calculated Fermi surface of
Bi2212\cite{BansilMatrixBi2212}. }\label{Bi2212BS}
\end{figure}

(ii). Momentum resolution;\\
The introduction of the angular mode operation in the new Scienta
analyzers has also greatly improved the angular resolution, from a
previous $\sim$2 degrees to 0.1$\sim$0.3 degree. This improvement of
the momentum resolution allows one to observe detailed structures in
the band structure and Fermi surface, as well as subtle but
important many-body effects. As an example, recent identification of
two Fermi surface sheets (so-called ``bilayer splitting")in
Bi$_2$Sr$_2$CaCu$_2$O$_8$ (Bi2212) (Fig.\ref{Bi2212BS}) is largely
due to such an improvement of momentum
resolution\cite{FengBilayersplitting,ChuangBilayerSplitting,BogdanovBilayerSplitting},
combined with the advancement of theoretical
calculations\cite{BansilMatrixBi2212}.

\begin{figure}[tbp]
\begin{center}
\includegraphics[width=1.0\linewidth,angle=0]{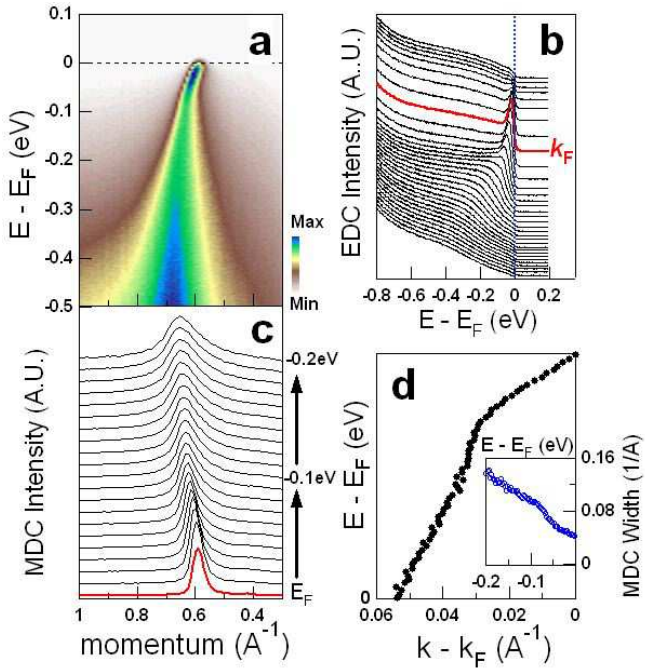}
\end{center}
\caption{Illustration of the MDC method for extracting the
electron self-energy. (a) Raw photoemission data for LSCO with
x=0.063 (T$_c$$\sim$12K) along the (0,0)-($\pi$,$\pi$) nodal
direction at 20 K\cite{XJZhouJSR}. The two-dimensional data
represent the photoelectron intensity (denoted by false color) as
a function of energy and momentum. (b) Energy distribution curves
(EDCs) at different momenta. The EDC colored red corresponds to
the Fermi momentum k$_F$. (c) Momentum distribution curves (MDCs)
at different binding energies. The MDC colored red corresponds to
the Fermi level. (d) Energy-momentum dispersion relation extracted
by the MDC method. The inset shows the MDC width as a function of
energy.}\label{MDCMethod}
\end{figure}

(iii). Two-dimensional multiple angle detection;\\
Traditionally, the electron energy analyzer collects one
photoemission spectrum, i.e., energy distribution curve (EDC),  at
one measurement for each emission angle. Modern electron energy
analyzers collect multiple angles simultaneously. As shown in
Fig.\ref{EKResolution}b, the latest Scienta R4000 analyzer can
collect photoemitted electrons in the angle range of 30 degrees
simultaneously. Therefore, at one measurement, the raw data thus
obtained, shown in Fig.\ref{MDCMethod}a, is a  2-dimensional image
of the photoelectron intensity (represented by false color) as a
function electron kinetic energy and emission angle (and hence
momentum). This 2-dimensionality greatly enhances data collection
efficiency and  provides a convenient way of analyzing the
photoemission data.

As shown in Fig.\ref{MDCMethod} , the traditional way to visualize
the photoemission data is by means of so-called energy distribution
curves (EDCs), which represent photoelectron intensity as a function
of energy for a given momentum. The 2D image comprising the raw data
is then equivalent to a number of EDCs at different momenta
(Fig.\ref{MDCMethod}b).  The peak position at different momenta will
give the energy-momentum dispersion relation determining the real
part of electron self-energy Re$\Sigma$. The EDC linewidth
determines the quasiparticle lifetime, or the imaginary part of
electron self-energy Im$\Sigma$. However, the EDC lineshape is
usually complicated by a background at higher binding energy, the
Fermi function cutoff near the Fermi level, and an undetermined bare
band energy which make it difficult to extract the electron
self-energy precisely.

An alternative way to visualize the 2D data is to analyze
photoelectron intensity as a function of momentum for a given
electron kinetic energy\cite{AebiMDC} by means of momentum
distribution curves (MDCs)\cite{VallaScience,LashellBe}. This
approach provides a different way of extracting the electron
self-energy. As shown in Fig.\ref{MDCMethod}c, the MDCs exhibit
well-defined peaks with flat backgrounds; moreover, they can be
fitted by a Lorentzian lineshape. When the bandwidth is large, the
band dispersion  $\epsilon$$_k$ can be approximated as
$\epsilon$$_k$ = v$_0$k in the vicinity of the Fermi level. Under
the condition that the electron self-energy shows weak momentum
dependence, A(k,$\omega$) indeed exhibits a Lorentzian lineshape
as a function of k for a given binding energy . By fitting a
series of MDCs at different binding energies to obtain the MDC
position $\tilde{k}$ and width $\Gamma$ (full-width at half
maximum, FWHM) (Fig.\ref{MDCMethod}d)\cite{XJZhouJSR}, one can
extract the electron self-energy directly as:
Re$\Sigma$=$\hbar$$\omega$-$\tilde{k}$v$_0$  and
Im$\Sigma$=$\Gamma$v$_0$/2.

It is worthwhile to point out the latest effort in attempting to
overcome the surface sensitivity issue related with photoemission.
As seen from Fig.\ref{EscapeDepth}, in the usual photon energy range
used for valence band photoemission,  the photoemitted electron
escape depth is on the order on 5$\sim$10 $\AA$. Therefore, it is
always an issue whether the photoemission results obtained in this
energy range represents the bulk properties. To overcome such a
problem, there have been two approaches by employing either high
photon energy or lower photon energy. As seen from
Fig.\ref{EscapeDepth}, when the photon energy is on the order of 1
KeV, the electron escape depth can be increased to $\sim$20
$\AA$\cite{SugahigheV}. However, this modest enhancement of the bulk
sensitivity comes at a price of sacrificing both the energy
resolution and momentum resolution. On the other hand, when the
photon energy is low, one can see that the electron escape depth
increases dramatically. Note that this ``universal" curve is
obtained from metals, whether the same curve can be applied to oxide
materials remains unclear yet. In addition to the potential
engancement of the bulk sensitivity,  one may further improve the
energy and momentum resolution  by going to lower photon energy..

\section{Electronic Structures of High Temperature Superconductors}

\subsection{Basic Crystal Structure and Electronic Structure}

\begin{figure}[tbp]
\begin{center}
\includegraphics[width=1.0\linewidth,angle=0]{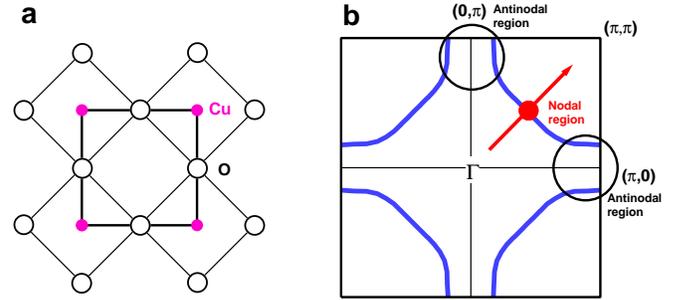}
\end{center}
\caption{(a) Schematic of the real-space CuO$_2$ plane. The CuO$_2$
plane consists of copper (pink solid circles) and oxygen (black open
circles). (b) The corresponding Brillouin zone in a reciprocal
space. In the first Brillouin zone, the area near ($\pi$/2, $\pi$/2)
(denoted as red circle)  is referred to as nodal region, and the
(0,0)-($\pi$,$\pi$) direction is the nodal direction (red arrow).
The area near ($\pi$,0) and (0,$\pi$) is referred to as the
antinodal region (shaded circles).  The blue solid line shows a
schematic Fermi surface.}\label{CuO2Plane}
\end{figure}

\begin{figure}[tbp]
\begin{center}
\includegraphics[width=0.65\linewidth]{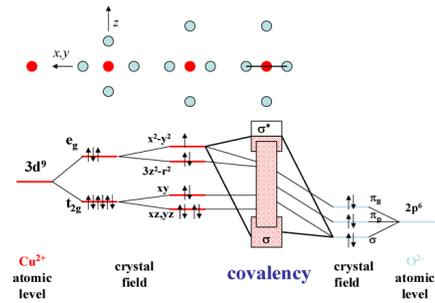}
\end{center}
\caption{Bonding in CuO$_2$ plane\cite{FinkBond}.  The atomic Cu 3d
level is split due to the cubic crystal field into e$_g$ and
t$_{2g}$ states. There is a further splitting due to an octahedral
crystal field into x$^2$ - y$^2$, 3z$^2$ - r$^2$, xy, and xz, yz
states. For divalent Cu which has nine 3d electrons, the uppermost
x$^2$ - y$^2$ level is half filled, while all other levels are
completely filled. There is a strong hybridization of the Cu states,
particularly the x$^2$ - y$^2$ states, with the O 2p states thus
forming a half-filled two-dimensional Cu 3d$_{x^2-y^2}$-O 2p$_{x,y}$
antibonding dp$\sigma$ band. The hybridization of the other 3d
levels is smaller and is indicated in Figure  only by a broadening.
}\label{Bonding}
\end{figure}

A common structural feature of all cuprate superconductors is the
CuO$_2$ plane (Fig.\ref{CuO2Plane}a) which is responsible for the
low lying electronic structure; the CuO$_2$ planes are sandwiched
between various block layers which serve as charge reservoirs to
dope CuO$_2$ planes\cite{Tokura,RJCava}. For the undoped parent
compound, such as La$_2$CuO$_4$, the valence of Cu is 2+,
corresponding to 3d$^9$ electronic configuration. Since the
Cu$^{2+}$ is surrounded by four oxygens in the CuO$_2$ plane and
apical oxygen(s)or halogen(s) perpendicular to the plane, the
crystal field splits the otherwise degenerate five {\it d}-orbitals,
as schematically shown in Fig.\ref{Bonding}\cite{PickettRMP}. The
four lower energy orbitals, including xy, xz, yz and 3z$^2$ - r$^2$,
are fully occupied, while the orbital with highest energy,
x$^2$-y$^2$, is half-filled. since the energies of the Cu d-orbitals
and O 2p-orbitals are close, there is a strong hybridization between
them. As a result, the topmost energy level has both Cu
d$_{x^2-y^2}$ and O 2p$_{x,y}$ character.

\begin{figure}[tbp]
\begin{center}
\includegraphics[width=0.85\linewidth,angle=0]{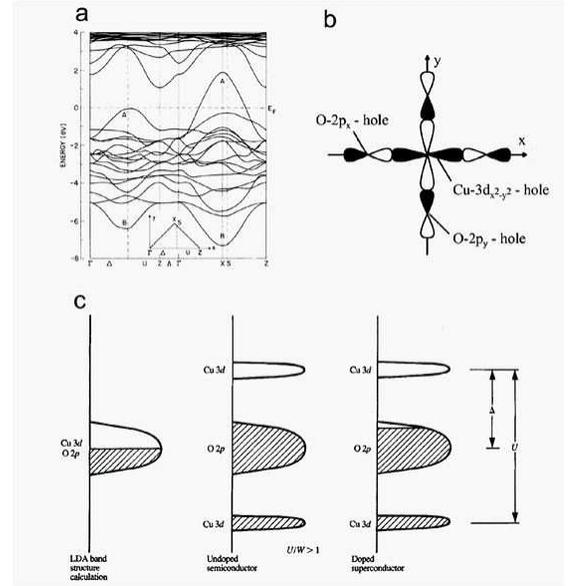}
\end{center}
\caption{(a). LDA calculated band structure of
La$_2$CuO$_4$\cite{MattheissLCO}. The band labeled B is bonding band
between Cu 3d$_{x^2-y^2}$ and O 2p states while the band labeled A
is the corresponding antibonding band that is half-filled; (b).
Schematic of Zhang-Rice singlet state\cite{ZhangRice,CDurr}. (c).
Schematic energy diagrams for undoped and doped CuO$_2$
planes\cite{SUchidaGap}. (c1). Band picture for a half-filled
(undoped) CuO$_2$ plane (Fermi liquid); (c2). Charge-transfer
insulating state of the CuO$_2$ plane with split Cu 3d bands due to
on-site Coulomb repulsive interaction $U$. The O 2p band is
separated by a charge transfer energy $\Delta$ from the upper Cu 3d
band; (c3) and (c4) show rigid charge transfer energy bands doped
with holes and electrons, respectively; (c4). Formation of mid-gap
states inside the charge transfer gap. }\label{LDALCO}
\end{figure}

The same conclusion is also drawn from band structure calculations
(Fig.\ref{LDALCO}a)\cite{PickettRMP}. According to both simple
valence counting (Fig.\ref{Bonding}) and band structure calculation
(Fig.\ref{LDALCO}a),  the undoped parent compound is supposed to be
a metal. However, strong Coulomb interactions between electrons on
the same Cu site makes it an antiferromagnetic insulator with an
energy gap of 2 eV\cite{SUchidaGap,VakninNeutron}. The basic
theoretical model for the electronic structure most relevant to our
discussion is the multi-band Hubbard
Hamiltonian\cite{Emery,VarmaModel} containing d states on Cu sites,
p states on O sites, hybridization between Cu-O states,
hybridization between O-O states, and Coulomb repulsion terms. In
terms of hole notation, i.e., starting from the filled-shell
configuration (3d$^{10}$,2p$^6$) corresponding to a formal valence
of Cu$^{1+}$ and O$^{2-}$, the general form of the model can be
written as\cite{RiceLecture}:\\

\begin{displaymath}
H=\sum_{i\sigma}\varepsilon_dd^{+}_{i\sigma}d_{i\sigma}+\sum_{l\sigma}\varepsilon_pp^{+}_{l\sigma}p_{l\sigma}
+\sum_{<li>\sigma}t_{pd}p^{+}_{l\sigma}d_{i\sigma} + h.c.\\
\end{displaymath}
\begin{displaymath}
+\sum_{i}U_dn_{i\uparrow}n_{i\downarrow}+\sum_{<ll'>\sigma}t_{O-O}p^{+}_{l\sigma}p_{l'\sigma}+h.c.
\end{displaymath}
\begin{equation}
+\sum_{<il>\sigma\sigma'}U_{pd}n_{l\sigma}n_{i\sigma'}+\sum_{l}U_pn_{l\uparrow}n_{l\downarrow}
\label{eq:Emery}
\end{equation}

\noindent where the operator $d^{+}_{i\sigma}$ creates Cu
(3d$_{x^2-y^2}$)holes at site $i$, and $p^{+}_{l\sigma}$ creates
O(2p) holes at the site $l$. U$_d$ is the on-site Coulomb repulsion
between two holes on a Cu site. The third term accounts for the
direct overlap between Cu-O orbitals. The fifth terms describes
direct hopping between nearest-neighbor oxygens, and U$_{pd}$ in the
sixth term is the nearest neighbor Coulomb repulsion between holes
on Cu and O atoms. Qualitatively, this model gives the energy
diagram in Fig.\ref{LDALCO}c.

Simplified versions of model Hamiltonians have also been proposed.
Notably among them are the single-band Hubbard
model\cite{AndersonRVB} and t-J model\cite{ZhangRice}. The t-J
Hamiltonian can be written in the following
form\cite{DagottoRMP,RiceLecture}:

\begin{equation}
H_{tJ}=-t\sum_{<ij>,\sigma}(\tilde{c}_{i\sigma}^{\dagger}\tilde{c}_{j\sigma}+H.c.)
+ J\sum_{<ij>}({\bf S_i}\cdot{\bf
S_j}-\hat{n}_{i\uparrow}\hat{n}_{j\downarrow}/4) \label{eq:tJmodel}
\end{equation}

\noindent where the operator
$\tilde{c}_{i\sigma}^{\dagger}$=$c_{i\sigma}^{\dagger}(1-\hat{n}_{i-\sigma}$)
excludes double occupancy, $J=4t^{2}/U$ is the antiferromagnetic
exchange coupling constant, and {\bf S$_i$}  is the spin operator.
Since the hopping process may also involve the second (t$^{'}$) and
third (t$^{"}$) nearest neighbor, an extended t-J model, the t$-$
t$^{'}$$-$t$^{"}$$-$J model, has also been proposed\cite{TTohyama}.

\subsection{Brief Summary of Some Latest ARPES Results}

ARPES has provided key information on the electronic structure of
high temperature superconductors, including the band structure,
Fermi surface, superconducting gap, and pseudogap. These topics are
well covered in recent
reviews\cite{DamascelliReview,CampuzanoReview} that we will not
repeat here. Instead, we briefly summarize some of the latest
developments not included before.

{\it Band structure and Fermi surface:} The bi-layer splitting of
the Fermi surface is well established in the overdoped
Bi2212\cite{FengBilayersplitting,ChuangBilayerSplitting,BogdanovBilayerSplitting},
as shown in Fig.\ref{Bi2212BS} and also suggested to exist in
underdoped and optimally doped
Bi2212\cite{ChuangBS,FengBS,GromkoAntinodalKink,KordyukBS}. Recent
measurements also show that there is a slight splitting along the
(0,0)-($\pi$,$\pi$) nodal direction\cite{KordyukNodalBS}. The
measurement on four-layered Ba$_2$Ca$_3$Cu$_4$O$_8$F$_2$ has
identified at least two clear Fermi surface sheets\cite{YLChenBCCO}.

{\it Superconducting gap and pseudogap:} Since the first
identification of an anisotropic superconducting gap in
Bi2212\cite{ShenSCGap}, subsequent measurements on the
superconductors such as
Bi2212\cite{Bi2212SCGap1,Bi2212SCGap2,Bi2212SCGap3,Bi2212SCGap4},
Bi2201\cite{Bi2201SCGap1,Bi2201SCGap2},
Bi2223\cite{Bi2223SCGap1,Bi2223SCGap2,Bi2223SCGap3},
YBa$_2$Cu$_3$O$_{7-\delta}$\cite{YBCOSCGap}, LSCO\cite{LSCOSCGap}
have established a universal behavior of the anisotropic
superconducting gap in these hole-doped superconductors which is
consistent with d-wave pairing symmetry (although it is still an
open question whether the gap form is a simple d-wave-like
$\Delta(k)$=$\Delta_0$[cos(k$_x$a)-cos(k$_y$a)] or higher harmonics
of the expansion should be included). The measurements on
electron-doped superconductors also reveal an anisotropic
superconducting gap\cite{ElectronSCGap1,ElectronSCGap2}.

One interesting issue is, if a material has multiple Fermi surface
sheets, whether the superconducting gap on different Fermi surface
sheets is the same. This issue traces back to superconducting
SrTiO$_3$ where it was shown from tunneling measurements that
different Fermi surface sheets may show different Fermi surface
gaps\cite{BednorzSrTiO3}. With the dramatic advancement of the ARPES
technique, different superconducting gaps on different Fermi surface
sheets have been observed in 2H-NbSe$_2$\cite{TokoyaNbSe2} and
MgB$_2$\cite{SoumaMgB2}. For high-T$_c$ materials, Bi2212 shows two
clear FS sheets, but no obvious difference of the superconducting
gas has been resolved\cite{Bi2212SCGap4}.  In
Ba$_2$Ca$_3$Cu$_4$O$_8$F$_2$, it has been clearly observed that the
two Fermi surface sheets have different superconducting
gaps\cite{YLChenBCCO}.

{\it Time reversal symmetry breaking:} It has been proposed
theoretically that, by utilizing circularly polarized light for
ARPES, it is possible to probe time-reversal symmetry breaking that
may be associated with the pseudogap state in the underdoped
samples\cite{VarmaTRS,SimonTRS}.  Kaminski et al. first reported the
observation of such an effect\cite{KaminskiTRS}.  However, this
observation is not reproduced by another group\cite{BorisenkoTRS}
and the subject remain controversial\cite{TRSControversy}.

\section{Electron-Phonon Coupling in High Temperature Superconductors}

The many-body effect refers to interactions of electrons with other
entities, such as other electrons, or collective excitations like
phonons, magnons, and so on. It has been recognized from the very
beginning that many-body effects are key to understanding cuprate
physics. Due to its proximity to the antiferromagnetic Mott
insulating state, electron-electron interactions are extensively
discussed in the literature\cite{DamascelliReview,CampuzanoReview}.
In this treatise, we will mostly review the recent progress in our
understanding of electrons interacting with bosonic modes, such as
phonons. This progress stems from improved sample quality,
instrumental resolution, as well as theoretical development. In a
complex system like the cuprates, it is not possible to isolate
various degrees of freedom as the interactions mix them together. We
will discuss the electron-boson interactions in this spirit, and
will comment on the interplay between electron-phonon and
electron-electron interactions whenever appropriate. Here by bosonic
modes, we are referring to collective modes with sharp collective
energy scale such as the optical phonons and the famous magnetic
resonance mode seen in some
cuprates\cite{ResonanceModeYBCO,ResonanceModeBi2212,RMHeTl2201}, but
not the broad excitation spectra such as those from the broad
electron/spin excitations as these issues have been discussed in
previous reviews. Furthermore, we believe the effects due to sharp
mode coupling seen in cuprates are caused by phonons rather than the
magnetic resonance. Our reason for not attributing the observed
effect to magnetic resonance will become apparent from the rest of
the manuscript. With more limited data, other groups have taken the
view that the magnetic resonance is the origin of the boson coupling
effect. For this reason, we will focus more on our own results in
reviewing the issues of electron-phonon interaction in cuprates.

The electron-phonon interactions can be characterized into two
categories: (i). Weak coupling where one can still use the
perturbative self-energy approach to describe the quasiparticle and
its lifetime and mass; (ii). Strong coupling and polaron regime
where this picture breaks down.

\subsection{Brief Survey of Electron-Phonon Coupling in High-Temperature Superconductors}

It is well-known that, in conventional superconductors,
electron-phonon (el-ph) coupling is responsible for the formation of
Cooper pairs\cite{Schrieffer}. The discovery of high temperature
superconductivity in cuprates was actually inspired by possible
strong electron-phonon interaction in oxides owing to polaron
formation or in mixed-valence systems\cite{BednorzMuller}. However,
shortly after the discovery, a number of experiments lead some
people to believe that electron-phonon coupling may not be relevant
to high temperature superconductivity. Among them
are\cite{PBAllen}:\\
(1). High critical transition temperature T$_c$\\
\indent So far, the highest T$_c$ achieved is 135 K in
HgBa$_2$Ca$_2$Cu$_3$O$_8$\cite{SchillingNature} at ambient pressure
and $\sim$160 K under high pressure\cite{HgHighPressure}. Such a
high T$_c$ was not expected in simple materials using the strongly
coupled version of BCS theory, or the McMillan equations.\\
(2). Small isotope effect on T$_c$\\
\indent It was found that the isotope effect in optimally-doped
samples is rather small, much less than that expected for
strongly-coupled
phonon-mediated superconductivity\cite{BatloggIsotope}. \\
(3). Transport measurement\\
\indent The linear resistivity-temperature dependence in optimally
doped samples and the lack of a saturation in resistivity over a
wide temperature range have been taken as an evidence of weak
electron-phonon coupling in the cuprate superconductors\cite{GurvitchMartin}.\\
(4).{\it d}-wave symmetry of the superconducting gap\\
\indent It is generally believed that electron-phonon coupling is
favorable to s-wave coupling.\\
(5). Structural instability.\\
\indent It is generally believed that sufficiently strong
electron-phonon coupling to yield high T$_c$
will result in structural instability\cite{CohenAndersonTcLimit}.\\

Although none of these observations can decisively rule out the
electron-phonon coupling mechanism in high-T$_c$ superconductors,
overall they suggest looking elsewhere.  Instead, strong
electron-electron correlation has been proposed to be the mechanism
of high-T$_c$ superconductivity \cite{PAndersonBook}. This approach
is attractive since {\it d}$-$wave pairing is a natural consequence.
Furthermore, the high temperature superconductors evolve from
antiferromagnetic insulating compounds where the electron-electron
interactions are strong \cite{Scalapino,DPines}

\begin{figure}[tbp]
\begin{center}
\includegraphics[width=0.9\linewidth,angle=0]{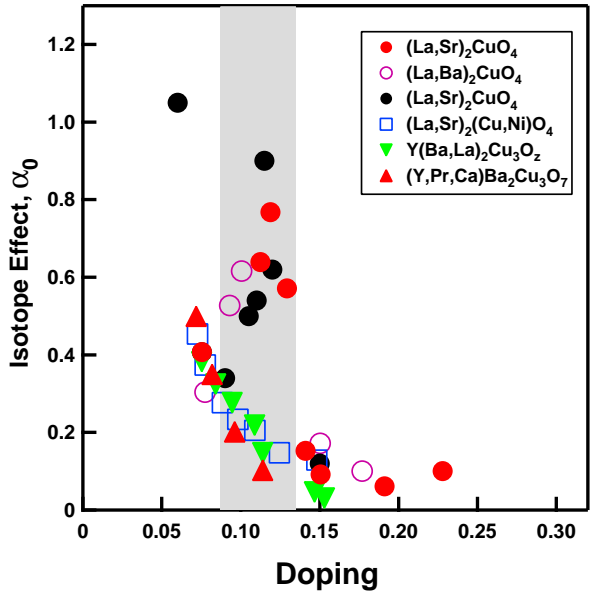}
\end{center}
\caption{Doping dependence of the oxygen isotope effect $\alpha$$_0$
on T$_c$ in several classes of
cuprates\cite{IsoGMZhao,CrawfordScience,TSchneider}. The "1/8
anomaly" data found in LSCO system is highlighted in the shaded
region. }\label{IsotopeTc}
\end{figure}

However, there is a large body of experimental evidence also showing
strong electron-phonon coupling in high-temperature
superconductors\cite{KAMuller,Kulic,MottAlexandrov}. Among them are:

\noindent (1). Isotope effect;\\
\indent As seen in Fig.\ref{IsotopeTc}, although at the optimal
doping, the oxygen isotope effect on T$_c$ is indeed small, it gets
larger and becomes significant with reduced doping\cite{TSchneider}.
In particular, near the ``1/8" doping level, the isotope effect in
(La$_{2-x}$Sr$_x$)CuO$_4$ and (La$_{2-x}$Ba$_x$)CuO$_4$ is
anomalously strong, which is related to the structural
instability\cite{CrawfordScience}. Furthermore, the measurement of
an oxygen isotope effect on the in-plane penetration depth also
suggests the importance of lattice vibration for
high-T$_c$ superconductivty\cite{JHofer}.\\

\begin{figure}[tbp]
\begin{center}
\includegraphics[width=1.0\linewidth,angle=0]{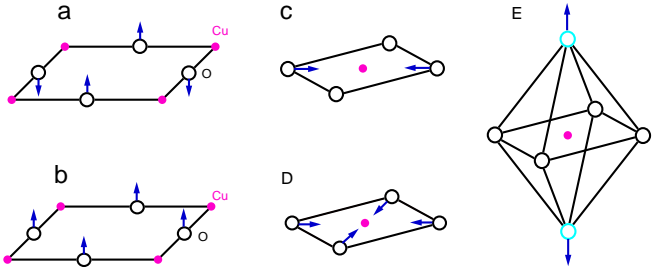}
\end{center}
\caption{Schematic of B$_{1g}$ mode (a), A$_{1g}$ mode (b),
half-breathing mode (c), full-breathing mode (d) and  apical oxygen
mode (e).}\label{PhononModes}
\end{figure}

\begin{figure}[tbp]
\begin{center}
\includegraphics[width=1.0\linewidth,angle=0]{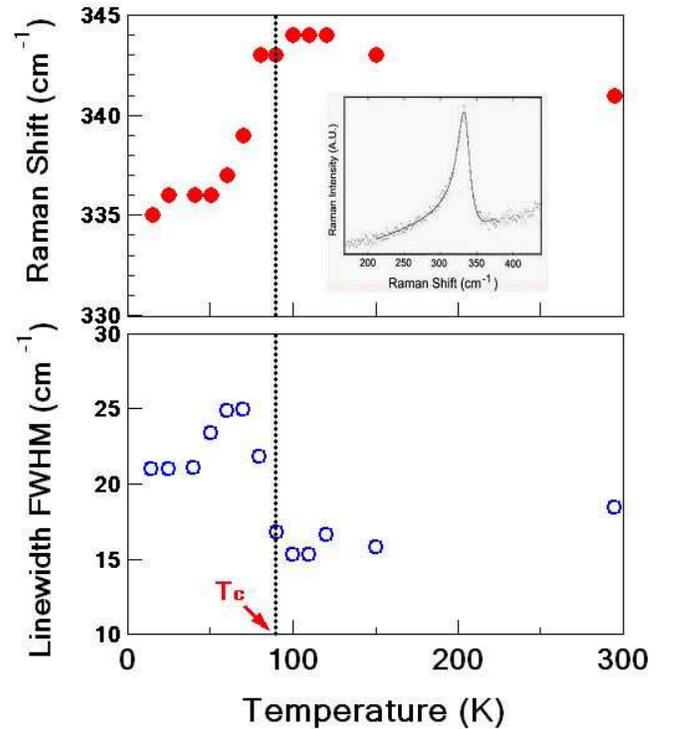}
\end{center}
\caption{Anomalous softening of the B$_{1g}$ phonon when YBCO is
cooled below T$_c$\cite{AltendorfB1g}. The inset shows the fit of a
Fano function to the phonon peak at
T=72K\cite{CThomson_B1g}}\label{YBCOB1g}
\end{figure}

\begin{figure}[b!]
\begin{center}
\includegraphics[width=1.0\linewidth,angle=0]{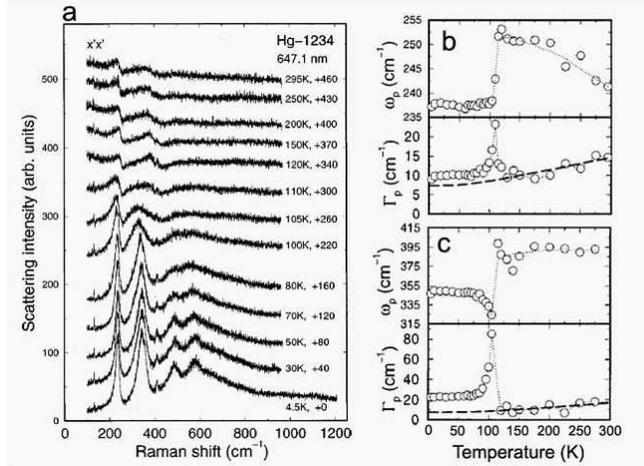}
\end{center}
\caption{Raman spectra of Hg1234 showing a giant
superconductivity-induced mode softening across T$_c$=123
K\cite{ZhouRaman}. The modes at 240 cm$^{-1}$ and 390 cm$^{-1}$
correspond to A$_{1g}$ out-of-plane, in-phase vibration of oxygens
in the CuO$_2$ planes. Upon cooling from room temperature to 4.5 K,
the 240 cm$^{-1}$ A$_{1g}$ mode shows a abrupt drop in frequency at
T$_c$ from 253 to 237 cm$^{-1}$ and the 390 cm$^{-1}$ mode drops
from 395 to 317 cm$^{-1}$\cite{ZhouRaman}.}\label{Hg1234A1g}
\end{figure}

\noindent (2).Optical spectroscopy and Raman scattering;\\
\indent Raman scattering\cite{MCardonaPhonon} and infrared
spectroscopy\cite{TajimaIR} reveal strong electron-phonon
interaction for certain phonon modes. Some typical vibrations
related to the in-plane and apical oxygens are depicted in
Fig.\ref{PhononModes}. In YBa$_2$Cu$_3$O$_{7-\delta}$, it has been
found that, the B$_{1g}$ phonon, which is related to the
out-of-plane, out-of-phase, in-plane oxygen vibrations (see
Fig.\ref{PhononModes}), exhibits a Fano-like lineshape
(Fig.\ref{YBCOB1g}) and shows an abrupt softening upon entering the
superconducting state\cite{CThomson_B1g,Friedl,AltendorfB1g}. The
A$_{1g}$ modes, as found in HgBa$_2$Ca$_3$Cu$_4$O$_{10}$
(Hg1234)\cite{ZhouRaman} and in HgBa$_2$Ca$_2$Cu$_3$O$_{8}$
(Hg1223)\cite{ZhouRaman1223}, exhibit especially strong
superconductivity-induced phonon softening(Fig.\ref{Hg1234A1g}).
Infrared reflectance measurements on various cuprates found that the
frequency of the Cu-O stretching mode in the CuO$_2$ plane is very
sensitive to the distance between copper and oxygen\cite{TajimaIR}.

\begin{figure}[tbp]
\begin{center}
\includegraphics[width=1.0\linewidth,angle=0]{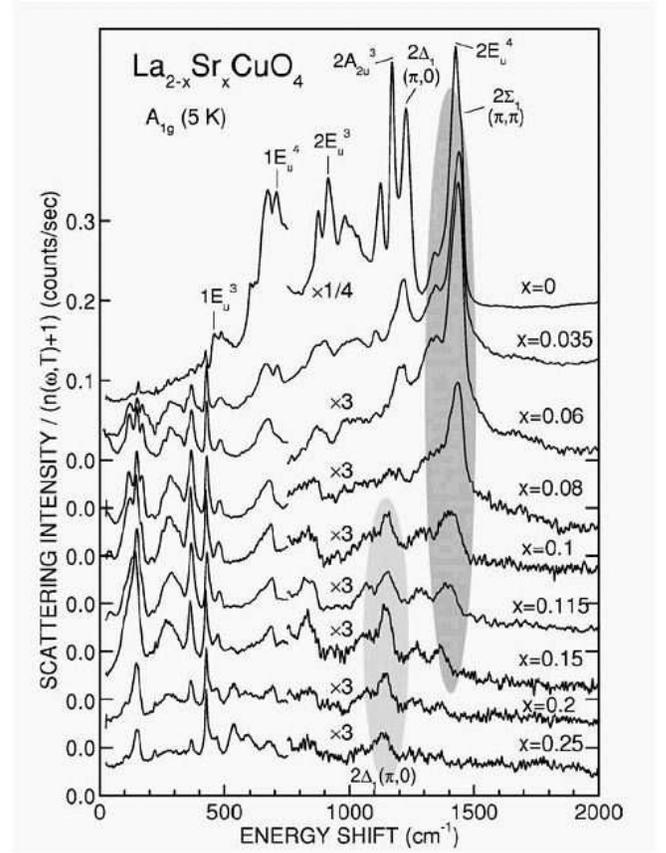}
\end{center}
\caption{A$_{1g}$ two-phonon Raman spectra in LSCO at different
dopings. The dark gray area indicates that the two-phonon peak of
the ($\pi$,$\pi$) LO mode is strong and the light gray area
indicates that the two-phonon peak of the ( $\pi$,0) LO mode is
strong\cite{SugaiLSCO}.}\label{SugaiRaman}
\end{figure}

Fig. \ref{SugaiRaman} shows Raman data as a function of doping in
LSCO\cite{SugaiLSCO}.  The sharp structures at high frequency are
signals from multiphonon processes, which can only occur if the
electron-phonon interaction is very strong.  One can see that this
effect is very strong in undoped and deeply underdoped regime, and
gets weaker with doping increase.

\begin{figure}[b!]
\begin{center}
\includegraphics[width=0.9\linewidth,angle=0]{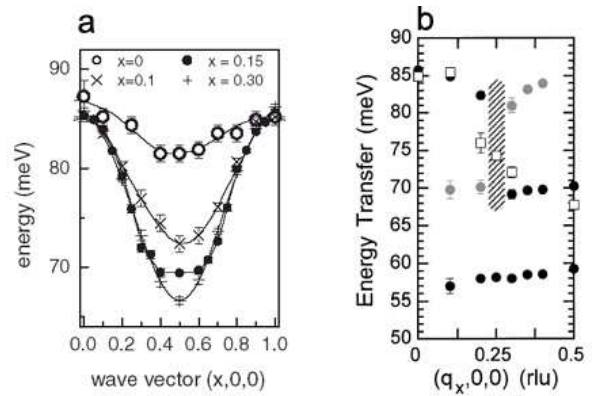}
\end{center}
\caption{(a). Dispersion of the Cu-O bond-stretching vibrations in
the (100)-direction in
(La$_{2-x}$Sr$_x$)CuO$_4$\cite{PintschoviusReview}. (b). Anomalous
dispersion of LO phonons in La$_{1.85}$Sr$_{0.15}$CuO$_4$. 10K data
are filled circles and room temperature data are empty squares. Grey
shaded circles indicate the frequency of the weak extra branch seen
at 10K\cite{McQueeneyBreathingMode}.}\label{NeutronHB}
\end{figure}

\begin{figure}[tbp]
\begin{center}
\includegraphics[width=1.0\linewidth,angle=0]{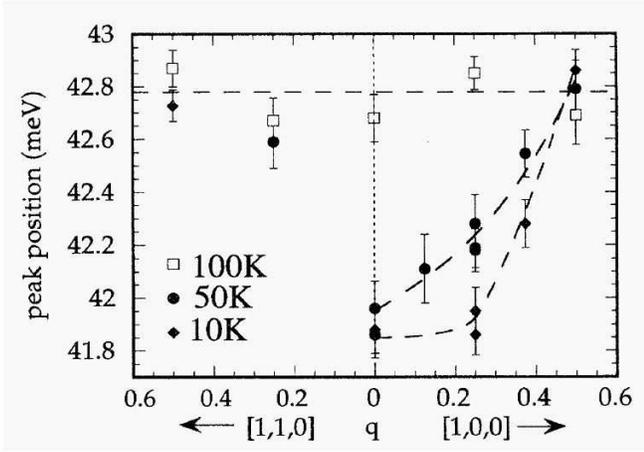}
\end{center}
\caption{q dependence of B$_{1g}$ mode peak position at different
temperatures in YBCO. Dashed lines are guides to the
eye\cite{YBCOB1gNeutron}}\label{NeutronB1g}
\end{figure}

\noindent (3). Neutron scattering\\
\indent Neutron scattering measurements have provided rich
information about electron-phonon coupling in high temperature
superconductors\cite{PintschoviusReview,EgamiReview,YBCOB1gNeutron}.
As seen from Fig.\ref{NeutronHB}a, the in-plane ``half-breathing"
mode exhibits strong frequency renormalizations upon doping along
(001) direction\cite{PintschoviusBreathingMode,PintschoviusReview}.
In (La$_{1.85}$Sr$_{0.15}$)CuO$_4$, it is reported that, at low
temperature, the half-breathing mode shows a discontinuity in
dispersion (Fig.\ref{NeutronHB}b)\cite{McQueeneyBreathingMode}. In
YBCO, neutron scattering indicates that the softening of the
B$_{1g}$ mode upon entering the superconducting state is not just
restricted near  q=0, as indicated by Raman scattering
(Fig.\ref{YBCOB1g}), but can be observed in a large part of the
Brillouin zone (Fig.\ref{NeutronB1g})\cite{YBCOB1gNeutron}.

\begin{figure}[b!]
\begin{center}
\includegraphics[width=1.0\linewidth,angle=0]{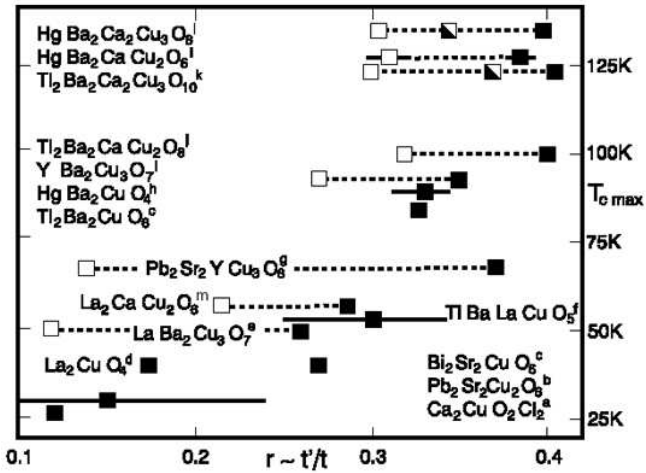}
\end{center}
\caption{Correlation between calculated range parameter r and
observed T$_{c max}$ where r is controlled by the energy of the
axial orbital, a hybrid between Cu 4s, apical-oxygen 2p$_z$ , and
farther orbitals\cite{MaterialTc}.  Filled squares: single-layer
materials and most bonding subband for multilayers. Empty squares:
most antibonding subband. Half-filled squares: nonbonding subband.
Dotted lines connect subband values. Bars give k$_z$ dispersion of r
in primitive tetragonal materials, For reference for a-m, refer to
\cite{MaterialTc}.} \label{TcMaterial}
\end{figure}

\noindent (4). Material and structural dependence;\\
\indent  There is a strong material and structural dependence to the
high-T$_c$ superconductivity, as exemplified in
Fig.\ref{TcMaterial})\cite{MaterialTc,ShenPM}. Empirically it is
found that, for a given homologous series of materials, the optimal
T$_c$ varies with the number of adjacent CuO$_2$ planes, n, in a
unit cell: T$_c$ goes up first with n, reaching a maximum at n=3,
and goes down as n further increases. For the cuprates with the same
number of CuO$_2$ layers, T$_c$ also varies significantly among
different classes. For example, the optimal T$_c$ for one-layered
(La$_{2-x}$Sr$_x$)CuO$_4$ is 40K while it is 95K for one-layered
HgBa$_2$CuO$_4$. These behaviors are clearly beyond simplified
models that consider CuO$_2$ planes only, such as the t$-$J model.
In fact, such effects were taken as evidence against theoretical
models based on such simple models and in favor of the interlayer
tunneling model\cite{AndersonILT}.  Although the interlayer
tunneling model has inconsistencies with some experiments, the issue
that the material dependence cannot be explained by single band
Hubbard and t-J model remains to be true.

The above results suggest that the lattice degree of freedom plays
an essential role. However, the role of phonons has not been
scrutinized as much, in particular in regard to the intriguing
question of whether high-T$_c$ superconductivity involves a special
type of electron-phonon coupling.  In other words, the complexity of
electron-phonon interaction has not been as carefully examined as
some of the electronic models.  As a result, many naive arguments
are used to argue against electron-phonon coupling as if the
conclusions based on simple metals are applicable here.  Recently,
a large body of experimental results from  angle-resolved
photoemission, as we review below, suggest that electron-phonon
coupling in cuprates is not only strong but shows  behaviors
distinct from conventional electron-phonon coupling. In particular,
the momemtum dependence and the interaction between electron-phonon
interaction and electron-electron interaction are very important.

\subsection{Electron-Phonon Coupling: Theory}

\subsubsection{General}

Theory of electron-phonon interaction in the presence of strong
electron correlation has not been developed.  Given both
interactions are important in cuprates, it is difficult {\it a
priori} to have a good way to address these issues.  In fact, we
believe that an important outcome of our research is the stimulus to
develop such a theory.  In the mean time, our strategy to separate
the problem in different regimes and see to what extent we can
develop a heuristic understanding of the experimental data.  Such
empirical findings can serve as a guide for comprehensive theory. We
now start our discussion with an overview of existing theories of
electron-phonon physics.

The theories of electron-phonon coupling in condensed matter have
been developed rather separately for metals and insulators. In the
former case, the dominant energy scale is the kinetic energy or the
Fermi energy $\varepsilon_F$ on order of $1-10$eV, and the phonon
frequency $\Omega \sim 1-100$meV is much smaller. The Fermi
degeneracy protects the many-body fermion system from perturbations
and only the small energy window near the Fermi surface responds.
Therefore even if the lattice relaxation energy $E_{LR} =
g^2/\omega$ for the localized electron is comparable to the kinetic
energy $\varepsilon_F$ the el-ph coupling is essentially weak and
the perturbative treatment is justified. The dimensionless coupling
constant $\lambda$ is basically the ratio of $E_{LR}/\varepsilon_F$,
which ranges $\lambda \cong 0.1-2$ in the usual metals. In the
diagrammatic language, the physics described above is formulated
within the framework of the Fermi liquid theory\cite{AGD}. The el-el
interaction is taken care of by the formation of the quasi-particle,
which is well-defined near the Fermi surface, and the el-ph vertex
correction is shown to be smaller by the factor of
$\Omega/\varepsilon_F$ and can be neglected. Therefore the
multi-phonon excitations are reduced and the single-loop
approximation or at most the self-consistent Born approximation is
enough to capture the physics well, i.e., Migdal-Eliashberg
formalism.

When a carrier is put into an insulator, on the other hand, it stays
near the bottom of the quadratic dispersion and its velocity is very
small. The kinetic energy is much smaller than the phonon energy,
and the carrier can be dressed by a thick phonon cloud and its
effective mass can be very large. This is called the phonon polaron.
Historically the single carrier problem coupled to the optical
phonon through the long range Coulomb interaction, i.e., Fr\"ohlich
polaron, is the first studied model, which is defined in the
continuum. When one considers the tight-binding models, which is
more relevant to the Bloch electron, the bandwidth $W$ plays the
role of $\varepsilon_F$ in the above metallic case. Then again we
have three energy scales, $W$, $E_{RL}$, and $\Omega$. Compared with
the metallic case, the dominance of the kinetic energy is not
trivial, and the competition between the itinerancy and the
localization is the key issue in the polaron problem, which is
controlled by the dimensionless coupling constant $\lambda =
E_{RL}/W$. Another dimensionless coupling constant is $S =
E_{RL}/\Omega$, which counts the number of phonon quanta in the
phonon cloud around the localized electron. This appears in the
overlap integral of the two phonon wavefunctions with and without
the phonon cloud as:
\begin{equation}
< {\rm phonon \ \ vacuum}| {\rm phonon \ \ cloud}> \propto e^{-S}
\label{eq:EPgeneral}
\end{equation}
\noindent This factor appears in the weight of the zero-phonon line
of the spectral function of the localized electron, and $S$ can be
regarded as the maximum value for the number of phonons $N_{\rm ph}$
near the electron. In a generic situation, $N_{\rm ph}$ is
controlled by $\lambda$, and there are cases where $N_{\rm ph}$
shows an (almost) discontinuous change from the itinerant undressed
large polaron to the heavily dressed small polaron as $\lambda$
increases. This is called the self-trapping transition.  Here a
remark on the terminology ``self-trapping" is in order. Even for the
heavy mass polaron, the ground state is the extended Bloch state
over the whole sample and there is no localization. However a small
amount of disorder can cause the localization. Therefore in the
usual situation, the formation of the small polaron implies the
self-trapping, and we use this language to represent the formation
of the thick phonon clouds and huge mass enhancement. In cuprates,
it is still a mystery why the transport properties of the heavily
underdoped samples do not show the strong localization behavior even
though the ARPES shows the small polaron formation as will be
discussed in D.1.

Now the most serious question is what is the picture for the el-ph
coupling in cuprates ? The answer seems not so simple, and depends
both on the hole doping concentration, momentum and energy. The
half-filled undoped cuprate is a Mott insulator with
antiferromagnetic ordering, and a single hole doped into it can be
regarded as the polaron subjected to the hole-magnon and hole-phonon
interactions. At finite doping, but still in the antiferromagnetic
(AF) order, the small hole pockets are formed and the hole kinetic
energy can be still smaller than the phonon energy. In this case the
polaron picture still persists. The main issue is to what range this
continues. One scenario is that once the antiferromagnetic order
disappears the metallic Fermi surface is formed and the system
enters the Migdal-Eliashberg regime. However, there are several
physical quantities such as the resistivity, Hall constant, optical
conductivity, which strongly suggest that the physics still bears a
strong characteristics of doped holes in an insulator rather than a
simple metal with large Fermi surface.  Therefore the crossover hole
concentration $x_c$ between the polaron picture and the
Migdal-Eliashberg picture remains an open issue. Probably, it
depends on the momentum/energy of the spectrum. For example, the
electrons have smaller velocity and are more strongly coupled to the
phonons in the anti-nodal region near $(\pm \pi,0)$, $(0,\pm \pi)$,
remaining polaronic up to higher doping, while in the nodal region,
the electrons behave more like the conventional metallic ones since
the velocity is large along this direction. Furthermore, the low
energy states near the Fermi energy are well described by Landau's
quasi-particle and Migdal-Eliashberg theory, while the higher energy
states do not change much with doping even at $x \cong 0.1$
\cite{KShenPolaron} suggestive of polaronic behavior. In any event,
the dichotomy between the hole doping picture and the metallic
(large) Fermi surface picture is the key issue in the research of
high T$_c$ superconductors.

\subsubsection{Weak Coupling -- Perturbative and Self-Energy Description}

We review  first the Migdal-Eliashberg regime, in which the
electron-phonon interaction results in single-phonon excitations and
can be considered as a perturbation to the bare band dispersion. In
this case, dominant features of the mode coupling behavior can be
captured using the following form for the self-energy:

\begin{equation}
\widehat{\Sigma}(k,\omega) = T/N\sum_{q,\nu} g^{2}(k,q) D(q,i\nu)
\tau_3\widehat{G}(k-q,i\omega - i\nu)\tau_3 \label{eq:selfenergy}
\end{equation}

\noindent where $D(q,\omega) = {\frac {2\Omega_{q}}{\omega^{2} -
\Omega_{q}^{2}}}$ is the phonon propagator,  $\Omega_{q}$ is the
phonon energy, T is temperature, N is the number of particles and
$\tau_3$ is the Pauli matrix.

\begin{figure}[tbp]
\begin{center}
\includegraphics[width=0.9\linewidth,angle=0]{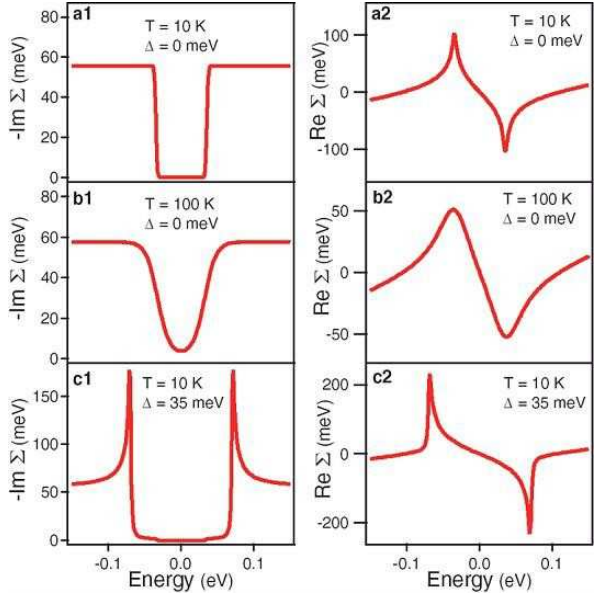}
\end{center}
\caption{Self-energy for electrons coupled to an Einstein mode with
$\Omega$ = 35 meV and electron-phonon vertex $g =0.15$
eV\cite{TanjaReview}. (a1), (b1), and (c1) plots Im$\Sigma$= -$Z_2
\omega +\chi_2$  for a normal state electron at 10 K, for a normal
state electron at 100 K, and for an electron in an s-wave
superconducting state at 10 K, respectively. (a2), (b2), and (c2)
plots the corresponding real parts, Re$\Sigma$ , obtained using the
Kramers-Kronig relation.}\label{SelfEnergySimulation}
\end{figure}

In this form of the self-energy, corrections to the electron-phonon
vertex, $g$, are neglected as mentioned above\cite{Migdal}.
Furthermore, we assume only one-iteration of the coupled self-energy
and Green's function equations.  In other words, in the equation for
the self- energy, $\Sigma$, we assume bare electron and phonon
propagators, G$_{0}$ and D$_{0}$. With these assumptions, the
imaginary parts of the functions $Z$, $\chi$, and $\phi$, denoted as
$Z_{2}$, $\chi_{2}$, and $\phi_{2}$, are:

\begin{displaymath}
Z_{2}(k,\omega)\omega = \sum_{q} g^{2}(k,q)(\pi/2)\{[\delta(\omega -
\Omega_{q} - E_{k-q})
\end{displaymath}
\begin{displaymath}
+ \delta(\omega - \Omega_{q} + E_{k-q})][f(\Omega_{q} - \omega) +
n(\Omega_{q})]
\end{displaymath}
\begin{equation}
 +[-\delta(\omega + \Omega_{q} - E_{k-q}) - \delta(\omega +
\Omega_{q} + E_{k-q})][f(\Omega_{q} + \omega) + n(\Omega_{q})]\}
\label{eq:TanjaZ2}
\end{equation}

\begin{displaymath}
\chi_{2}(k,\omega)= \sum_{q} g^{2}(k,q)(\pi\varepsilon_{k-q}
/2E_{k-q})\{[-\delta(\omega - \Omega_{q} - E_{k-q})
\end{displaymath}
\begin{displaymath}
+ \delta(\omega - \Omega_{q} + E_{k-q})][f(\Omega_{q} - \omega) +
n(\Omega_{q})]
\end{displaymath}
\begin{equation}
+ [\delta(\omega + \Omega_{q} - E_{k-q}) - \delta(\omega +
\Omega_{q} + E_{k-q})][f(\Omega_{q} + \omega) + n(\Omega_{q})]\}
\label{eq:Tanjax2}
\end{equation}

\begin{displaymath}
\phi_{2}(k,\omega)= \sum_{q} g^{2}(k,q)(\pi\Delta_{k-q}
/2E_{k-q})\{[\delta(\omega - \Omega_{q} - E_{k-q})
\end{displaymath}
\begin{displaymath}
- \delta(\omega - \Omega_{q} + E_{k-q})][f(\Omega_{q} - \omega) +
n(\Omega_{q})]
\end{displaymath}
\begin{equation}
+ [-\delta(\omega + \Omega_{q} - E_{k-q}) + \delta(\omega +
\Omega_{q} + E_{k-q})][f(\Omega_{q} + \omega) + n(\Omega_{q}]\}
\label{eq:Tanjap2}
\end{equation}

\noindent where $f(x)$, $n(x)$, are the Fermi, Bose distribution
functions and $E_{k}$ is the superconducting state dispersion,
$E_{k}^2 = \varepsilon_{k}^{2} + \Delta_{k}^2$.

The above equations are essentially those of Eliashberg theory for
strongly-coupled superconductors. Although $\lambda$ can be large
($> 1$), i.e.,  ``strongly-coupled",  the vertex corrections and
multi-phonon processes are still negligible due to the Fermi
degeneracy and small $\Omega/E_{F}$\cite{Eliashberg}. To illustrate
the essential features of mode coupling, we consider an Einstein
phonon coupled isotropically to a parabolic band.  We present this
calculation in the spirit of Engelsberg and Schrieffer, who first
calculated the spectral function for an electron-phonon coupled
system\cite{Englesberg} and which provided the foundation for the
later work by Scalapino, Schrieffer, and Wilkins\cite{PbTheory} in
the superconducting state.  Fig.\ref{SelfEnergySimulation}  plots
$-Z_{2}\omega + \chi_{2}$, the imaginary part of the phonon
self-energy, Im$\Sigma$, that represents the renormalization to the
diagonal channel of the electron propagator, or the one in which the
charge number density is subjected to electron-phonon interactions.
This part of the self-energy gives a finite lifetime to the
electron, and consequently broadens the peak in the spectra
(Im$\Sigma$ in $A(k,\omega)$  (Eq.\,\ref{eq:SpecFunc}) is the
half-width-at-half-maximum, HWHM of the peak). In the normal state,
$-Z_{2}\omega + \chi_{2}$ takes the familiar
form:\\

\begin{displaymath}
{\rm Im}\Sigma(k,\omega) = \Sigma_{q} -\pi g^{2}(k,q)[2n(\Omega_{q})
+
\end{displaymath}
\begin{equation}
f(\Omega_{q}+\omega) + f(\Omega_{q}-\omega)]\delta(\omega - E_{k-q})
\label{eq:SEimG}
\end{equation}

\noindent which when integrated over q becomes:
\begin{equation}
{\rm Im}\Sigma(k,\omega) = \int
d\Omega\alpha^{2}_{k}F(\Omega)[2n(\Omega) + f(\Omega+\omega) +
f(\Omega-\omega)] \label{eq:SEimS}
\end{equation}

\noindent where $\alpha^{2}_{k}F(\Omega)$, the Eliashberg function,
represents the coupling of the electron with Fermi surface momentum
$k$, to all $\Omega$ phonons connecting that electron to other
points on the Fermi surface.

For the normal state electron at 10K
(Fig.\ref{SelfEnergySimulation}a1), there is a sharp onset of the
self-energy that broadens the spectra beyond the mode energy; for
the normal state electron at 100K (Fig.
\ref{SelfEnergySimulation}b1), the onset of the self-energy is much
smoother and occurs over $\sim$ 50meV; for the superconducting state
electron (Fig. \ref{SelfEnergySimulation}c1), there is a singularity
that causes a much more abrupt broadening of the spectra at the
energy $\Omega$ + $\Delta$. The superconducting state singularity is
due to the density of states pile-up at the gap energy; the energy
at which the decay onsets shift by $\Delta$, since below the gap
energy there are no states to which a hole created by photoemission
can decay. For each of these imaginary parts of the self-energy, one
can use the Kramers-Kronig transform to obtain the real part of the
self energy, which renormalizes the peak position (${\rm Re} \Sigma$
in  $A(k,\omega)$  (Eq.\,\ref{eq:SpecFunc}) changes the position of
the peak in the spectral function). The real self energies thereby
obtained are also plotted in Fig. \ref{SelfEnergySimulation}a2, Fig.
\ref{SelfEnergySimulation}b2, and Fig. \ref{SelfEnergySimulation}c2.
In the superconducting state, again there is a singularity that
causes a more abrupt break from the bare-band dispersion at the
energy $\Omega$ + $\Delta$.

\begin{figure*}[Floatfix]
\begin{center}
\includegraphics[width=0.95\linewidth,angle=0]{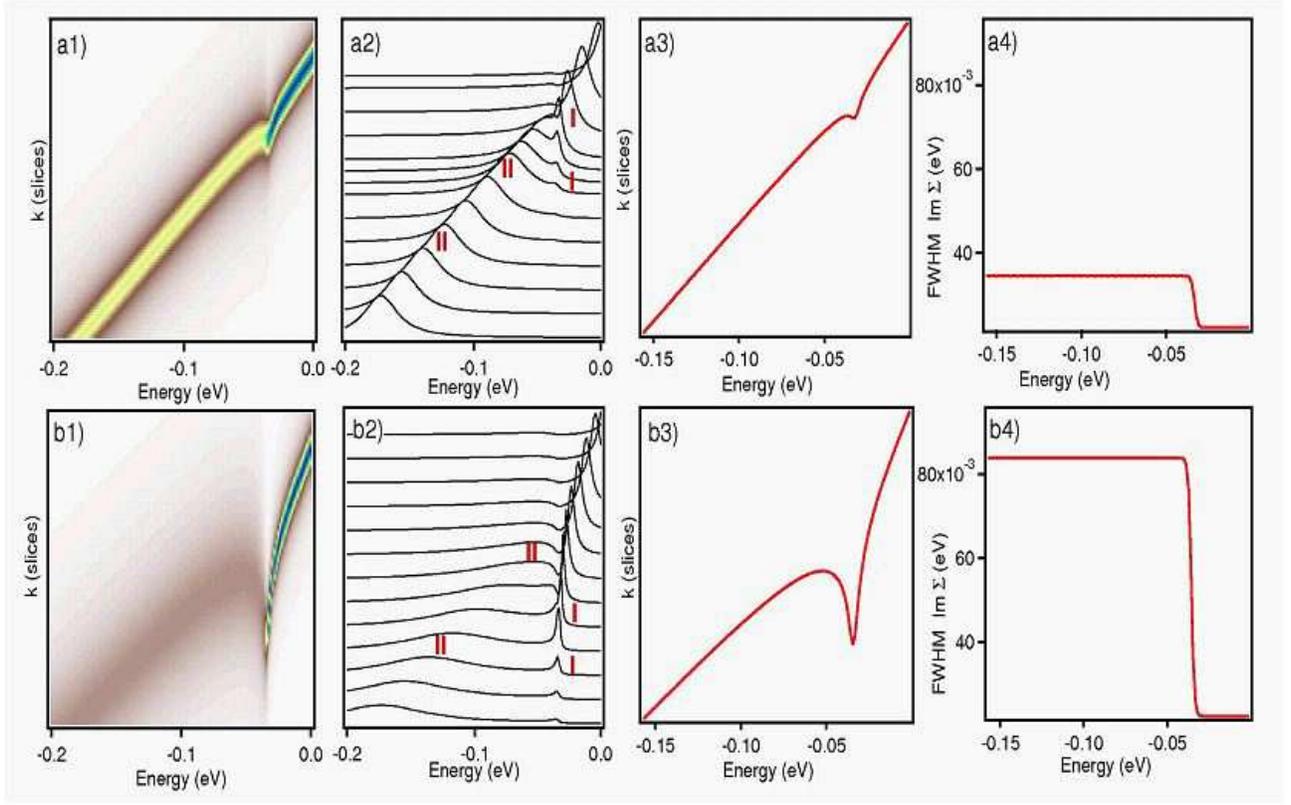}
\end{center}
\caption{Simulated electron-phonon coupling using Einstein model.
Spectral function (a1, b1), EDCs (a2, b2), MDC-derived dispersion
(a3, b3), and the MDC-derived width (a4, b4) (imaginary part of
self-energy) for two different couplings (a weak, b five times
stronger) to a linear bare band.}\label{EPSimulation}
\end{figure*}

For most metals, where the electrons are weakly interacting, and
therefore the poles of the spectral function are well-defined, one
would expect such a treatment to hold and indeed it does, as
evidenced by several cases including
Beryllium\cite{Jensen,Hensberger} and Molybdenum\cite{VallaMo}. {\it
A priori}, one might not expect the same to hold in ceramic
materials such as the copper-oxides, where the copper d-wave
electrons are localized and subject to strong electron-electron and
electron-phonon interactions. Nonetheless, in the superconducting
state of the copper-oxides at optimal and overdoped regime, one
recovers narrow peaks (20$\sim$30meV) of the spectral function. The
above self-energy, then, is able to describe the phenomenology of
the mode-coupling behavior for the superconducting state.  The
difference between the self-energy induced for a particular mode and
coupling constant in the normal state at T=100K
(Fig.\ref{SelfEnergySimulation}) and the superconducting state at
T=10K (Fig.\ref{SelfEnergySimulation}) also shows the extent to
which one can expect a temperature-dependent mode coupling in the
high-T$_c$ cuprates.

To illustrate the salient features of mode-coupling on the
dispersion, we consider a linear bare band coupled to an Einstein
phonon in the normal state at T=10K. The effect of electron-phonon
interaction on the one-electron spectral weight $A(k, \omega)$  of a
d$_{x^2-y^2}$ superconductor has been simulated by Sandvik et
al.\cite{Sandvik}.  In Fig.\ref{EPSimulation}, we show image plots,
EDCs, MDC derived dispersions, and the MDC extracted widths for two
different coupling constants (the case of stronger coupling is a
factor of five increase in the vertex, $g^{2}$).

There are three characteristic signatures of mode coupling behavior
evident:

1) A break up of a single dispersing peak into two
branches(Fig.\ref{EPSimulation}a1 and b1)---a peak that decays as it
asymptotically approaches the mode energy (I in Fig.
\ref{EPSimulation}a2 and b2), and a hump that traces out a
dispersing band (II in Fig.\ref{EPSimulation}a2 and b2).

2) In the image plots (Fig.\ref{EPSimulation}a1 and b1), a
significant broadening of the spectra beyond the mode energy is
readily apparent.  This is also the origin of the broad hump of the
dispersing band seen in the EDCs (Fig.\ref{EPSimulation}a2 and b2)
and the step in the extracted widths (or lifetime)
(Fig.\ref{EPSimulation}a4 and b4).

3) At the mode energy itself, there is a ``dip" between the peak and
the hump in the EDCs (Fig. \ref{EPSimulation}a2 and b2) leading to
the ``peak-dip-hump" structure often discussed in the literature.

From these generic features of electron-phonon coupling, one could
ascertain the mode energy and coupling strength.  Theoretically, the
mode energy should be the energy to which the peak in the EDC curve
decays. If there is a well-defined peak that has enough phase space
range to decay, the last point at which it can be measured is the
best indication of the mode energy. Otherwise, estimates can be made
from the EDC, MDC-derived dispersions, and the position of the step
in the MDC widths.   The coupling strength is indicated by the
extent of the break up of the spectra into a peak and a hump, the
sharpness of the ``kink" in the MDC-derived dispersion, and the
magnitude of the step in the MDC-derived widths. Quantitative
assessments of the coupling strength, however, require either a full
model calculation or an extraction procedure to invert the phonon
density of states coupled to the electronic spectra.

\subsubsection{Strong Coupling -- Polaron}

When the kinetic energy of the particles is less than the phonon
energy, the dressing of the phonon cloud could be large and the
el-ph coupling enters into the polaron regime. A single particle
coupled to the phonon is the typical case, on which extensive
theoretical studies have been done. Let $g(q)$ be the coupling
constant of the phonon with wavenumber $q$ to  the electrons, and
the lattice relaxation energy $E_{LR}$ is estimated as $E_{LR} \cong
<|g(q)|^2>/\Omega$. When this $E_{LR}$ is smaller than the
bandwidth, the effective reduction of the el-ph coupling due to the
rapid motion of the electron, i.e., the motional narrowing, occurs
and the weight of the one-phonon side-band is of the order of
$g(q)/W$ with the number of the phonon quanta $N_{\rm ph}$ being
estimated as $N_{\rm ph} \sim <|g(q)|^2>/W^2 \sim S (\Omega/W)^2$
where $S = E_{LR}/\Omega$.  As the el-ph coupling constant
increases, the polaron state evolves from this weak coupling large
polaron to the strong coupling small polaron. This behavior is
non-perturbative in nature, and the theoretical analysis is rather
difficult. One useful method is the adiabatic approximation where
the frequency of the phonon is set to be zero while $E_{LR}$ remains
finite. In this limit, one can regard the phonon as a classical
lattice displacement, whose Fourier component is denoted by $Q_q$.
Then one can investigate the stability of the weak coupling large
polaron state, i.e., zero distortion state in the present
approximation, by the perturbative way. Namely the energy gain
second order in $g(q)$ reads:

\begin{equation}
\delta E = - { 1 \over N} \sum_{q,\Omega} {{g(q)^2} \over { E(q) -
E(0)}} Q_q Q_{-q} \label{eq:NaotoGain}
\end{equation}

\noindent with the energy dispersion $E(k)$ of the electron. Here
the electron is at the ground state with the energy $E(0)$ in the
unperturbed state. Introducing the index $\ell$ characterizing the
range of the coupling as $g(q) \propto q^{-\ell}$,  and considering
the smallest possible wavenumber $q_{\rm min} \propto N^{-1/d}$ for
the linear sample size $L = N^{1/d}$ in spatial dimension $d$, one
can see that the index
\begin{equation}
s=d- 2(1 + \ell) \label{eq:NaotoS}
\end{equation}
separates the two different behavior of $\delta E$. For $s>0$,
$\delta E$ for $q=q_{\rm min}$ goes to zero as $N \to \infty$, which
suggests  that the weak coupling state is always locally stable,
separated by an energy barrier from the strong coupled small polaron
state. This means that a discontinuous change from the weak to
strong coupling polaron states occurs where the mass becomes so
heavy that the carrier is easily localized by impurities. Namely,
the self-trapping transition occurs. For $s<0$, on the other hand,
the zero distortion state is always unstable for infinitesimal
$g(q)$ and hence the lowest energy state continues smoothly as the
coupling increases, i.e., no self-trapping transition.  The most
relevant case of the short range el-ph coupling in two-dimensions,
i.e., $d=2$, $\ell=0$, corresponds to $s=0$, and hence is the
marginal class. Therefore whether the self-trapping transition
occurs or not is determined by the model of interest, and is
nontrivial.

For the study of the polaron in the intermediate to strong coupling
region, one needs to invent a reliable theoretical method to
calculate the energy, phonon cloud, effective mass, and the spectral
function. Up to very recently, it has been missing but the
diagrammatic quantum Monte Carlo method\cite{Prok} combined with the
stochastic analytic continuation \cite{Analytic} enabled the
``numerically exact" solution to this difficult problem. By this
method, the crossover from the weak to strong coupling regions have
been analyzed accurately for various models \cite{Pekar,optical}
With this method, the polaron problem in the t-J model has been
studied, and detailed information on the spectral function is now
available which can be directly compared with experimental results.
It is found that the self-trapping transition occurs in the
two-dimensional t-J polaron model, and in comparison with
experiment, the realistic coupling constant for the undoped case
corresponds to the strong coupling region. Namely the single hole
doped into the undoped cuprates is self-trapped. See below (IV. D)
for more details of how the polaron model relates to such
experimentally determinable quantities as the lineshape, dispersion,
and the chemical potential shift with doping.

Now we turn to the ARPES measurements that can be related to the two
regimes of electron-phonon coupling. We will first review the band
renormalization effects along the (0,0)-($\pi$,$\pi$) nodal
direction and near the ($\pi$,0) antinodal region. The weak
electron-phonon coupling picture is useful in accounting for many
observations. However, there are experimental indications that defy
the conventional electron-phonon coupling picture. Then we will move
on to review the polaron issue which manifests in undoped and
heavily underdoped samples.

\subsection{Band Renormalization and Quasiparticle Lifetime Effects}

\subsubsection{El-Ph Coupling Along the (0,0)-($\pi$,$\pi$) Nodal Direction}

\begin{figure}[tbp]
\begin{center}
\includegraphics[width=1.0\linewidth,angle=0]{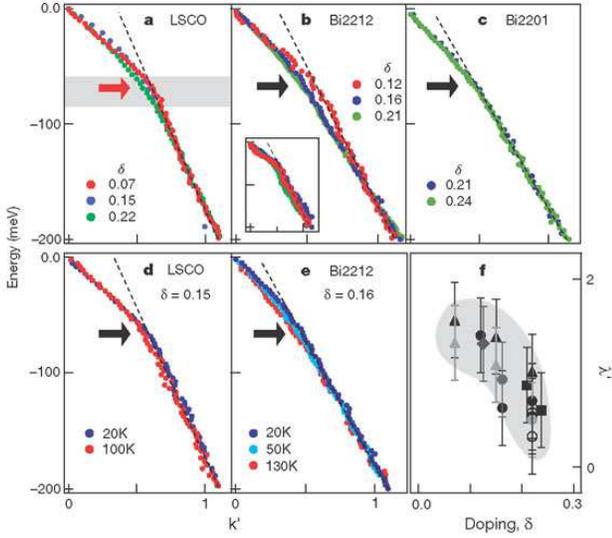}
\end{center}
\caption{Ubiquitous existence of a kink in the nodal dispersion of
various cuprate materials\cite{LanzaraKink}. Top panels (a, b, c)
plot dispersions along (0, 0)-($\pi$,$\pi$) direction (except for
panel b inset, which is off this direction) as a function of the
rescaled momentum k' for different samples and at different doping
levels ($\delta$): (a) LSCO at 20 K, (b) Bi2212 in superconducting
state at 20 K, and (c) Bi2201 in normal state at 30 K. Dotted lines
are guides to the eye. Inset in (b) shows that the kinks in the
dispersions off the (0, 0)-($\pi$,$\pi$) direction sharpen upon
moving away from the nodal direction. The black arrows indicate the
position of the kink in the dispersions. Panels (d) and (e) show the
temperature dependence of the dispersions for (d) optimally doped
LSCO (x=0.15) and (e) optimally-doped Bi2212, respectively. Panel
(f) shows the doping dependence of the effective electron-phonon
coupling strength $\lambda$' along the (0, 0)-($\pi$,$\pi$)
direction. Data are shown for LSCO (filled triangles), Nd-doped LSCO
(1/8 doping; filled diamonds), Bi2201 (filled squares), and Bi2212
(filled circles in the first Brillouin zone and unfilled circles in
the second zone). The different shadings represent data obtained in
different experimental runs. Shaded area is a guide to the
eye.}\label{ALPhonon}
\end{figure}

\begin{figure}[tbp]
\begin{center}
\includegraphics[width=1.0\linewidth,angle=0]{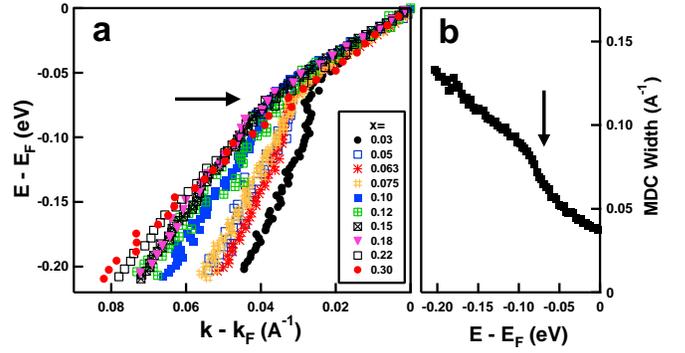}
\end{center}
\caption{Doping dependence of the nodal electron dynamics in LSCO
and universal nodal Fermi velocity\cite{XJZhouUniversalVF}. (a)
Dispersion of LSCO with various doping levels (x=0.03 to 0.30)
measured at 20 K along the (0,0)-($\pi$,$\pi$) nodal direction. The
arrow indicates the position of kink that separates the dispersion
into high-energy and low-energy parts with different slopes. (b).
Scattering rate as measured by MDC width
(full-width-at-half-maximum, FWHM) of the LSCO (x = 0.063) sample
measured at 20 K. The arrow indicates a drop at an energy $\sim$70
meV. }\label{UVelocity}
\end{figure}

The nodal direction denotes the (0,0)-($\pi$,$\pi$) direction in the
Brillouin zone (Fig. \ref{CuO2Plane}b). The {\it d}-wave
superconducting gap is zero along this particular direction. As
shown in Fig.\ref{ALPhonon} and Fig.\ref{UVelocity}a, the
energy-momentum dispersion curves from MDC method exhibit an abrupt
slope change (``kink") near 70 meV. The kink is accompanied by an
accelerated drop in the MDC width at a similar energy scale (Fig.
\ref{UVelocity}b). The existence of the kink has been well
established as ubiquitous in hole-doped cuprate
materials\cite{BogdanovKink,LanzaraKink,KaminskiKink,PJohnsonKink,BorisenkoNodalKink,XJZhouUniversalVF,GweonIsotope}:

1.  It is present in various hole-doped cuprate materials, including
Bi$_2$Sr$_2$CaCu$_2$O$_8$ (Bi2212), Bi$_2$Sr$_2$CuO$_6$ (Bi2201),
(La$_{2-x}$Sr$_x$)CuO$_4$ (LSCO) and others. The energy scale (in
the range of 50-70 meV) at which the kink occurs is similar for
various systems.

2.  It is present both below T$_c$ and above T$_c$.

3.  It is present over an entire doping range (Fig.
\ref{UVelocity}a). The kink effect is stronger in the underdoped
region and gets weaker with increasing doping.

    While there is a consensus on the data, the exact meaning of the
data is still under discussion. The first issue concerns whether
the kink in the normal state is related to an energy scale. Valla
et al. argued that the system is quantum critical and thus has no
energy scale, even though a band renormalization is present in the
data\cite{VallaScience}. Since their data do not show a sudden
change in the scattering rate at the corresponding energy, they
attributed the kink in Bi2212 above T$_c$ to the marginal Fermi
liquid (MFL) behavior without an energy scale\cite{PJohnsonKink}.
Others believe the existence of energy scale in the normal and
superconducting states has a common origin, i.e., coupling of
quasiparticles with low-energy collective excitations
(bosons)\cite{BogdanovKink,LanzaraKink,KaminskiKink}. The sharp
kink structure in dispersion and concomitant existence of a drop
in the scattering rate which is becoming increasingly clear with
the improvement of signal to noise in the data, as exemplified in
underdoped LSCO (x=0.063) in the normal state (Fig.
\ref{UVelocity}b)\cite{XJZhouDichotomy}, are apparently hard to
reconcile with the MFL behavior.

\begin{figure}[tbp]
\begin{center}
\includegraphics[width=1.0\linewidth,angle=0]{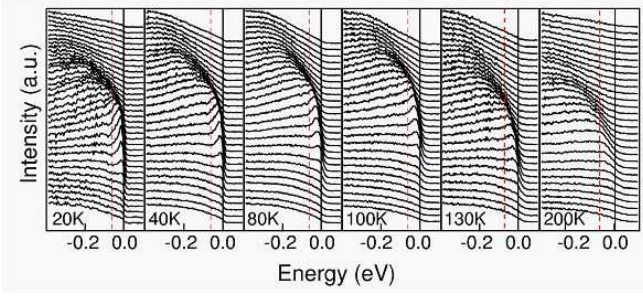}
\end{center}
\caption{Energy Distribution curves (EDC) in the normal state of
underdoped Bi2201 (T$_c$= 10K) at several temperatures (from 20K to
200K)\cite{LanzaraBi2201}.  A dip in the EDCs can be clearly
observed almost for all the temperatures. The dip position (dotted
line) is ~60meV and is roughly temperature independent.
}\label{LanzaraBi2201}
\end{figure}

The existence of a well-defined energy scale over an extended
temperature range is best seen in Bi2201
compound\cite{LanzaraBi2201}. As shown in Fig.\ref{LanzaraBi2201},
the spectra reveal a ``peak-dip-hump" structure up to temperatures
near 130K, almost ten times the superconducting critical temperature
T$_c$. Such a ``peak-dip-hump" structure is very natural in an
electron-phonon coupled system, but will not be there if there is no
energy scale in the problem as argued by Valla et
al.\cite{VallaScience}.

    A further issue concerns the origin of the bosons involved in
the coupling, with a magnetic resonance
mode\cite{KaminskiKink,PJohnsonKink} and optical
phonons\cite{LanzaraKink} being possible candidates considered. The
phonon interpretation is based on the fact that the sudden band
renormalization (or ``kink") effect is seen for different cuprate
materials, at different temperatures, and at a similar energy scale
over the entire doping range\cite{LanzaraKink}. For the nodal kink,
the phonon considered in the early work was the half-breathing mode,
which shows an anomaly in neutron
experiments\cite{PintschoviusBreathingMode,McQueeneyBreathingMode}.
Unlike the phonons, which are similar in all cuprates, the magnetic
resonance (at correct energy) is observed only in certain materials
and only below T$_c$. The absence of the magnetic mode in LSCO and
the appearance of magnetic mode only below T$_c$ in some cuprate
materials are not consistent with its being the cause of the
universal presence of the kink effect. Whether the magnetic
resonance can cause any additional effect is still an active
research topic\cite{KeeMagneticMode,AbanovMagneticMode}.

\begin{figure}[tbp]
\begin{center}
\includegraphics[width=0.9\linewidth,angle=0]{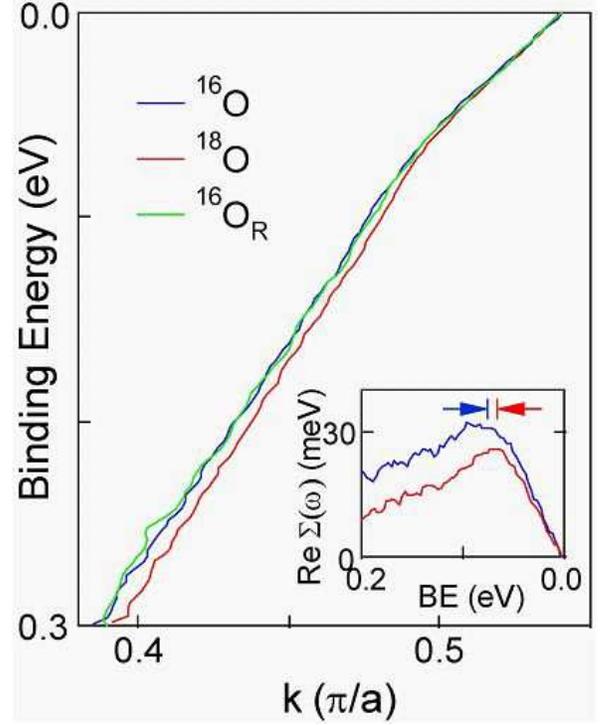}
\end{center}
\caption{Isotope-induced changes of the nodal
dispersion\cite{GweonIsotope}. The data were taken on optimally
doped Bi$_2$Sr$_2$CaCu$_2$O$_8$ samples (T$_c$ $\sim$ 91 to 92 K)
with different oxygen isotopes $^{16}$O and $^{18}$O at T $\sim$ 25
K along the nodal direction. The low energy dispersion is nearly
isotope-independent, while the high energy dispersion is
isotope-dependent. The effect is reversible by isotope
re-substitution (green). Inset shows the real part of the electron
self-energy, Re$\Sigma$ , obtained from the dispersion by
subtracting a line approximation for the one-electron band $E_k$,
connecting two points (one at E$_F$ and the other at a 300-meV
binding energy) of the $^{18}$O dispersion. }\label{EPIsotope}
\end{figure}

    To test the idea of electron-phonon coupling, an isotope exchange
experiment has been carried out\cite{GweonIsotope}. When exchanging
$^{18}$O and $^{16}$O in Bi2212, a strong isotope effect has been
reported in the nodal dispersions (Fig.\ref{EPIsotope}).
Surprisingly, however, the isotope effect mainly appears in the high
binding energy region above the kink energy; at the lower binding
energy near the Fermi level, the effect is minimal. This is quite
different from the conventional electron-phonon coupling where
isotope substitution will result in a small shift of phonon energy
while keeping most of the dispersion intact. The origin of this
behavior is still being investigated.

\begin{figure}[tbp]
\begin{center}
\includegraphics[width=1.0\linewidth,angle=0]{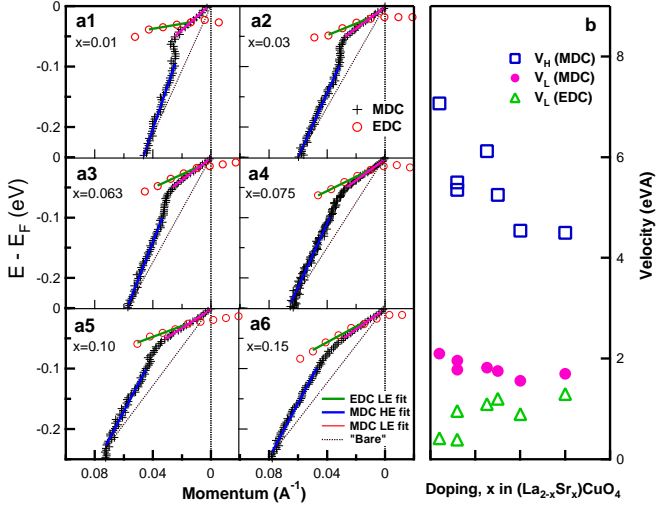}
\end{center}
\caption{(a). Energy-momentum dispersions for LSCO with different
dopings, using both EDC and MDC methods\cite{AndreiEDCMDC}.  The EDC
low-energy velocity is obtained by fitting the EDC dispersion
linearly in the intermediate energy range because the data points
very close to Fermi level is affected by the Fermi cutoff while the
data at higher energy have large uncertainty because the EDCs are
broader.  The MDC low (high) energy velocity v$_H$ is obtained by
fitting MDC dispersion at binding energy 0$\sim$50meV  (100~250meV)
using a linear line. (b). Low and high-energy velocities as a
function of doping obtained from MDC and EDC
dispersions.}\label{UVEDC}
\end{figure}

It is interesting to note in Fig. \ref{UVelocity}a that the energy
scale of the kink also serves as a dividing point where the high and
low energy dispersions display different doping
dependence\cite{XJZhouUniversalVF}. The dispersion in this Figure
were obtained by the MDC method. In Fig. \ref{UVEDC}a, we reproduce
some of these MDC-extracted dispersions,  but we also plot the
dispersion extracted using EDCs by following the EDC peak position.
Since the first derivative of the dispersion,
$\partial$E/$\partial$k, corresponds to velocity, the dispersions at
high binding energy (-0.1$\sim$-0.25eV) and low binding energy
(0$\sim$-0.05eV) are fitted by  straight lines to quantitatively
extract velocities, as plotted in Fig.
\ref{UVEDC}b\cite{AndreiEDCMDC}.

While nodal data clearly reveal the presence of coupling to
collective modes with well-defined energy scale, there are a couple
of peculiar behaviors associated with the doping evolution of the
nodal dispersion.  One obvious anomaly is the difference of low
energy velocity obtained from MDC and EDC methods(Fig.
\ref{UVEDC}b).  As seen from Fig. \ref{UVelocity}a and Fig.
\ref{UVEDC}b, the low-energy dispersion and velocity from the MDC
method is insensitive to doping over the entire doping range, giving
the so-called ``universal nodal Fermi velocity"
behavior\cite{XJZhouUniversalVF}. Similar behavior was also reported
in Bi2212\cite{PJohnsonKink}.  However, improved LSCO data where we
can resolve a well-defined quasiparticle peak to extract dispersion
using EDC method reveal a dichotomy in EDC and MDC derived
dispersions, particularly for low doping (Fig. \ref{UVEDC}), like
x=0.01\cite{XJZhou001}.  This discrepancy between EDC and MDC cannot
be reconciled within the conventional el-ph interaction picture, as
simulations considering experimental resolutions show.

In terms of conventional electron-phonon coupling, if one considers
that the ``bare band" does not change with doping but the
electron-phonon coupling strength increases with decreasing doping,
as it is probably the case for LSCO,  one would expect that the low
energy dispersion and velocity show strong doping dependence, while
the high-energy ones converge. However, one sees that the
high-energy velocity is highly doping dependent. Moreover, its trend
is anomalous if one takes electron-electron interaction into
account. It is known in cuprates that, with decreasing doping, the
electron-electron interaction gets stronger. According to
conventional wisdom, this would result in a larger effective mass
and smaller velocity. However, the doping dependence of the
high-energy velocity is just opposite to this expectation, as seen
from Fig. \ref{UVEDC}b.

Therefore, these anomalies indicate a potential deviation from the
standard Migdal-Eliashberg theory and the possibility of a complex
interplay between electron-electron and electron-phonon
interactions. As we discuss later, this phenomenon is a hint of
polaronic effect where the traditional analysis fails.  Such a
polaron effect gets stronger in deeply underdoped system even along
the nodal direction.  Under such a condition, one needs to use EDC
derived dispersion when the peaks are resolved.

\subsubsection{Multiple Modes in the Electron
Self-Energy}

In conventional superconductors, the successful extraction of the
phonon spectral function, or the Eliashberg function,
$\alpha^{2}F(\omega)$,  from electron tunneling data played a
decisive role in cementing the consensus on the phonon-mediated
mechanism of superconductivity\cite{Rowell}. For high temperature
superconductors, the extraction of the bosonic spectral function can
provide fingerprints for more definitive identification of the
nature of bosons involved in the coupling.

In principle, the ability to directly measure the dispersion, and
therefore, the electron self-energy, would make ARPES the most
direct way of extracting the bosonic spectral function. This is
because, in metals, the real part of the electron self-energy
Re$\Sigma$ is related to the Eliashberg function
$\alpha^{2}F(\Omega;\epsilon,\mathbf{\hat{k}})$ by:
\begin{equation}
\mathrm{Re}\Sigma(\mathbf{\hat{k}},\epsilon;T)=\int_{0}^{\infty}d\Omega\alpha^{2}F(\Omega;\epsilon,\mathbf{\hat{k}})K\left(\frac{\epsilon}{kT},\frac{\hbar\Omega}{kT}\right)\,,\label{eq:ReSigma}
\end{equation}

\noindent where

\begin{equation}
K(y,z)=\int_{-\infty}^{\infty}dx\frac{2z}{x^{2}-z^{2}}f(x+y)\,,\label{eq:MEMk}
\end{equation}
with $f(x)$ being the Fermi distribution function. Such a relation
can be extended to any electron-boson coupling system and the
function  $\alpha^{2}F(\omega)$ then describes the underlying
bosonic spectral function. We note that the form of
$\mathrm{Re}\Sigma(\mathbf{\hat{k}},\epsilon;T)$
(Eq.\,\ref{eq:ReSigma}) can be derived by taking the Kramers-Kronig
transformation of $\mathrm{Im}\Sigma$ for the normal state as shown
above (Eq.\,\ref{eq:SEimS}).  Unfortunately, given that the
experimental data inevitably have noise, the traditional
least-square method to invert an integral problem is mathematically
unstable.

Very recently,  Shi et al. have made an important advance in
extracting the Eliashberg function from ARPES data by employing the
maximum entropy method (MEM) and successfully applied the method to
Be surface states\cite{JRShiMEM}.  The MEM approach\cite{JRShiMEM}
is advantageous over the least squares method in that: (i) It treats
the bosonic spectral function to be extracted as a probability
function and tries to obtain the most probable one. (ii) More
importantly, it is a natural way to incorporate the {\it priori}
knowledge as a constraint into the fitting process. In practice, to
achieve an unbiased interpretation of data,  only  a few basic
physical constraints to the bosonic spectral function are imposed:
(a) It is positive. (b) It vanishes at the limit
$\omega$$\rightarrow$0. (c) It vanishes above the maximum energy of
the self-energy features. As shown in the case of Be surface state,
this method is robust in extracting the Eliasberg
function\cite{JRShiMEM}.

\begin{figure}[htb]
\includegraphics[width=0.9\linewidth]{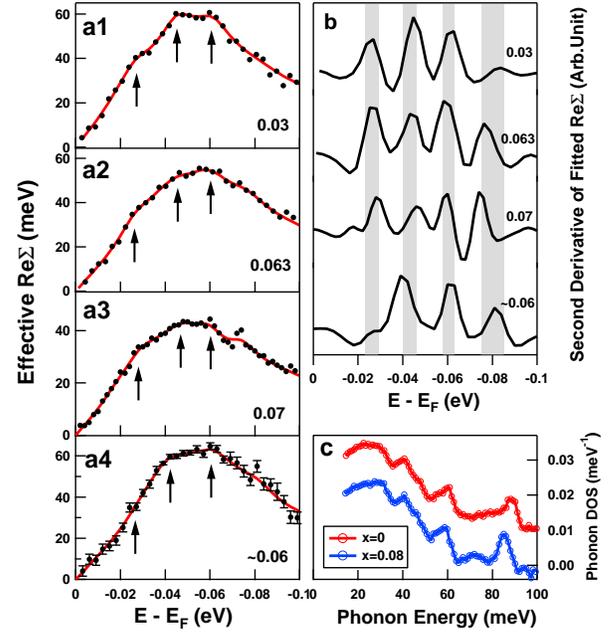}
\caption{Multiple modes coupling in electron self-energy in
LSCO\cite{XJZhouMultipleMode}.   (a).The effective real part of the
electron self-energy for LSCO x=0.03 (a1), 0.063 (a2), 0.07 (a3)and
$\sim$0.06 (a4) samples. Data (a1-a3) were taken using Scienta 2002
analyzer while data (a4) were taken using Scienta R4000 analyzer.
Data (a1-a3) were taking using 10eV pass energy at an energy
resolution of $\sim$18meV. Data (a4) were taken a $x \sim$0.06
sample using 5eV pass energy with an energy resolution of
$\sim$12meV. For clarity, the error bar is only shown for data (a4)
which becomes larger with increasing binding energy. The arrows in
the figure mark possible fine structures in the self-energy. The
data are fitted using the maximum entropy method (solid red lines).
The values of ($\alpha$$_1$, $\alpha$$_2$) (the unit of $\alpha$$_1$
and $\alpha$$_2$ are eV$\cdot$$\AA$ and eV$\cdot$$\AA$$^2$,
respectively) for bare band are (-4.25,0) for (a1), (-4.25, 13) for
(a2), (-3.7,7) for (a3) and (-4.3, 0) for (a4). (b). The
second-order derivative of the calculated Re$\Sigma$. The ruggedness
in the curves is due to limited discrete data points. The four
shaded areas correspond to energies of (23$\sim$29), (40$\sim$46),
(58$\sim$63) and (75$\sim$85) meV where the fine features fall in.
((c) The phonon density of state $F(\omega)$ for LSCO $x=0$ (red)
and $x=0.08$  (blue) measured from neutron
scattering\cite{McQueeneyDOS}. }\label{Multimode}
\end{figure}

Initial efforts have been made to extend this approach to underdoped
LSCO and evidence for electron coupling to several phonon modes has
been revealed\cite{XJZhouMultipleMode}. As seen from Fig.
\ref{Multimode}, from both the electron self-energy(Fig.
\ref{Multimode}a), and the derivative of their fitted curves ((Fig.
\ref{Multimode}a), one can identify two dominant features near
$\sim$ 40 meV and $\sim$60 meV. In addition, two addition modes may
also be present near $\sim$25meV and $\sim$75
meV\cite{XJZhouMultipleMode}.  The multiple features in Fig.
\ref{Multimode}b show marked difference from the magnetic excitation
spectra measured in LSCO which is mostly featureless and doping
dependent\cite{HaydenTranquada}. In comparison, they show more
resemblance to the phonon density-of-states (DOS), measured from
neutron scattering on LSCO
(Fig.\ref{Multimode}c)\cite{McQueeneyDOS}, in the sense of the
number of modes and their positions. This similarity between the
extracted fine structure and the measured phonon features favors
phonons as the bosons coupling to the electrons. In this case, in
addition to the half-breathing mode at 70$\sim$80 meV that we
previously considered strongly coupled to
electrons\cite{LanzaraKink}, the present results suggest that
several lower energy optical phonons of oxygens are also actively
involved. Particularly we note that the mode at $\sim$ 60meV
corresponds to the vibration of apical oxygens.

We note that, in order to be able to identify fine structure in the
electron self-energy, it is imperative to have both high energy
resolution and high statistics\cite{ZhouValla}.  These requirements
have made the experiment highly challenging because of the necessity
to compromise between two conflicting requirements for the
synchrotron light source: high energy resolution and high photon
flux. Further improvements in photoemission experiments will likely
enable a detailed understanding of the boson modes coupled to
electrons, and provide critical information for the pairing
mechanism.

One would like to extend this method to the superconducting state,
in momentum around the BZ, and to higher temperatures. The
superconducting state could, in principle, be achieved by using the
BCS dispersion of the quasiparticles rather than the normal state
dispersion and is currently under study. However, considering the
anisotropy of the el-ph coupling detailed below, the anisotropy of
the underlying band structure, and the d-wave superconducting gap,
extending the procedure in momentum may be somewhat more difficult.
The $\alpha^{2}F(\omega,\varepsilon,{\hat k})$ used for the above
form of the real part of the self-energy is assumed to be only
weakly dependent on the initial energy $\varepsilon$ and momentum
$k$ of the electron. But again, one in principle could begin to
consider a different form of the calculated Re$\Sigma$ and then
apply the MEM method with it instead. Extending the method to higher
temperatures, for example $\sim$ 100K for normal state Bi2212 data,
may be, however, a limitation that cannot be overcome. The method's
strength is in resolving fine structures due to the phonon density
of states. Those fine structures occur predominantly at lower
temperatures. At higher temperatures of $\sim$ 100K, the imaginary
and real parts of the self energy get broadened on the order of the
phonon energy itself. In that case, two or more neighboring phonons
would contribute to the electronic renormalization at a given
energy, both broadening the fine structures in the data and
weakening the resolving power of the method itself. So, while the
MEM method can directly extract fine features from ARPES data in
agreement with neutron scattering without implicitly assuming a
phonon model, it does not have the freedom to incorporate the
temperature and momentum dependence needed to describe the ARPES
data in both superconducting and normal states, near the vHS and
near the node. Both modelling of the data and direct extraction,
then, are needed, to gain a complete picture of the mode-coupling
features in the data.

\subsubsection{El-Ph Coupling Near the ($\pi$,0) Antinodal Region}

\begin{figure*}[floatfix]
\begin{center}
\includegraphics[width=0.95\linewidth,angle=0]{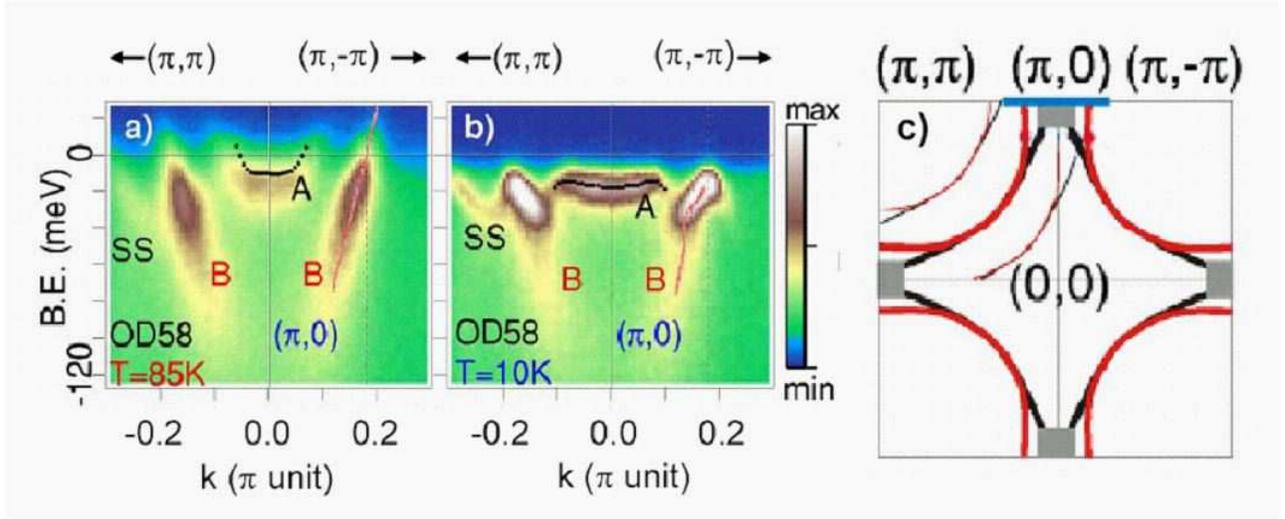}
\end{center}
\caption{Antinodal kink near ($\pi$,0) in a heavily overdoped Bi2212
sample (T$_c$$\sim$58 K). (a) Normal-state data (T=85 K) near the
antinodal region. (b) Superconducting-state data from the same
sample at 10 K, showing the emergence of a dispersion kink in the
bilayer split-B band. The B band dispersions (red curves) were
determined by fitting MDC peak positions. The black dots represent A
band EDC peak positions. (c) Brillouin zone with bonding band B
(thick red) and antibonding band A (thick black) Fermi surfaces, as
well as momentum-cut locations for panels (a) and (b) (blue bars).
The two sets of thin curves are replicas of Fermi surface
originating from the superstructure in Bi2212.}\label{DessauKink}
\end{figure*}

The antinodal region refers to the ($\pi$,0) region in the Brillouin
zone where the {\it d}-wave superconducting gap has a maximum (Fig.
\ref{CuO2Plane}b). Recently, a low-energy kink was also identified
near the ($\pi$,0) antinodal region in
Bi2212\cite{KaminskiKink,GromkoAntinodalKink,KimAntinodalKink,TCukAntinodalKink}.
This observation was made possible thanks to the clear resolution of
the bi-layer
splitting\cite{FengBilayersplitting,ChuangBilayerSplitting,BogdanovBilayerSplitting}.
As there are two adjacent CuO$_2$ planes in a unit cell of Bi2212,
these give rise to two Fermi surface sheets from the
higher-binding-energy bonding band (B) (thick red curves in Fig.
\ref{DessauKink}c) and the lower-binding- energy antibonding band
(A) (thick black curves in Fig. \ref{DessauKink}c).

Consider a cut along ($\pi$,$\pi$)-(-$\pi$,$\pi$) across ($\pi$,0)
in Bi2212, both above T$_c$ (Fig. \ref{DessauKink}a) and below T$_c$
(Fig. \ref{DessauKink}b)\cite{GromkoAntinodalKink}. Superimposed are
the dispersion of the bonding band determined from the MDC (red
lines) and antibonding band from the EDC (black dots). When the
bandwidth is narrow, the applicability of the MDC method in
obtaining dispersion becomes questionable so one has to resort to
the traditional EDC method. In the normal state, the bonding-band
dispersion (Fig. \ref{DessauKink}a) is nearly linear and featureless
in the energy range of interest. Upon cooling to 10 K (Fig.
\ref{DessauKink}b), the dispersion, as well as the near-$E_F$
spectral weight, is radically changed. In addition to the opening of
a superconducting gap, there is a clear kink in the dispersion
around 40 meV.

\begin{figure}[b!]
\begin{center}
\includegraphics[width=0.95\linewidth,angle=0]{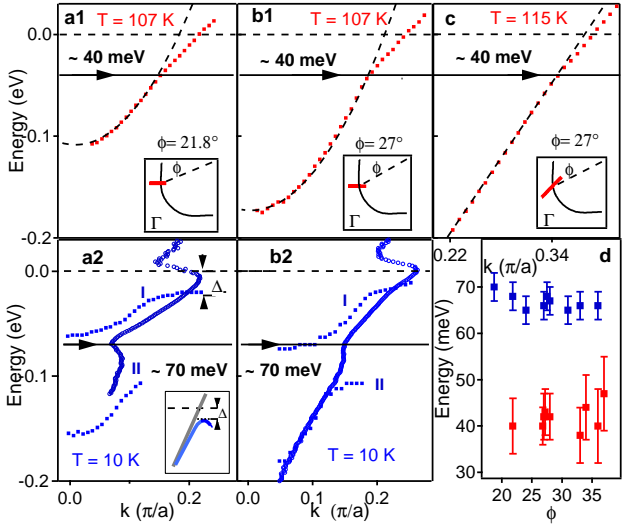}
\end{center}
\caption{Antinodal kink near ($\pi$,0) in the normal state (a1,b1,c)
and superconducting state (a2,b2) in an optimally-doped
Bi2212\cite{GromkoAntinodalKink}. The dispersions in a1, b1, and c
were derived by the EDC method; the position of the momentum cuts is
labelled in the insets. The red dots are the data; the fit to the
curve (black dashed line) below the 40-meV line is a guide to the
eye. The dispersions at the same location in the superconducting
state (10 K) are shown in (a2) and (b2), which were derived by the
MDC method (blue circles). Also plotted in (a2) and (b2) are the
peak (blue squares, I) and hump positions (blue squares, II) of the
EDCs for comparison. The inset of (a2) shows the expected behavior
of a Bogoliubov-type gap opening. The s-like shape below the gap
energy is an artifact of the way the MDC method handles the backbend
of the Bogoliubov quasiparticle. (d). Kink positions as a function
of momentum cuts in the antinodal region.}\label{CukKink}
\end{figure}

Gromko et al.\cite{GromkoAntinodalKink} reported that the antinodal
kink effect appears only in the superconducting state and gets
stronger with decreasing temperature. Their momentum-dependence
measurements show that the kink effect is strong near ($\pi$,0) and
weakens dramatically when the momentum moves away from the ($\pi$,0)
point. Excluding the possibility that this is a by-product of a
superconducting-gap opening, they attributed the antinodal kink to
the coupling of electrons to a bosonic excitation, such as a phonon
or a collective magnetic excitation. The prime candidate they
considered is the magnetic-resonance mode observed in inelastic
neutron scattering experiments.

The temperature and momentum dependence identified for a range of
doping levels has also led others to attribute the effect to the
magnetic resonance \cite{KaminskiKink,KimAntinodalKink}. However,
there are some inconsistencies with this interpretation: (1) the
magnetic resonance has not yet been observed by neutron scattering
in such a heavily doped cuprate and (2) the magnetic resonance has
little spectral weight, and may be too weak to cause the effect seen
by ARPES. Furthermore, the electron-phonon coupling in the early
tunneling spectra, such as Pb, appeared prominently only in the
superconducting state.  The linear MDC-derived dispersion in the
normal state of Bi2212 at (pi,0) that Gromko et. al.
reports\cite{GromkoAntinodalKink} is not conclusive enough proof
that the same mode does not couple to the electrons in the normal
state.   On the other hand, the clear determination of mode-coupling
by Gromko et. al. in the anti-nodal region, where the gap is
maximum, without the complication of bilayer splitting or
superstructure, suggests that the renormalization effects seen by
ARPES in the cuprates may indeed by related to the microscopic
mechanism of superconductivity.

\begin{figure*}[tbp]
\begin{center}
\includegraphics[width=0.95\linewidth,angle=0]{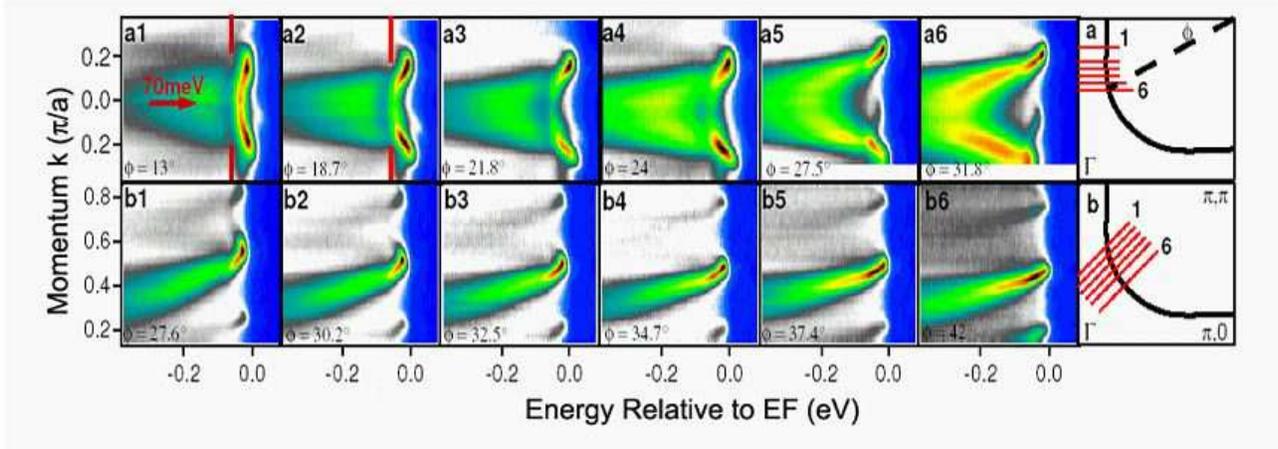}
\end{center}
\caption{Anisotropic electron-boson coupling in
Bi2212\cite{TCukAntinodalKink}.  Image plots in (a1-a6) and (b1-b6)
are cuts taken parallel to (0, $\pi$)-($\pi$, $\pi$) and (0,
0)-($\pi$, $\pi$)  respectively at the locations indicated in the
Brillouin zone ((a) and (b)) at 15 K for an optimally doped sample
(T$_c$ = 94 K).}\label{Bi2212kDependence}
\end{figure*}

\begin{figure}[tbp]
\begin{center}
\includegraphics[width=0.95\linewidth,angle=0]{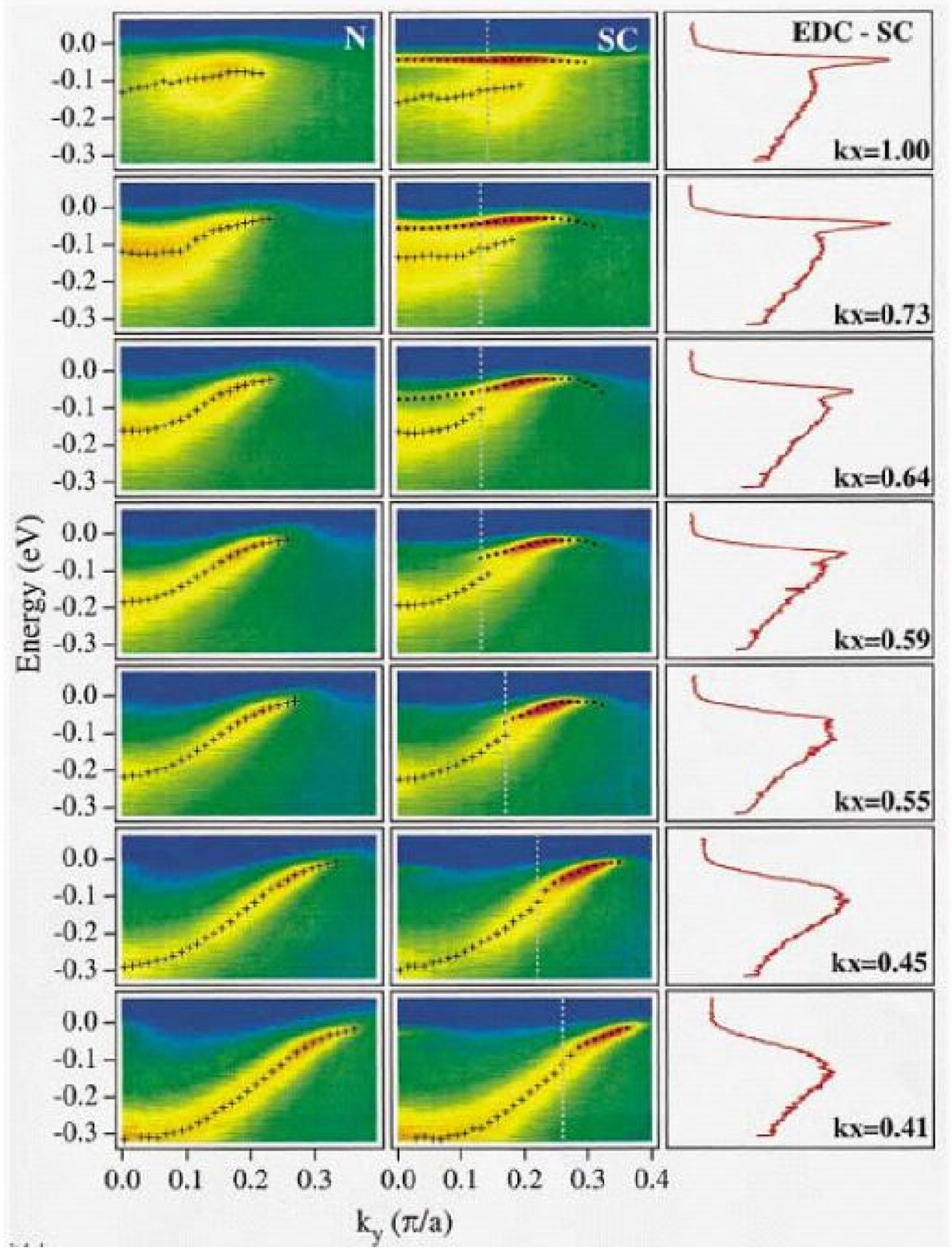}
\end{center}
\caption{Momentum dependence of photoemission data in
optimally-doped Bi2212\cite{KaminskiKink}.  Left panels:
Photoemission data in the normal state (T=140 K) along selected cuts
parallel to M($\pi$,0)-Y($\pi$,$\pi$). EDC peak positions are
indicated by crosses. Middle panels: Photoemission data in the
superconducting state (T=40 K) at the same cuts as for left panels.
Crosses indicate positions of broad high energy peaks, dots sharp
low energy peaks. Right panels: EDCs at locations marked by the
vertical lines in the middle panels.}\label{KaminskiKink}
\end{figure}

Cuk $\textit{et. al.}$ \cite{TCukAntinodalKink} and Devereaux
$\textit{et. al.}$ \cite{DevereauxAntinodalKink} have recently
proposed a new interpretation of the renormalization effects seen in
Bi2212. Specifically, the key observation that prompted them to rule
out the magnetic resonance interpretation is the unraveling of the
existence of the antinodal kink even in the normal state. This
observation was made possible by utilizing the EDC method because
the MDC method is not appropriate when the assumed linear
approximation of the bare band fails near ($\pi$,0) where the band
bottom is close to Fermi level E$_F$. Figs. \ref{CukKink}a1,
\ref{CukKink}b1, and \ref{CukKink}c show dispersions in the normal
state of an optimally-doped sample which consistently reveal a ~40
meV energy scale that has eluded detection previously. Upon entering
the superconducting state,  the energy scale shifts to 70meV
consistent with a gap opening of 35$\sim$40 meV. This coupling is
also found to be more extended in a Brillouin zone than previously
reported\cite{GromkoAntinodalKink}.   In Fig.
\ref{Bi2212kDependence}, we show data from the optimally-doped
Bi2212 sample for a large portion of the BZ in the superconducting
state\cite{TCukAntinodalKink}. The renormalization occurs at 70 meV
throughout the BZ and increases in strength from the nodal to
anti-nodal points. Similar behaviors are also noted by
others\cite{KaminskiKink} (Fig. \ref{KaminskiKink}). The increase in
coupling strength can be seen in the following ways: Near ($\pi$,0),
the band breaks up into two bands (peak and hump) as seen in Fig.
\ref{Bi2212kDependence}a2 and a3. For cuts taken in the (0,0) -
($\pi$, $\pi$) direction, the band dispersion is steeper and the
effects of mode-coupling, though significant, are less pronounced.

\begin{figure*}[Floatfix]
\begin{center}
\includegraphics[width=0.95\linewidth,angle=0]{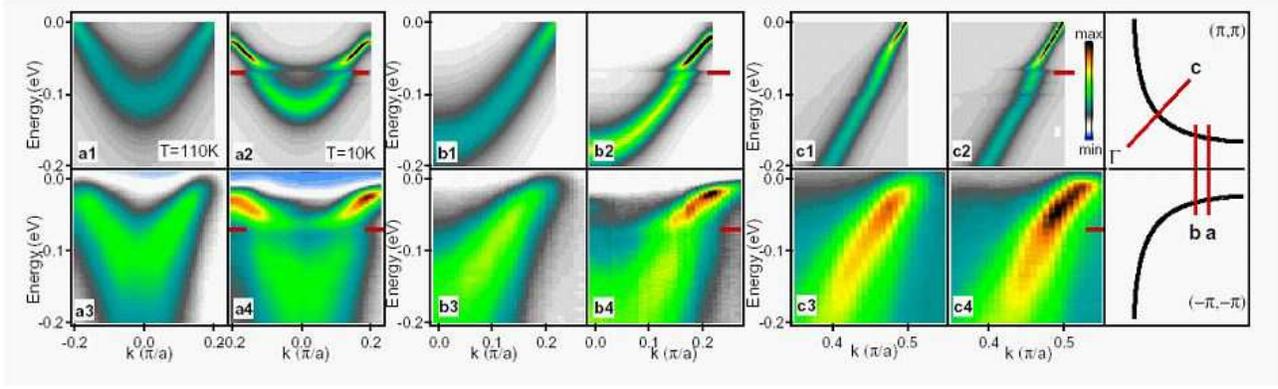}
\end{center}
\caption{Comparison between the calculated and measured spectral
function in Bi2212 including electron-phonon coupling for three
different momentum cuts (a, b, c) through the Brillouin zone.
(a1,b1,c1) and (a2,b2,c2) show the calculated spectral functions in
the normal and superconducting states,
respectively\cite{DevereauxAntinodalKink}. The measured spectral
functions are shown in (a3,b3,c3) for the normal state and in
(a4,b4,c4) for the superconducting state. The corresponding momentum
cuts a, b, and c are shown in the rightmost panel. The red markers
in the superconducting state indicate 70 meV. The simulation
includes B1g oxygen buckling mode and half-breathing
mode.}\label{TomSimu}
\end{figure*}

It is quite natural that phonon modes of different origin and energy
preferentially couple to electrons in certain k-space regions. While
the detection of multiple modes in the normal state of
LSCO(\cite{XJZhouMultipleMode} suggests that several phonons may be
involved, only one has the correct energy and momentum dependence to
understand the prominent signature seen in the superconducting
state.  This new interpretation\cite{TCukAntinodalKink} attributes
the renormalization seen in the superconducting state to the
``bond-buckling" B$_{1g}$ phonon mode involving the out-of-plane,
out-of-phase motion of the in-plane oxygens. The bond-buckling
phonon is observed at 35 meV in the B$_{1g}$ polarization of Raman
scattering on an optimally doped sample, the same channel in which
the $\sim$ 35-40 meV d-wave superconducting gap shows up
\cite{Friedl, Devereaux3, Opel}.  Applying simple symmetry
considerations and kinematic constraints, it is found that this
B$_{1g}$ buckling mode involves small momentum transfers and couples
strongly to electronic states near the
antinode\cite{DevereauxAntinodalKink}. In contrast, the in-plane
Cu-O breathing modes involve large momentum transfers and couple
strongly to nodal electronic states.  Band renormalization effects
are also found to be strongest in the superconducting state near the
antinode, in full agreement with angle-resolved photoemission
spectroscopy data (Fig. \ref{TomSimu}). The dramatic temperature
dependence stems from a substantial change in the electronic
occupation distribution and the opening of the superconducting
gap\cite{TCukAntinodalKink,DevereauxAntinodalKink}.  It is important
to note that the electron-phonon coupling, especially the one with
B$_{1g}$ phonon, explains the temperature and momentum dependence of
the self-energy effects that were taken as key evidence to support
the magnetic resonance interpretation of the data. Compounded with
the findings that cannot be explained by the magnetic resonance as
discussed earlier, this development makes the phonon interpretation
of the kink effect self-contained.

\subsubsection{Anisotropic El-Ph Coupling}

The full Migdal-Eliashberg-based calculation consists of a
tight-binding band structure and el-ph coupling to the breathing
mode as well as the B$_{1g}$ bond-buckling mode and is based on an
earlier calculation \cite{TDevereaux2}. The electron-phonon coupling
vertex $g(k,q)$, where $k$ represents the initial momentum of the
electron and $q$ the momentum of the phonon is determined on the
basis of the oxygen displacements for each mode in the presence of
the underlying band-structure.  In the case of the breathing mode,
the in-plane displacements of the oxygen modulate the CuO$_2$
nearest neighbor hopping integral as well as the site energies.  In
the case of the bond-buckling mode, one must suppose that the mirror
plane symmetry across the CuO$_2$ plane is broken in order for
electrons to couple linearly to phonons. The mirror plane symmetry
can be broken by the presence of a crystal field perpendicular to
the plane, tilting of the Cu-O octahedral, static in-plane buckling,
or may be dynamically generated.

\begin{figure}[tbp]
\begin{center}
\includegraphics[width=0.9\linewidth,angle=0]{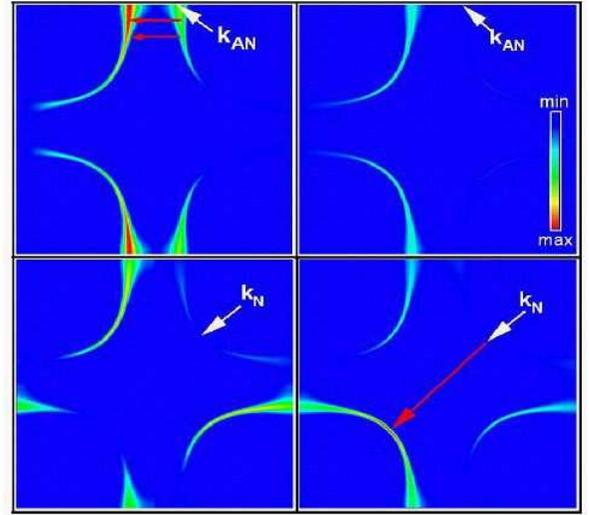}
\end{center}
\caption{Plots of the electron-phonon coupling $\mid
g(\textbf{k},\textbf{q})\mid^{2}$ for initial $\textbf{k}$ and
scattered $\bf{k^{\prime}=k-q}$ states on the Fermi surface for the
buckling mode (left panels) and breathing mode (right panels) for
initial fermion $\textbf{k}$ at an anti-nodal (top panels) and nodal
(bottom panels) point on the Fermi surface, as indicated by the
arrows. The red/blue color indicates the maximum/minimum of the
el-ph coupling vertex in the BZ for each
phonon\cite{DevereauxAntinodalKink}.}\label{Anisolambda}
\end{figure}

The $g_{B_{1g}}(k,q)$ form factor leads to preferential $q \sim
2k_{f}$ scattering between the parallel pieces of Fermi surface in
the anti-nodal region, as shown in Fig. \ref{Anisolambda} depicting
$g(k,k')$ for the buckling mode (where $k'= k - q$) for an electron
initially at the anti-node ($k_{AN}$; upper-left) and for an
electron initially at the node ($k_{N}$; bottom-left). This coupling
anisotropy partially accounts for the strong manifestation of
electron-phonon coupling in the anti-nodal region where one sees a
break up into two bands. The breathing mode, in contrast, modulates
the hopping integral and has a form factor, $g_{br}(k,q)$, that
leads to preferential scattering for large $q$ and couples opposing
nodal states.  This coupling anisotropy then accounts for the 70 meV
energy scale seen most prominently in a narrow $k$-space region near
the nodal direction in the normal state of LSCO. Fig.
\ref{Anisolambda} also shows that the magnitude of the
electron-phonon vertex is largest for an electron initially sitting
at the node, $k_{N}$, that scatters to the opposing nodal state. For
more details on this calculation, see Devereaux $\textit{et. al.}$
\cite{DevereauxAntinodalKink}.

The anisotropy of the mode-coupling in both the superconducting
state data and the calculation is peculiar to the cuprates.  Such a
strong anisotropy in the electron-phonon coupling is not
traditionally expected.  In cuprates, the sources of the anisotropy
are: 1) an electron-phonon vertex for the B$_{1g}$ bond-buckling
mode and the breathing mode that depends both on the electron
momentum $k$ as well as the phonon momentum $q$. This comes from a
preferential scattering in the Brillouin zone, in which nodal states
couple to other nodal states and anti-nodal states to other
anti-nodal states.  2) a strongly anisotropic electronic band
structure characterized by a van Hove Singularity (vHS) at
($\pi$,0).  In the anti-nodal region and along the
$(\pi,0)-(\pi,\pi)$ direction in which 2$k_{F}$ scattering is
preferred, the bands are narrow, giving rise to a larger electronic
density of states near the phonon energy and therefore a stronger
manifestation of electron-phonon coupling. 3) a d-wave
superconducting gap and 4) a collusion of energy scales in the
anti-nodal region that resonate to enhance the above effects---the
vHS at $\sim$ 35meV in the tight-binding model that best fits the
data, the maximum d-wave gap at $\sim$ 35meV, and the bond-buckling
phonon energy at $\sim$ 35meV.  All these three factors collide to
give the anisotropy of the mode-coupling behavior in the
superconducting and normal states. For a detailed look at how each
plays a role in the agreement with the data, please see Cuk
\textit{et. al.} \cite{TCukAntinodalKink}. The coincidence of energy
scales, along with the dominance of the renormalization near the
anti-node, indicates the potential importance of the B$_{1g}$ phonon
to the pairing mechanism, which is consistent with some theory on
the B$_{1g}$ phonon\cite{Jepsen, KAMuller, Scalapino2, Nazarenko}
but remains to be investigated.

\begin{figure}[tbp]
\begin{center}
\includegraphics[width=0.95\linewidth,angle=0]{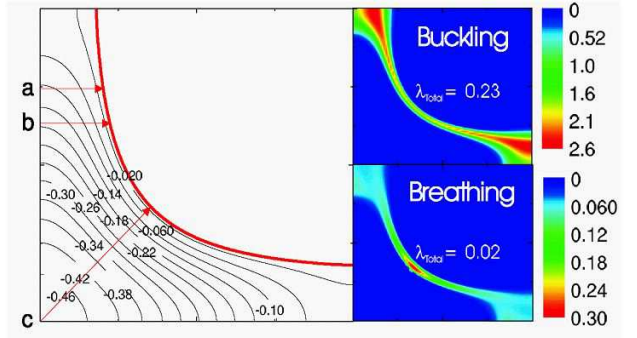}
\end{center}
\caption{Plots of the electron-phonon coupling $\lambda_{\bf k}$ in
the first quadrant of the BZ for the buckling mode (right top panel)
and breathing mode (right bottom panel). The color scale is shown on
the right for each phonon. The left panel shows energy contours for
the band structure
used\cite{DevereauxAntinodalKink}.}\label{Modelambda}
\end{figure}

The cuprates provide an excellent platform on which to study
anisotropic electron-phonon interaction.  In one material, such as
optimally doped Bi2212, the effective coupling can span $\lambda$ of
order $\sim$1 at the node to 3 at the anti-node
(Fig.\ref{Modelambda})\cite{TCukAntinodalKink,DevereauxAntinodalKink}.
In addition to the large variation of coupling strength, there is a
strong variation in the kinematic considerations for electron-phonon
coupling.  In the nodal direction, the band bottom is far from the
relevant phonon energy scales.  However, at the anti-node, the
relevant phonon frequencies approach the bandwidth. Indeed the
approximation of Migdal, in which higher order vertex corrections to
the el-ph coupling are neglected due to the smallness of
($\lambda*\Omega_{_{ph}}/E_{F}$), may be breaking down in the
anti-nodal region.  Non-adiabatic effect beyond the Migdal
approximation have been considered and are under continuing study
\cite{Pietrano}.

\subsection{Polaronic Behavior}

\subsubsection{Polaronic Behavior in Parent Compounds}

The parent compounds of the cuprate superconductors, being
antiferromagnetic Mott insulators, provide an ideal testing ground
for investigating the dynamics of one hole in an antiferromagnetic
background.  Indeed, many theories have been formed and tested by
ARPES on a number of compounds, among them are
Sr$_2$CuO$_2$Cl$_2$\cite{WellsSCOC,LaRosa,CKimSCOC,Pothuizen,RonningUniver},
Ca$_2$CuO$_2$Cl$_2$\cite{RonningScience,RonningUniver,RonningPRB,KShenPolaron,KShenScience},
Nd$_2$CuO$_4$\cite{ArmitageNCCO}, and
La$_2$CuO$_4$\cite{InoSITrans,YoshidaArc,RoschLCO}. However, several
aspects of the data can only be explained by invoking polaron
physics, as we will now discuss.

\begin{figure}[tbp]
\begin{center}
\includegraphics[width=0.95\linewidth,angle=0]{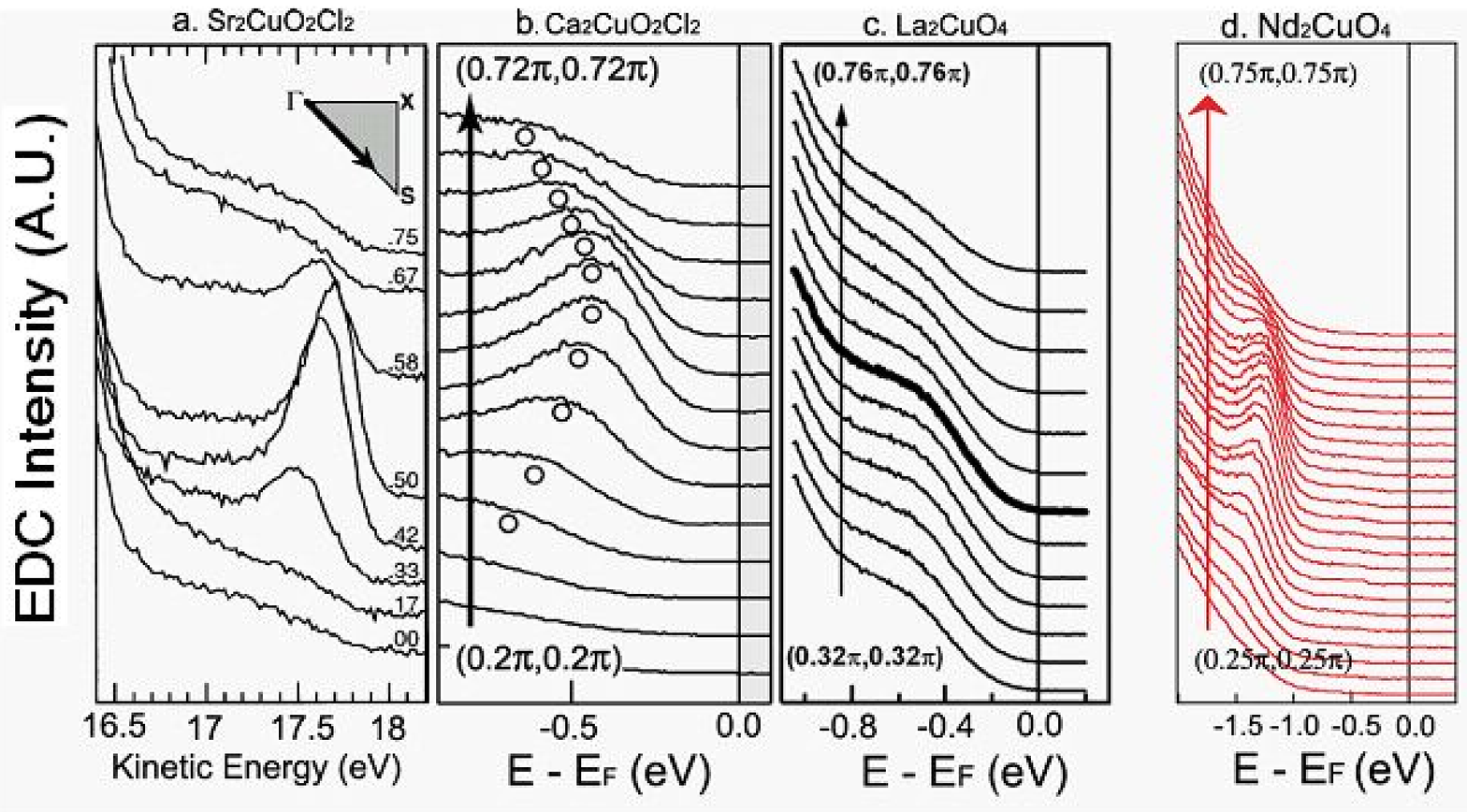}
\end{center}
\caption{(a). Photoemission spectra along the (0,0) and
($\pi$,$\pi$) direction in
Sr$_2$CuO$_2$Cl$_2$\cite{WellsSCOC,CKimSCOC},
Ca$_2$CuO$_2$Cl$_2$\cite{RonningPRB,KShenPolaron},
La$_2$CuO$_4$\cite{YoshidaArc,RoschLCO} and
Nd$_2$CuO$_4$\cite{ArmitageNCCO}.}\label{UndopedARPES}
\end{figure}

\begin{figure}[tbp]
\begin{center}
\includegraphics[width=0.95\linewidth,angle=0]{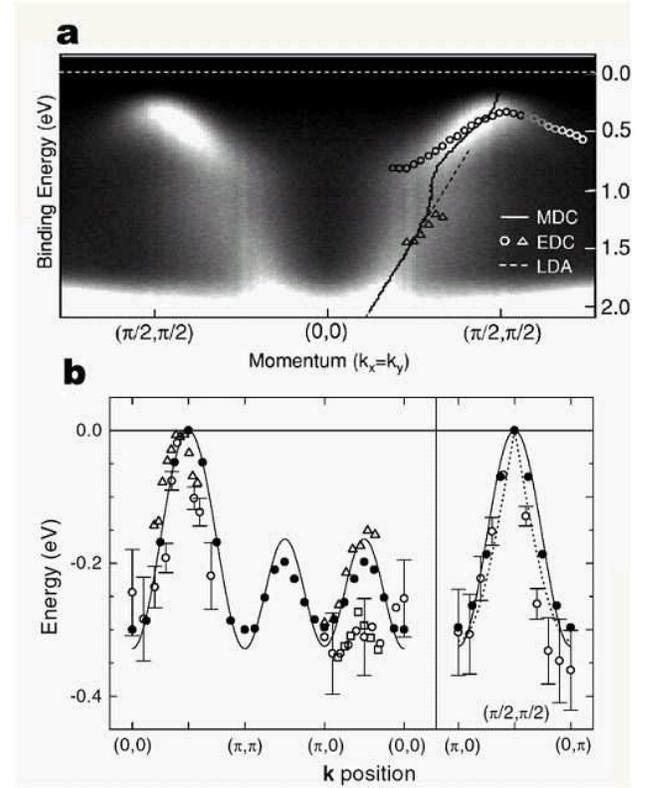}
\end{center}
\caption{(a). Intensity plot of ARPES data as functions of the
binding energy and momentum for Ca$_2$CuO$_2$Cl$_2$ along the
$\Gamma$(0,0)-($\pi$,$\pi$) direction\cite{RonningPRB}. The data was
symmetrized around the $\Gamma$ point. Also shown on the plot are
the dispersions obtained by following the peak positions of the MDCs
(solid line) and the EDCs (circle and triangles). The results are
compared with the shifted dispersion from the LDA calculation
(dashed line). (b) Energy dispersion of quasiparticle for insulating
Sr$_2$CuO$_2$Cl$_2$ measured from the top of the band. Experimental
data are taken from \cite{WellsSCOC} (open circles),
\cite{LaRosa}(open triangles) and \cite{CKimSCOC}(open squares).
Solid circles: the results of the self-consistent Born approximation
(SCBA) for the t$-$t$^{'}$$-$t$^{"}$$-$J model with t = 0.35 eV,
$t^{'}$ = -0.12 eV, $t^{"}$ = 0.08 eV and J = 0.14 eV. The solid
lines are obtained by fitting the SCBA data to a dispersion relation
given by E$_0$(k) + E$_1$(k), being t$^{'}_{eff}$ = -0.038 eV and
t$^{"}_{eff}$ = 0.022 eV. The dashed line along the ($\pi$,
0)¨C(0,$\pi$) direction represents the spinon dispersion from
\cite{LaughlinSCOC}.}\label{ExttJcalculation}
\end{figure}

The ARPES measurements on SCOC\cite{WellsSCOC,CKimSCOC} and
CCOC\cite{RonningPRB,KShenPolaron} give essentially similar results.
As seen in Fig. \ref{UndopedARPES}a and \ref{UndopedARPES}b, along
the (0,0)-($\pi$,$\pi$) direction, the lowest energy feature
disperses toward lower binding energy with increasing momentum,
reaches its lowest binding energy position near ($\pi$/2,$\pi$/2)
where it becomes sharpest in its lineshape, and then suddenly loses
intensity after passing ($\pi$/2,$\pi$/2) and bends back to high
binding energy. This behavior can be more clearly seen in the image
plot of Fig. \ref{ExttJcalculation}a\cite{RonningPRB} where the
``band" breaks into two branches. The lowest binding energy feature
shows a dispersion of $\sim$0.35 eV while an additional band at high
binding energy (Fig. \ref{ExttJcalculation}a) is very close to the
unrenormalized band predicted by band theory\cite{RonningPRB}. The
dispersion of low binding-energy band along the (0,0)-($\pi$,$\pi$)
direction and other high symmetry directions are shown in Fig.
\ref{ExttJcalculation}b by keeping track of the EDC peak
position\cite{TTohyama}. The total dispersion of the peak is
$\sim$0.35eV. This is in contrast to the predictions of one-electron
band calculations which gives an occupied band width of $\sim$ 1.5
eV and total bandwidth of $\sim$3.5 eV\cite{CCOCBandCal}.
Nevertheless, it is consistent with the calculations from the t-J
model where the predicted occupied bandwidth is
$\sim$2.2J\cite{ZPLiu,DagottoRMP}. This indicates that the dynamics
of one-hole in an antiferromagnetic background is renormalized from
scale $t$ to scale $J$.

While the t-J model and experiments show agreement along the
(0,0)-($\pi$,$\pi$) direction, there are discrepancies along other
directions, such as the (0,0)-($\pi$,0) and
(0,$\pi$)-($\pi$/2,$\pi$/2)-($\pi$,0) directions\cite{WellsSCOC}.
The later intensive theoretical effort resolved this issue by
incorporating the hopping to the second (t$^{'}$)  and third
(t$^{"}$) nearest-neighbors\cite{SCOCExttJ}.  More precise
calculations of the dispersion in the t$-$t$^{'}$$-$t$^{"}$$-$J
model are performed by using a self-consistent Born approximation
(SCBA)\cite{SCB}. These calculations show a satisfactory agreement
with experimentally derived dispersion, as shown in Fig.
\ref{ExttJcalculation}b\cite{TTohyama}.

\begin{figure}[tbp]
\begin{center}
\includegraphics[width=0.95\linewidth,angle=0]{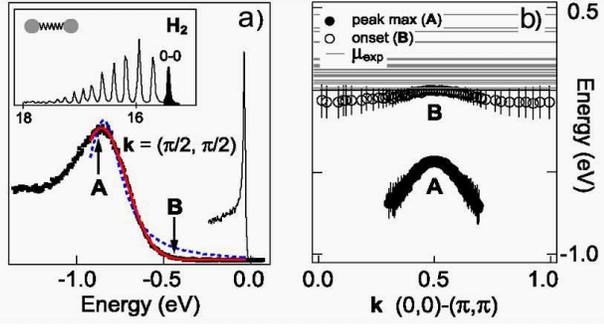}
\end{center}
\caption{(a) Photoemision spectrum of Ca$_2$CuO$_2$Cl$_2$ at
k=($\pi$,$\pi$) with fits to a Lorentzian spectral function (dashed)
and Gaussian (red or gray)\cite{KShenPolaron}. A and B denote the
peak maximum and the onset of spectral weight, respectively.
Comparison with Fermi-liquid system Sr$_2$RuO$_4$ is shown (thin
black). Upper inset shows photoemission spectra from
H$_2$\cite{HydrogenFC}. (c) Dispersion of A and B along
(0,0)-($\pi$,$\pi$), along with experimental values for the chemical
potential $\mu$ (lines).}\label{CCOCPolaron}
\end{figure}

However, there remain a few prominent puzzles related to the
interpretation of the photoemission data in undoped parent
compounds\cite{KShenPolaron}. The first prominent issue is the
linewidth of the peak near ($\pi$/2,$\pi$/2). As highlighted in Fig.
\ref{CCOCPolaron}a, the width of the sharpest peak near
($\pi$/2,$\pi$/2) is $\sim$300 meV which is comparable with the
entire occupied bandwidth $2J \approx$350 meV \cite{KShenPolaron}.
This is much broader than that from t-J model calculations and too
broad to be considered as a coherent quasiparticle peak for which
the quasiparticle peak is basically resolution limited, as
exemplified by the data on Sr$_2$RuO$_4$ in Fig. \ref{CCOCPolaron}a.
An early attempt interpreted this anomalously large linewidth to
additional interaction with a non$-$magnetic boson bath of
excitations, such as phonons\cite{Pothuizen}. But this
interpretation meets with difficulty in explaining little
renormalization in the dispersion from this ``extra interaction"
because dispersion and linewidth are closely related.  A
diagrammatic quantum Monte Carlo study\cite{AMtJPolaron} showed that
this problem can be resolved by considering the polaron effect in
the t-J model. Namely the dispersion for the center of mass of the
spectral function obeys that of the pure t-J model, while the
lineshape is strongly modified. The details of this will be given
below.

Another unresolved issue is the chemical potential $\mu$. For an
insulator, $\mu$ is not well defined, and may be pinned by surface
defects or impurities and will vary between different samples. If
one considers that the peak A in Fig.\ref{CCOCPolaron}a represents a
quasiparticle peak, one would expect the chemical potential to vary
anywhere above the top of this valence band. However, the
experimental chemical potential clearly sets a lower bound that is
$\sim$0.45eV apart from the peak A (Fig.
\ref{CCOCPolaron}b)\cite{KShenPolaron}.  Shen et
al.\cite{KShenPolaron} invented a new method to determine the
chemical potential using both the energy of the non-hybridized
oxygen orbital and the detailed line-shape of Na-CCOC.

The resolution of these discrepancies between experiment and
expectation leads to identifying polaron physics as responsible for
the bulk of the lineshape in underdoped cuprates. In fact, the
photoemission spectra in the underdoped cuprates resemble the
Frank-Condon effect seen in photoemission spectra of molecules such
as H$_2$\cite{HydrogenFC}(inset of Fig. \ref{CCOCPolaron}), where
only the ``0-0" peak (filled black) represents the H final state
with no excited vibrations and comprises only $\sim$10$\%$ of the
total intensity. In the solid state, this ``0-0" would correspond to
the quasiparticle or the coherent part of the spectral function,
A$_{coh}$, whereas the excited states comprise the incoherent part,
A$_{inc}$. This behavior is reminiscent of polarons, and such models
have been invoked in systems where strong couplings are
present\cite{ARPESPolaron}. In this picture, in the undoped
compound, the true QP (B) is hidden within the tail of spectral
intensity, with a quasiparticle residue $Z$ vanishingly small,
while feature A is simply incoherent weight associated with
shake-off excitations.

\begin{figure}[tbp]
\begin{center}
\includegraphics[width=0.95\linewidth,angle=0]{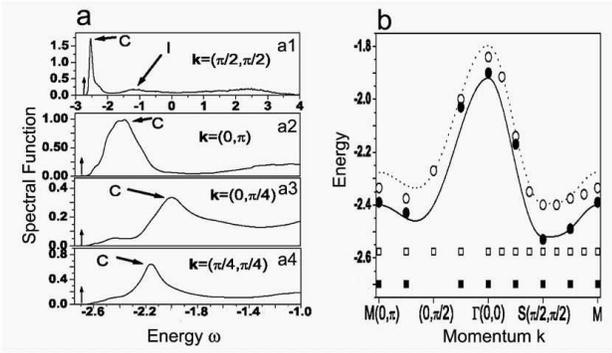}
\end{center}
\caption{(a). Calculated hole spectral function in ground state at
$J/t=0.3$ for different momenta\cite{AMtJPolaron}. (a1) Full
energy range for $k=(\pi/2,\pi/2)$.  (a2)$-$(a4) Low energy part
for different momenta. Slanted arrows show broad peaks which can
be interpreted in ARPES spectra as ``coherent" (C) and incoherent
(I) part. Vertical arrows indicate position of ground state
resonance which is not seen in the vertical scale of the figure.
(b). Dispersion of resonance energies at $J/t=0.3$.  Broad
resonance (filled circles) and lowest polaron resonance (filled
squares) at $g=0.231125$;  third broad resonance (open circles)
and lowest polaron resonance (open squares) at g=0.2. The solid
curves are dispersions of a hole in the pure t-J model at
$J/t=0.3$.}\label{tJPolaron}
\end{figure}

From the viewpoint of polaron physics, the cuprates offer a unique
and first opportunity to compare experimental spectra with theory in
detail. The single hole interacts both with magnons and phonons. The
hole-magnon interaction has been successfully analyzed in terms of
the self-consistent Born approximaiton\cite{SCB}. The success of the
Born approximation results from a ``saturation" effect; namely the
single spin 1/2 can flip only once, and hence magnon clouds do not
become large enough to induce the self-trapping transition to the
small polaron. On the other hand, phonon clouds can be larger and
larger as the coupling constant $g$ increases and can lead to a
self-trapping transition. The t-J model coupled to phonons  in the
polaronic regime has illuminated one-hole dynamics in the parent
compound in the following way\cite{AMtJPolaron}. (1). With
increasing electron-phonon coupling strength, the spectral function
experiences a transition from weak-coupling, to intermediate
coupling, and to strong-coupling regimes. (2). In the strong
coupling regime, the spectral function consists a ground state
resonance (as indicated by vertical arrows) with vanishing intensity
and a broad peak denoted as ``coherent C", as shown in
Fig.\ref{tJPolaron}a. (3). The broad peak C shows strong momentum
dependence while the lowest state is dispersionless. These results
are in good correspondence to experimental observations. The most
surprising result is that the broad resonance has the momentum
dependence of the t-J model without coupling to phonons (shown in
Fig. \ref{tJPolaron}b). In the Franck-Condon effect for molecules a
similar result occurs.  The center of the shake-off band corresponds
to the hole motion in the background of the frozen lattice
configuration, i.e. the dispersion of the hole remains that of the
non-interacting limit, while the line-width broadens.  A more
elaborated analytic treatment of the t-J polaron model in the
Franck-Condon approximation\cite{GunnEur} successfully reproduced
this Monte Carlo results.  The calculated spectral function
line-shape most consistent with experiment has a $\lambda \cong
0.9-1.3$, well within the strong-coupling, small-polaron regime.
Recent realistic shell model calculation\cite{RoschLCO} also
concluded $\lambda = 1.2$ for La$_2$CuO$_4$.

\begin{figure}[tbp]
\begin{center}
\includegraphics[width=0.95\linewidth,angle=0]{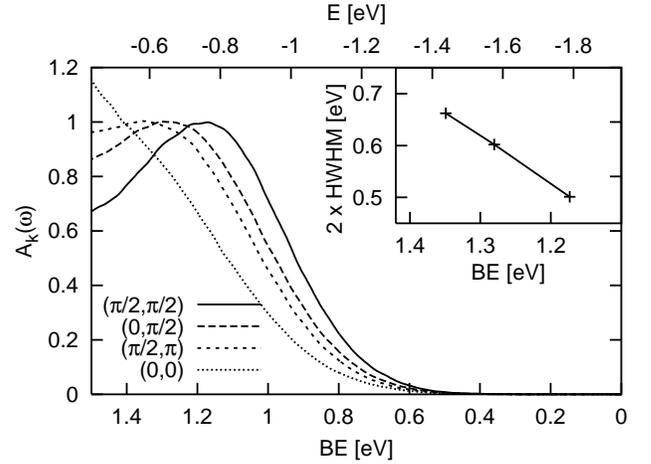}
\end{center}
\caption{Polarons in La$_2$CuO$_4$\cite{RoschLCO}.  Calculated ARPES
spectra for the undoped La$_2$CuO$_4$ system at T=0 for different
$\bf {k}$ normalized to the height of the phonon side band. The
lower abscissa shows binding energies (BE) and the upper abscissa
the energies of the final states corresponding to the spectral
features. The inset shows the dependence of the width of the phonon
side band on its binding energy.  The width of the $(0,0)$ spectrum
is poorly defined and not shown.}\label{LCOPolaron}
\end{figure}

In La$_2$CuO$_4$, a broad feature near -0.5eV (Fig.
\ref{LCOPolaron}) was identified as the lower Hubbard
band\cite{InoSITrans,YoshidaArc}. The electron-phonon coupling
strength, calculated using a shell model,  puts La$_2$CuO$_4$ in the
polaron regime, similar to Ca$_2$CuO$_2$Cl$_2$. In this picture, the
-0.5eV feature corresponds to the phonon side-band while the real
quasiparticle residue is very weak. As shown in Fig.
\ref{LCOPolaron}, the calculated spectral function agrees well with
the measured data\cite{RoschLCO}.

\subsubsection{Doping Dependence: From Z$\sim$0 Polaron to Finite Z
Quasiparticles}

We next turn to the question of how the small polaron state evolves
as a function of doping, connecting to the Migdal-Eliashberg regime
discussed in section C. There are two possible ways to dope the Mott
insulator, schematically shown in Fig.
\ref{LDALCO}c\cite{SUchidaGap,Veenendaal}:  (1). Upon doping, the
chemical potential shifts to the top of the valence band for hole
doping (Fig. \ref{LDALCO}c3) or to the bottom of the conduction band
for electron doping (Fig. \ref{LDALCO}c4). (2).The chemical
potential is pinned inside the charge transfer gap. Upon doping, new
states will form inside the gap (Fig. \ref{LDALCO}c5).

\begin{figure}[tbp]
\begin{center}
\includegraphics[width=0.9\linewidth,angle=0]{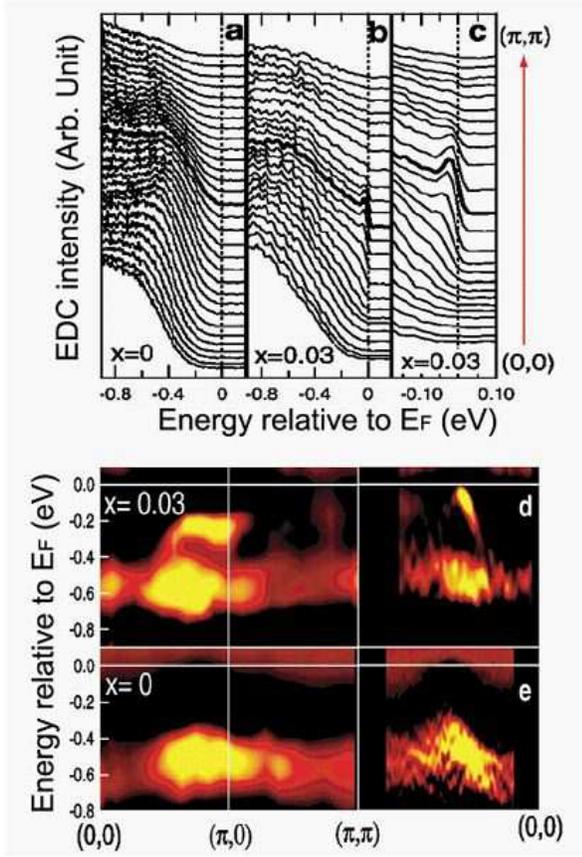}
\end{center}
\caption{Creation of nodal quasiparticles in lightly-doped
LSCO\cite{YoshidaArc}. ARPES spectra for LSCO with $x$=0 and
$x$=0.03. Panels a and b are EDC's along the nodal direction
(0,0)-($\pi,\pi$) in the second Brillouin zone (BZ). The spectra for
$x$ = 0.03 are plotted on an expanded scale in panel c. Panels d and
e represent energy dispersions deduced from the second derivative of
the EDC's. }\label{CreationQP}
\end{figure}

Recent ARPES measurements on lightly-doped (La$_{2-x}$Sr$_x$)CuO$_4$
compounds provide a good window to look into this issue. As shown in
Fig. \ref{CreationQP}a and \ref{CreationQP}e, for undoped
La$_2$CuO$_4$, the main feature is the broad peak near -0.5 eV which
exhibits weak dispersion\cite{YoshidaArc}. There is also little
spectral weight present near the Fermi level.  However, upon only a
doping of $x=0.03$, the electronic structure undergoes a dramatic
change. A new dispersive band near the Fermi level develops along
the (0,0)-($\pi$,$\pi$) nodal direction(Fig. \ref{CreationQP}e,
right panel),  while along the (0,0)-($\pi$,0) direction a saddle
band residing -0.2eV below the fermi level develops. Even for more
lightly-doped samples, such as $x=0.01$, new states near the Fermi
level are created\cite{XJZhou001}.  Note that, for these
lightly-doped samples, the original -0.5 eV remains, although with
weakened spectral weight (Fig. \ref{CreationQP}d). So, the -0.5eV
peak and the new dispersive band coexist at doping levels close to
the parent compounds.

\begin{figure}[tbp]
\begin{center}
\includegraphics[width=1.0\linewidth,angle=0]{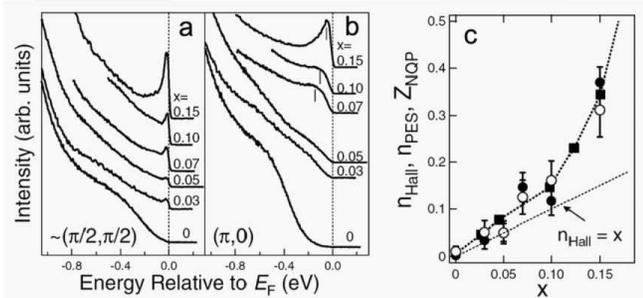}
\end{center}
\caption{ARPES spectra at $\mathbf{k} = \mathbf{k}_\mathrm{F}$ in
the nodal direction in the second BZ (a) and those at ($\pi,0$)(b)
for various doping levels\cite{YoshidaArc}. (c).Doping dependence of
the nodal QP spectral weight, $Z_{\rm NQP}$, and the spectral weight
integrated at $E_{\mathrm F}$ over the entire second Brillouin zone,
$n_{\rm PES}$\cite{YoshidaArc}. They show similar doping dependence
to the hole concentration evaluated from Hall coefficient ($n_{\rm
Hall}$) \cite{TakagiHall}. }\label{LSCOZ}
\end{figure}

The systematic evolution of the photoemission spectra near the nodal
and antinodal regions with doping in LSCO is shown in Fig.
\ref{LSCOZ}a and b\cite{YoshidaArc}.  The nodal quasiparticle
weight, Z$_{QP}$, integrated over a small energy window near the
Fermi level, is shown in Fig. \ref{LSCOZ}c. In the underdoped
region, it increases with increasing doping nearly linearly, and no
abrupt change occurs near the nonsuperconductor-superconductor
transition at $x \sim$0.05.

\begin{figure}[b!]
\begin{center}
\includegraphics[width=1.0\linewidth,angle=0]{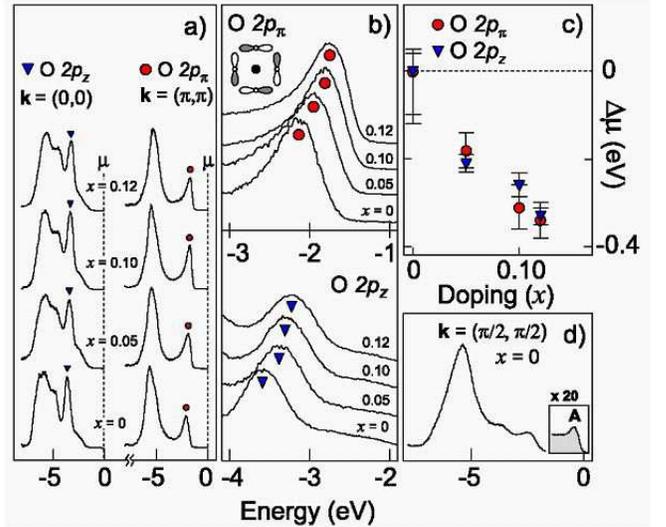}
\end{center}
\caption{Chemical potential shift in Na-CCOC\cite{KShenPolaron}. (a)
Valence band spectra for x=0, 0.05, 0.10, and 0.12 compositions at
k=(0,0) and ($\pi$,$\pi$). O 2p$_z$ and O 2p$_{pi}$ states are
marked by triangles and circles, respectively. (b) Shifts of the O
2p$_z$ and O 2p$_{pi}$ peaks shown on an expanded scale. (c) Doping
dependence of chemical potential $\Delta$$\mu$  determined from (b).
(d) Valence band at k=($\pi$/2,$\pi$/2), showing the lower Hubbard
band (A) on an expanded scale. }\label{CCOCChemP}
\end{figure}

(Ca$_{2-x}$Na$_x$)CuO$_2$Cl$_2$ (Na$-$CCOC) is another ideal system
to address the doping evolution of the electronic structure. The
precise measurement of the chemical potential (Fig.
\ref{CCOCChemP}a), in conjunction with the identification of polaron
physics in the under-doped compounds, provides a globally consistent
picture of the doping evolution of the cuprates\cite{KShenPolaron}.
Instead of measuring the chemical potential with deep core level
spectroscopy (the usual method)\cite{InoChemPotential}, one utilizes
orbitals in the valence band at lower energies(Fig. \ref{CCOCChemP}a
and b). The measured chemical shift, $\Delta$$\mu$, exhibits a
strong doping dependence, $\partial$$\mu$/$\partial$x=-1.8$\pm$0.5
eV/hole, comparable to the band structure estimation ($\sim$-1.3
eV/hole) (Fig. \ref{CCOCChemP}c).

\begin{figure}[tbp]
\begin{center}
\includegraphics[width=1.0\linewidth,angle=0]{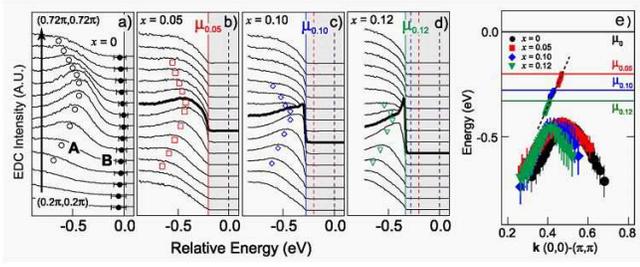}
\end{center}
\caption{(a-d). EDC spectra of Na-CCOC  x=0 (a), 0.05(b), 0.10(c),
and 0.12(d) from (0.2$\pi$,0.2$\pi$) to (0.72$\pi$,0.72$\pi$) with
hump positions marked by open symbols and the EDC at k$_F$ shown in
bold\cite{KShenPolaron}. Data are plotted on a relative energy scale
referenced to the shift in $\mu$ shown in Fig. \ref{CCOCChemP}c.
(e). Summary of hump (symbols)from Fig. (a-d) and MDC dispersions
(lines).
 }\label{CCOCEDC}
\end{figure}

Figs. \ref{CCOCEDC}(a-d) show the doping evolution of the near-E$_F$
EDCs plotted relative to $\mu$$_0$ of the undoped sample (determined
in Fig. \ref{CCOCChemP}c).  With doping, feature A evolves smoothly
into a broad, high energy hump with a backfolded dispersion similar
to the parent insulator (symbols), while B shifts downward relative
to its position in the un-doped compound. Spectral weight increases
with doping at B, and a well-defined peak emerges for the $x=$0.10
and 0.12 samples, resulting in a coherent, low-energy band. The
dispersion of the high-energy hump (A), tracked using the local
maxima or second derivative of the EDCs, shows little change as a
function of doping (Fig. \ref{CCOCChemP}e).  The lowest energy
excitations (feature B, -0.05eV$<$E$<$E$_F$), tracked using MDC
analysis, evolve with doping in such a way that the quasiparticle
dispersion ($v_F$) and Fermi wave vectors ($k_F$) virtually collapse
onto a single straight line.

\subsubsection{Doping Evolution of Fermi Surface: Nodal-Antinodal
Dichotomy}

\begin{figure}[tbp]
\begin{center}
\includegraphics[width=1.0\linewidth,angle=0]{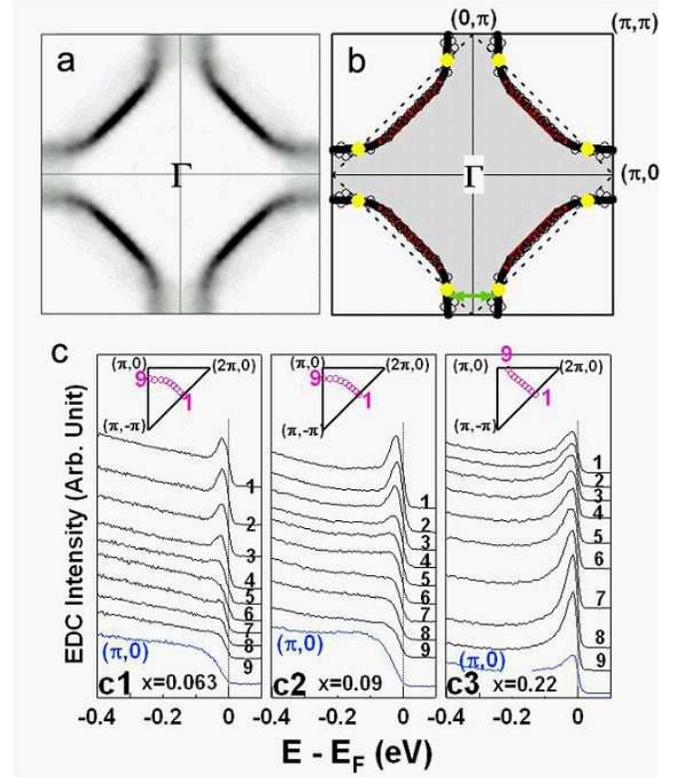}
\end{center}
\caption{Dichotomy between nodal and antinodal excitations in
LSCO\cite{XJZhouDichotomy}. (a). Spectral weight near a small energy
window of Fermi level as a function of $k_x$ and $k_y$ for LSCO
$x=0.063$ sample measured at $\sim$20K. The original data was taken
in the second Brillouin zone and converted into the first Brillouin
zone and symmetrized under four-fold symmetry. (b). Experimental
Fermi surface for LSCO $x=0.063$ sample. The black open circles are
obtained from the MDC peak position at $E_F$. The solid lines are
guides to the eye for the measured Fermi surface. The red lines
represent the portion of Fermi surface where one can see
quasiparticle peaks. The dotted black line represents the
antiferromagnetic Brillouin zone boundary; its intersection with the
Fermi surface gives eight ¡®¡®hot spots¡¯¡¯ (solid yellow circles)
from ($\pi$,$\pi$) magnetic excitations. The double-arrow-ended
green line represents a nesting vector between the antinodal part of
the Fermi surface (c). EDCs on Fermi surface for LSCO $x=0.063$
(c1), 0.09 (c2), and 0.22 (c3) samples. All samples are measured at
$\sim$20 K. The corresponding momentum position is marked in the
upper inset of each panel. Also included are the spectra at
($\pi$,0) points, colored as blue. }\label{LSCODichotomy}
\end{figure}

\begin{figure*}[floatfix]
\begin{center}
\includegraphics[width=1.0\linewidth,angle=0]{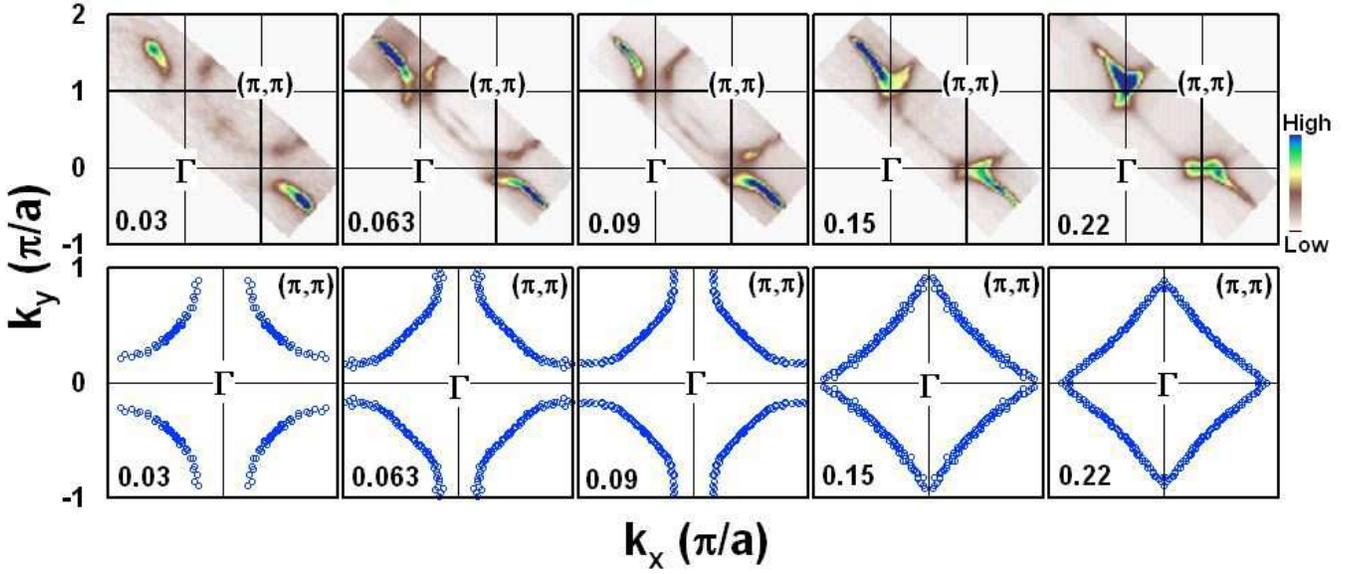}
\end{center}
\caption{Doping evolution of underlying "Fermi surface" in
(La$_{2-x}$Sr$_x$)CuO$_4$\cite{YoshidaArc,XJZhouDichotomy,XJZhouDuality,Yoshida22}.
The data were measured at a temperature of $\sim$20K.
}\label{LSCO_FS}
\end{figure*}

So far, we have discussed the doping evolution along the nodal
direction, and seen that a sharply defined quasiparticle peak
develops out of the small weight near the chemical potential in the
undoped samples.  We now discuss the doping evolution in other
directions of the Brillouin zone.  Surprisingly, one finds that the
coherent peak near the Fermi level in the lightly doped samples is
confined to the nodal region, and quickly disappears with momentum
around the Brillouin zone.  The spectral weight near the Fermi
level, confined to the ($\pi$/2,$\pi$/2) nodal region, forms a
so-called ``Fermi arc". This dichotomy between nodal and antinodal
excitations is shown in Fig.
\ref{LSCODichotomy}\cite{XJZhouDichotomy}. For the $x=0.063$ sample,
which is close to the nonsuperconductor-superconductor transition
and therefore heavily underdoped, the spectral weight near Fermi
level is mainly concentrated near the nodal region (Fig.
\ref{LSCODichotomy}a). The coherent peaks in the EDCs (Fig.
\ref{LSCODichotomy}c1) near the nodal region disappear as one
approaches the anti-nodal region, where the EDCs exhibit a step
rather than a peak.   The LSCO $x=0.09$ sample exhibits similar
behavior(Fig. \ref{LSCODichotomy}c2). In contrast, for overdoped
LSCO such as $x=0.22$ (Fig. \ref{LSCODichotomy}c3), sharp peaks are
observable along the entire Fermi surface. These observations
indicate that the electrons near the antinodal region experience
additional scattering.   Therefore, as shown in Fig. \ref{LSCO_FS},
the ``Fermi surface" in LSCO evolves from the ``Fermi arc" in
lightly-doped samples, to a hole-like Fermi surface in underdoped
samples, and to an electron-like Fermi surface in overdoped samples
(x$>$0.15).

\begin{figure}[tbp]
\begin{center}
\includegraphics[width=1.0\linewidth,angle=0]{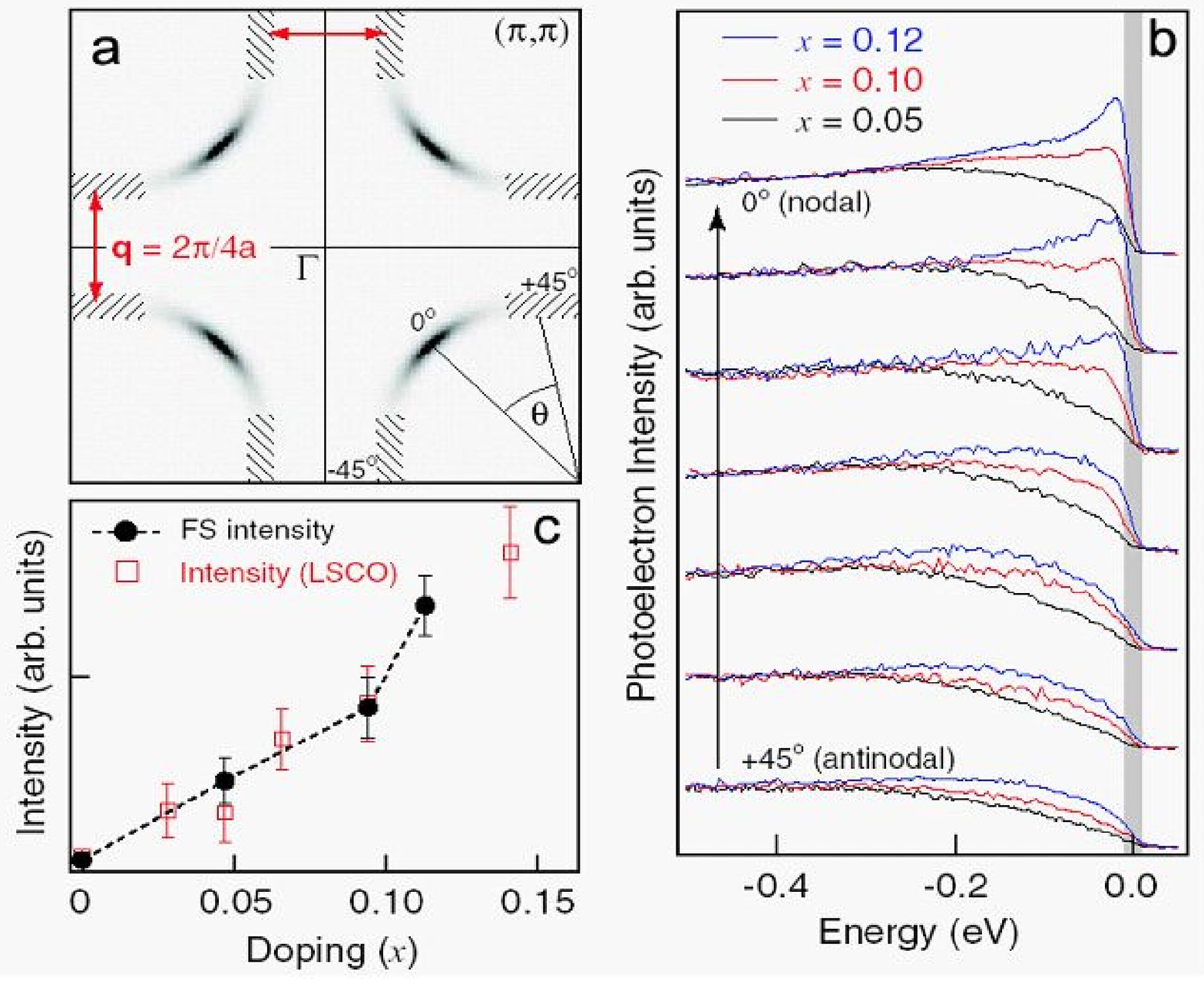}
\end{center}
\caption{Dichotomy between nodal and antinodal excitations in
Na-CCOC\cite{KShenScience}.  (a).  Schematic of the low-lying
spectral intensity for (Ca$_{2-x}$Na$_x$)CuO$_2$Cl$_2$ (x=0.10). The
hatched regions show the nested portions of Fermi surface, and the
Fermi surface angle is defined in the lower right quadrant. (b).
EDCs taken at equal increments along the FS contour from the nodal
direction (top) to the antinodal region (bottom) for x=0.05, 0.10,
and 0.12 at a temperature of 15 K. (c). The doping evolution of the
low-lying spectral weight (circles), along with corresponding data
from La$_{2¨Cx}$Sr$_x$CuO$_4$ (squares), with the error bars
representing the uncertainty in integrated weight as well as
sample-to-sample variations.}\label{CCOCDicho}
\end{figure}

\begin{figure}[tbp]
\begin{center}
\includegraphics[width=1.0\linewidth,angle=0]{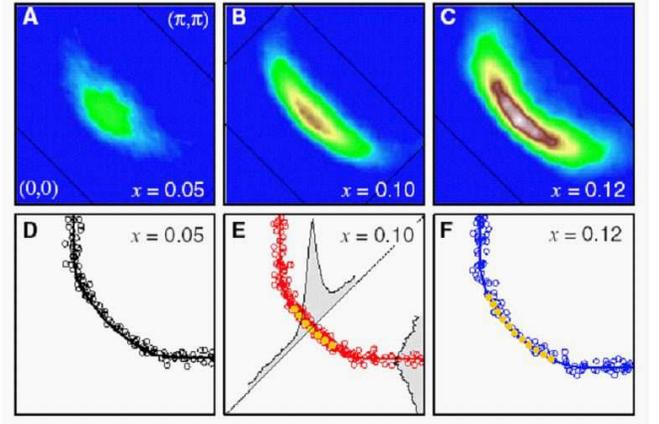}
\end{center}
\caption{Doping evolution of ``Fermi surface" in
Na-CCOC\cite{KShenScience}.  (A to C) The momentum distribution of
spectral weight within a $\pm$10-meV window around E$_{F}$ for
x=0.05, 0.10, and 0.12 in one quadrant of the first Brillouin zone.
Data were taken at 15 K and symmetrized along the (0,0)-
($\pi$,$\pi$) line. The data acquisition range is shown within the
black lines. The FS contours shown in (D to F) were compiled from
more than four samples for each composition with different photon
energies and photon polarizations. Data from these samples
constitute the individual points; the best fit is shown as a solid
line. The region in which a low-energy peak was typically observed
is marked by gold circles. The gray shaded areas in (E) represent
the momentum distribution of intensity at E$_F$$\pm$10 meV along the
(0,0)-($\pi$,$\pi$) and ($\pi$,0)-($\pi$,$\pi$) high-symmetry
directions. }\label{CCOCFermisurface}
\end{figure}

The evolution of electronic structure with doping in
(Ca$_{2-x}$Na$_x$)CuO$_2$Cl$_2$ exhibits marked resemblance to that
in (La$_{2-x}$Sr$_x$)CuO$_4$\cite{KShenScience}. As summarized in
Fig. \ref{CCOCDicho},  at low doping, the quasiparticle weight is
again confined to the nodal region and the weight in the
quasiparticle peak, Z$_{qp}$, increases with increasing doping,
consistent with LSCO.  In the Na$-$CCOC system, recent scanning
tunneling microscopy (STM) work has revealed a real space pattern of
4a$_0$ $\times$ 4a$_0$ two-dimensional charge
ordering\cite{CCOCSTM}. In momentum space, as seen from Fig.
\ref{CCOCFermisurface},  strong Fermi surface nesting exists in
Na$-$CCOC with a nesting vector insensitive to doping close to
2$\ast$$\pi$/4 that may account for the broad near-E$_F$ spectra in
the anti-nodal region.  In LSCO, neutron scattering has also
indicated the existence of dynamic stripes\cite{TranquadaStripe}.
These similarities suggest an intrinsic commonality between the
low-lying excitations across different cuprate families and may
imply a generic microscopic origin for these essential nodal states
irrespective of other ordering tendencies. At very low doping
levels, the nodal excitations should entirely dominate the transport
properties, consistent with the high-temperature metallic tendencies
observed in very lightly doped cuprates\cite{AndoMetalLSCO}. Thus
any microscopic models of charge ordering must simultaneously
explain and incorporate the existence of coherent nodal states and
broad antinodal excitations.

The nodal-antinodal dichotomy of quasiparticle dynamics in the
normal state also exists in Bi2212\cite{ShenSchrieffer}. A number
of possible mechanisms have been proposed to account for the
antinodal spectral broadening in the normal state. A prime
candidate is the ($\pi$,$\pi$) magnetic excitations observed in
various
cuprates\cite{ResonanceModeYBCO,ResonanceModeBi2212,RMHeTl2201}.
As schematically shown in Fig. \ref{LSCODichotomy}b, this
excitation will give rise to ``hot spots" on the Fermi surface
that can be connected by ($\pi$,$\pi$) momentum transfer.
Electrons around these hot spots experience additional scattering
from the ($\pi$,$\pi$) magnetic scattering. The same mechanism has
also been proposed for (Nd$_{2-x}$Ce$_x$)CuO$_4$ for which the
spectral broadening is localized to the expected ``hot
spot"\cite{NCCOHotSpot}.  However, in LSCO, the same magnetic
response, magnetic resonance mode, is not observed. Instead,
incommensurate magnetic peaks are observed at low energy (below 15
meV)\cite{TranquadaStripe}, which broaden rapidly with increasing
energy although the magnetic fluctuation can persist up to 280
meV\cite{HaydenPRL}. Intrigued by the fact that the extra
broadening sets in when the Fermi surface turns from the
($\pi$,0)-(0,$\pi$) diagonal direction to the (0,0)-($\pi$,0) or
the (0,0)-(0,$\pi$) direction (Fig. \ref{LSCODichotomy}b and c),
an alternative mechanism was proposed\cite{XJZhouDichotomy} in
which the scattering in question causes a pair of electrons on two
parallel antinodal segments to be scattered to the opposite ones
(Fig. \ref{LSCODichotomy}b).  In the normal state, this scattering
can cause a quasiparticle to decay into two quasiparticles and one
quasihole. The antinodal spectral broadening occurs as a result of
the frequent occurrence of such a decay which renders the normal
state quasiparticle ill defined.

Another potential explanation for the broad antinodal features may
come from models based on the polaron picture discussed
before\cite{KShenPolaron,AMtJPolaron,RoschLCO}. In such a scenario,
the strong coupling of the electrons to any bosonic excitations
would result in $Z \ll$1,  and spectral weight is transferred to
incoherent, multiboson excitations. An effective anisotropic
coupling could lead to a larger $Z$ (weaker coupling) along the
nodal direction and a much smaller, yet still finite $Z$, at the
antinodes (strong coupling). In this picture, the antinodal polaron
effect in LSCO (Fig. \ref{LSCODichotomy}c)\cite{XJZhouDichotomy} is
much weaker than Na-CCOC (Fig. \ref{CCOCDicho}b)\cite{KShenScience}
if one compares the spectral weight near Fermi level around the
antinodal region. Regardless of the microscopic explanation, the
broad and nested antinodal FS segments observed by ARPES are
consistent with the propensity for 2-dimensional charge ordering in
the lightly doped cuprates seen in STM experiments on
Na$-$CCOC\cite{CCOCSTM} and
Bi2212\cite{COBi2212,HoffmanVortex,McElroyUD}.  Furthermore, an
explanation based on an anisotropic coupling (coming from either
polaron physics or the magnetic resonance) may not be sufficient to
cause the 2-dimensional charge order; it may be a combination of
strong coupling and Fermi surface nesting which ultimately
stabilizes the antinodal charge-ordered state.

\subsection{Electron-Phonon Coupling and High Temperature  Superconductivity}

Much of the physics discussed in this review has attributed
essential features of the ARPES data to electron-phonon coupling,
and if not to electron-phonon coupling alone, to electron-phonon
coupling in an antiferromagnetic background.  The question remains
as to how this electron-phonon coupling can account for
high-temperature superconductivity with d-wave pairing seen in the
cuprates.  It is often assumed that el-ph coupling leads to s-wave
pairing, and that therefore such a mechanism contradicts with the
d-wave symmetry of the Cooper pairing in the cuprates. Instead,
electronic correlations have been thought to be consistent with
d-wave pairing.  However, while strong electronic correlations will
suppress the Cooper pair amplitude on the same orbital, and hence
induce a d-wave like symmetry,  they do not tell us much about the
explicit pairing mechanism.  One of the early studies on possible
phononic mechanisms of high T$_c$ superconductivity\cite{Nazarenko}
pointed out that the out-of-plane displacement of the oxygen, i.e.,
the buckling mode,  combined with antiferromagnetic correlations,
leads to $d_{x^2-y^2}$ pairing. Bulut and Scalapino \cite{Bulut}
studied the various phonon modes from the viewpoint of the possible
pairing force. They found that the interaction which becomes more
positive as the momentum transfer increases helps $d_{x^2-y^2}$
pairing (the case for buckling mode, but not the case for the apical
oxygen mode or the in-plane breathing mode).

One can understand the nature of the $q$ momentum dependence by
considering how the phonon couples to the electron density. For
deformation phonons, the coupling is dipolar driven and thus small
for small $q$, the case for the breathing modes. This also includes
infrared active phonons. Yet for Raman active modes, which couple
via the creation of isotropic and quadrupolar moments, the coupling
is generally strongest for small $q$. Specifically for the cuprates,
such strong $k,q$- dependencies occur explicitly for c-axis phonons,
which include the Raman active in-phase buckling $A_{1g}$,
out-of-phase buckling $B_{1g}$ and modes involving the apical oxygen
$A_{1g}$. The $k$ momentum dependence comes from the phonon
eigenvectors as well as the direction of charge-transfer induced by
the phonon. For example, for the $B_{1g}$ phonon the eigenvectors
enforce a change of sign when $k_x$ and $k_y$ are interchanged, a
factor $\sim$ $~\cos(k_{x}a)-\cos(k_{y}a)$, while for the apical
charge transfer coupling between Cu and the three oxygen orbitals, a
factor $\sim [\cos(k_{x}a)-\cos(k_{y}a)]^{2}$ emerges.

As discussed by Bulut and Scalapino\cite{Bulut}  among others, the
$q$ dependence of phonons can be important to give $d_{x^{2}-y^{2}}$
pairing. In particular, if the attractive electron-phonon
interaction falls off for momentum transfers $q$ along the diagonal,
then conceptually the interaction is of the same structure as the
magnetic pairing from antiferromagnetic spin fluctuations. This type
of structure occurs for both $B_{1g}$ and $A_{1g}$ c-axis
Raman-active phonons, and thus they contribute to the pairing
interaction in the d-wave channel, parameterized by $\lambda_{d}$
\begin{equation}
\lambda_{d}=\frac{2\sum_{k,k^{\prime}}d_{k}d_{k^{\prime}} \mid
g(k,k-k^{\prime})\mid^{2}\delta(\epsilon_{k})\delta(\epsilon_{k^{\prime}})}
{\Omega_{ph}\sum_{k}\delta(\epsilon_{k})d_{k}^{2}}
\end{equation}
with the d-wave form factor $d_{k}=[\cos(k_{x}a)-\cos(k_{y}a)]/2$.
However, the $A_{1g}$ phonons predominantly contribute to the
s-pairing channel (replacing $d_{k}$ by 1 in the above equation) in
the absence of any Coulomb interaction, leaving the $B_{1g}$ phonon
as the largest contributor to d-wave pairing, as found in LDA
studies\cite{OKA}.

However Coulomb interactions change this picture. They cannot be
neglected since they are necessary to screen the long-wavelength
nature of isotropic charge fluctuations. The screened
electron-phonon interaction $\bar g$ is of the form
\begin{equation}
\bar
g(k,q,\Omega)=g(k,q)+\frac{V(q)\Pi_{g,1}(q,\Omega)}{1-V(q)\Pi_{1,1}(q,\Omega)},
\end{equation}
where $V(q)=4\pi e^{2}/q^{2}$ is the 3D Coulomb interaction and
$\Pi_{a,b}(q,\Omega)$ is the frequency-dependent polarizability
calculated with vertices $a,b$ respectively. Note if $g$ were
independent of momentum, then the effective electron-phonon coupling
would be screened by the dielectric function
$\epsilon(q,\Omega)=1-V(q)\Pi_{1,1}(q,\Omega)$. Particularly in the
limit $q \rightarrow 0$ we recover complete screening and $\bar g=0$
for $\Omega=0$, restating particle number conservation, while for
$\Omega=\Omega_{ph}$ the renormalized coupling is of order
$\Omega_{ph}/\Omega_{pl}$. However, any fermion $k$-dependence of
the electron-phonon coupling survives screening {\it even at $q=0$}
as shown by Abrikosov and Genkin\cite{AG}, and the effective charge
vertex in this limit is $\bar g(k,q\rightarrow 0)=g(k,q \rightarrow
0)- \delta g$, with $\delta g=\langle g(k,q \rightarrow 0)\rangle$,
and $\langle \cdots \rangle$ denotes an average over the Fermi
surface, defined as
\begin{equation}
\langle A \rangle = \frac{\sum_k A(k)\delta(\epsilon(k))}{\sum_k
\delta(\epsilon(k))}.
\end{equation}
Thus screening removes the constant part of the electron-phonon
interaction and can highlight the d-wave channel. This is important
if the bare coupling is highly anisotropic with the Fermionic
momentum $k$, the case of the apical oxygen coupling.

Moreover, the issue of strong local correlations on electron-phonon
interactions has been recently readdressed by the Hubbard X operator
method \cite{Zeyher,Kulic} and quantum Monte Carlo simulations
\cite{Hanke}. Assuming no specific phonon and that phonons couple to
the on-site charge density, i.e., diagonal coupling, these works
found enhanced forward scattering (i.e., small momentum transfer),
while large momentum transfer process were suppressed. Therefore,
$d_{x^2-y^2}$ pairing can occur by el-ph coupling. Furthermore, the
vertex correction explains the absence of phonon features in the
resistivity, since the transport relaxation rate contains the factor
$1-\cos \theta$  ($\theta$: the angle between the initial and final
state momenta) which reduces the contribution for forward
scatterings.  There has been controversy as to whether the vertex
correction for the off-diagonal el-ph coupling, which modulates the
bond, also enhances forward scattering and suppresses large momentum
transfers\cite{ShenPM}.  The in-plane half-breathing mode, which
modulates the bond, exhibits a sharp softening with doping in
neutron scattering\cite{PintschoviusBreathingMode} and has been
studied in particular.  The  Zhang-Rice singlet couples to the
half-breathing mode much stronger than estimated in LDA
calculations, and the vertex correction leads to an effective
attractive interaction for $d_{x^2-y^2}$ pairing \cite{Ishihara}. On
the other hand, later analysis \cite{GunnarssonPRL} has shown that
the cancelation of terms reduces the off-diagonal coupling, and the
diagonal coupling dominates even after the vertex correction has
been taken into account.  In understanding the effects of this
vertex correction on experimental spectra, one should note that the
correction works differently for phononic and electronic
self-energies.  The sum rules\cite{GunnarssonSum} conclude that the
phononic self-energy is reduced by an additional factor of $x$ (hole
concentration) as compared to the electronic self-energy.
Intuitively, the difference between phononic and electronic
self-energies arises because a small number of holes cannot
influence phonons as much as phonons, in which atoms vibrate at
every site, can influence a single hole.

In summary, local coulomb repulsion suppresses charge density
modulations, which in turn decreases the strength of the
electron-phonon interaction at large momentum transfers. This has
two effects: first, as a consequence the contribution of {\it all}
phonons to the resistivity will be reduced by the correlation
effect. Second, and more relevant to pairing, small $q$ phonons will
have an accentuating $\lambda$ for d-wave pairing since the coupling
will decrease faster for large $q$ than without correlations. Thus
it appears that Coulomb interactions in general can have a dramatic
impact on electron-phonon driven $d_{x^{2}-y^{2}}$ pairing. However
theoretical developments are still needed in order to treat the
simultaneous importance of strong correlations and electron-phonon
coupling. This is a promising direction for future research.

\section{Summary}

ARPES experiments have been instrumental in identifying the
electronic structure, observing and detailing the electron-phonon
mode coupling behavior, and mapping the doping evolution of the
high-Tc cuprates. The spectra evolve from the strongly coupled,
polaronic spectra seen in underdoped cuprates to the
Migdal-Eliashberg like spectra seen in the optimally and overdoped
cuprates.  In addition to the marked doping dependence, the cuprates
exhibit pronounced anisotropy with direction in the Brillouin zone:
sharp quasiparticles along the nodal direction that broaden
significantly in the anti-nodal region of the underdoped cuprates,
an anisotropic electron-phonon coupling vertex for particular modes
identified in the optimal and overdoped compounds, and preferential
scattering across the two parallel pieces of Fermi surface in the
antinodal region for all doping levels.  This also contributes to
the pseudogap effect. To the extent that the Migdal-Eliashberg
picture applies, the spectra of the cuprates bear resemblance to
that seen in established strongly coupled electron-phonon
superconductors such as Pb.  On the other hand, the cuprates deviate
from this conventional picture. In the underdoped regime, the
carriers are best understood as small polarons in an
antiferromagnetic, highly electron correlated background, while the
doped compounds require an anisotropic electron-phonon vertex to
detail the prominent mode coupling signatures in the superconducting
state.  Electronic vertex corrections to the electron-phonon
coupling furthermore may enhance, and for certain phonons,
determine, the anisotropy of the electron-phonon coupling.  A
consistent picture emerges of the cuprates, combining strong,
anisotropic electron-phonon coupling, particular phonon modes that
could give rise to such a coupling, and an electron-electron
interaction modifying the el-ph vertex.  Such a combination, albeit
with further experimental and theoretical effort, may indeed lead to
an understanding of the high-critical transition temperature with
d-wave pairing in the cuprate superconductors.

\begin{acknowledgments}
We are grateful to A. S. Mishchenko, S. Ishihara, O. Gunnarsson, T.
Egami, J. Zaanen for discussions.

The work at the ALS and SSRL is supported by the DOE's Office of
BES, Division of Material Science, with contract
DE-FG03-01ER45929-A001 and DE-AC03-765F00515. The work at Stanford
was also supported by NSF grant DMR-0304981 and ONR grant
N00014-04-1-0048-P00002.  NN is supported by NAREGI project and
Grant-in-Aids from the Ministry of Education, Culture, Sports,
Science, and Technology.

\end{acknowledgments}


\end{document}